\documentclass[3p]{elsarticle}
\input{config.tex}
\begin{document}
\renewcommand*{\today}{January 03, 2020}
\begin{frontmatter}
\title{Contribution of high risk groups' unmet needs
may be underestimated in epidemic models without risk turnover:\linebreak[1]
a mechanistic modelling analysis\tnoteref{x}}
\journal{Epidemics}
\author[MAP]{Jesse~Knight}
\author[JHU]{Stefan~D.~Baral}
\author[JHU]{Sheree~Schwartz}
\author[MAP]{Linwei~Wang}
\author[MAP]{Huiting~Ma}
\author[THC]{Katherine~Young}
\author[THC]{Harry~Hausler}      
\author[MAP,IDM,IHP,IMS]{Sharmistha~Mishra\corref{c}}
\address[MAP]{MAP Centre for Urban Health Solutions, Unity Health Toronto}
\address[JHU]{Deptartment of Epidemiology, Johns Hopkins Bloomberg School of Public Health}
\address[THC]{TB HIV Care, South Africa}
\address[IDM]{Department of Medicine, Division of Infectious Disease, University of Toronto}
\address[IHP]{Institute of Health Policy, Management and Evaluation,
  Dalla Lana School of Public Health, University of Toronto}
\address[IMS]{Instituof Medical Sciences, University of Toronto}
\tnotetext[x]{On behalf of the Siyaphambili study team}
\cortext[c]{Corresponding author (\texttt{sharmistha.mishra@utoronto.ca})}
\begin{abstract}
\textsc{Background.}
Epidemic models of sexually transmitted infections (STIs)
are often used to characterize
the contribution of risk groups
to overall transmission
by projecting the transmission population attributable fraction (tPAF)
of unmet prevention and treatment needs within risk groups.
However, evidence suggests that
STI risk is dynamic over an individual's sexual life course,
which manifests as turnover between risk groups.
We sought to examine the mechanisms by which turnover
influences modelled projections of the tPAF of high risk groups.
\textsc{Methods.}
We developed a unifying, data-guided framework to simulate risk group turnover
in deterministic, compartmental transmission models.
We applied the framework to an illustrative model of an STI
and examined the mechanisms by which
risk group turnover influenced equilibrium prevalence across risk groups.
We then fit a model with and without turnover to
the same risk-stratified STI prevalence targets
and compared the inferred level of risk heterogeneity and
tPAF of the highest risk group projected by the two models.
\textsc{Results.}
The influence of turnover on group-specific prevalence
was mediated by three main phenomena:
movement of previously high risk individuals with the infection into lower risk groups;
changes to herd immunity in the highest risk group; and
changes in the number of partnerships where transmission can occur.
Faster turnover led to
a smaller ratio of STI prevalence between the highest and lowest risk groups.
Compared to the fitted model without turnover,
the fitted model with turnover inferred greater risk heterogeneity
and consistently projected a larger tPAF of the highest risk group over time.
\textsc{Implications.}
If turnover is not captured in epidemic models,
the projected contribution of high risk groups, and thus,
the potential impact of prioritizing interventions to address their needs, could be underestimated.
To aid the next generation of tPAF models,
data collection efforts to parameterize risk group turnover should be prioritized.
\end{abstract}
\begin{keyword}
  mathematical modelling \sep
  transmission \sep
  risk heterogeneity \sep
  turnover \sep
  sexually transmitted infection \sep
  population attributable fraction
\end{keyword}
\end{frontmatter}
\abbrfootnote{\textit{Abbreviations:}
  STI: sexually transmitted infection,
  HIV: human immunodeficiency virus,
  tPAF: transmission population attributable fraction}
\clearpage
\pagenumbering{gobble}
\section*{Highlights}
\begin{enumerate}
  \item A new framework for parameterizing turnover in risk groups is developed
  \item Mechanisms by which turnover influences STI prevalence in risk groups are examined
  \item Turnover reduces the ratio of equilibrium STI prevalence in high vs low risk groups
  \item Inferred risk heterogeneity is higher when fitting transmission models with turnover
  \item Ignoring turnover in risk could underestimate the tPAF of high risk groups
\end{enumerate}

\setcounter{tocdepth}{2}
\tableofcontents
\clearpage
\pagenumbering{arabic}
\setcounter{page}{1}
\section{Introduction}\label{s:intro}
Heterogeneity in transmission risk is a consistent characteristic of
epidemics of sexually transmitted infections (STI) \citep{Anderson1991}.
This heterogeneity is often demarcated by identifying
specific populations whose risks of acquisition and onward transmission of STI are the highest,
such that their specific unmet prevention and treatment needs
can sustain local epidemics of STI \citep{Yorke1978}.
Disproportionate risk can be conferred in several ways at the
individual-level (higher number of sexual partners), 
partnership-level (reduced condom use within specific partnership types), 
or structural-level (stigma as a barrier to accessing prevention and treatment services)
\citep{Baral2013}.
The contribution of high risk groups to the overall epidemic
can then be used as an indicator in the appraisal of STI epidemics,
helping to guide intervention priorities
\citep{Shubber2014,Mishra2016}.
\par
Traditionally, contribution to an epidemic was quantified using either:\ %
the classic \textit{population attributable fraction} (PAF)
via the relative risk of incident infections within a risk group
versus the rest of the population
and the relative size of the risk group \citep{Hanley2001};
or the distribution of new infections across subsets of a population
\citep{Case2012,Mishra2014}.
So when small risk groups experience disproportionately higher rate of
incident infections -- e.g. 5 percent of a population acquire 30 percent
of STI infections -- contribution is interpreted as 5 percent of the population contributing to
30 percent of all infections \citep{Pruss-Ustun2013}.
However, the classic PAF does not account for chains of (indirect) transmission, and has been
shown to underestimate the contribution of some higher-risk groups to cumulative
STI infections, especially over time \citep{Mishra2014}.
Thus, transmission models are increasingly being used to quantify
contribution by accounting for indirect transmission and projecting
the \textit{transmission population attributable fraction} (tPAF).
The tPAF is calculated by
simulating counterfactual scenarios where transmission
between specific subgroups is stopped, and
the relative difference in cumulative infections in the total population
over various time-periods is measured \citep{Mishra2014,Mukandavire2018}.
Transmission can be stopped by
setting susceptibility and/or infectiousness to zero in the model \citep{Mishra2014}.
The tPAF is then interpreted as
the fraction of all new infections that stem, directly and indirectly, from
a failure to prevent acquisition and/or to provide effective treatment
in a particular risk group \citep{Mishra2016,Mukandavire2018,Maheu-Giroux2017}.
\par
There is limited evidence on how model structure 
might influence the tPAF of higher risk groups
\citep{Mishra2016,Mukandavire2018,Maheu-Giroux2017},
especially movement of individuals between risk groups,
an epidemiologic phenomenon  that is well-described
in the context of sexual behaviour \citep{Watts2010}.
Such movement is often referred to in the STI epidemiology literature as
\textit{turnover} \citep{Watts2010}.
For example, turnover may reflect entry into or retirement from formal sex work,
or other periods associated with higher STI susceptibility and onward transmission
due to more partners and/or vulnerabilities
\citep{Marston2006,Watts2010}.
Risk group turnover has been shown to
influence the predicted equilibrium prevalence of an STI \citep{Stigum1994,Zhang2012};
the fraction of transmissions occurring during acute HIV infection \citep{Zhang2012};
the basic reproductive number $R_0$ \citep{Henry2015}; and
the coverage of antiretroviral therapy required to achieve HIV epidemic control \citep{Henry2015}.
Yet how, and the extent to which, turnover influences tPAF has yet to be examined.
\par
There is variability in how turnover has been previously implemented
\citep{Stigum1994,Koopman1997,Eaton2014,Boily2015},
in large part because of four main assumptions or epidemiologic
constraints surrounding movement between risk groups.
For example, in the context of turnover, the relative 
size of specific populations in the model 
may be constrained to remain constant over time
\citep{Stigum1994,Koopman1997,Eaton2014},
such as the proportion of individuals who sell sex.
Second, some individuals may enter into high risk groups at an early age,
and subsequently settle into lower risk groups;
thus the distribution of risks among individuals entering into the transmission model
may be assumed to be different from
the distribution of risks among individuals already in the transmission model
\citep{Eaton2014}.
Third, turnover may be constrained to reflect the average duration of time spent 
within a given risk group \citep{Boily2015},
such as duration engaged in formal sex work \citep{Watts2010}.
Finally, turnover could reflect data on how sexual behaviour changes
following exit from a given risk group \citep{Boily2015}.
Most prior models used some combination of these constraints,
based on their specific data or research question,
but to date there is no unified approach to modelling turnover.
\par
In this study, we explored the mechanisms by which turnover
may influence the tPAF of a high risk group
using an illustrative STI model
with treatment-induced immunity and without STI-attributable mortality.
First, we developed a unified approach to
implementing turnover based on epidemiologic constraints.
We then sought the following objectives:
1)~understand the mechanisms by which turnover
influences group-specific STI prevalence and ratios of prevalence between risk groups;
2)~examine how inclusion/exclusion of turnover in a model influences
the level of risk heterogeneity inferred during model fitting; and
3)~examine how inclusion/exclusion of turnover in a model influences
the projected tPAF of the highest risk group
after model fitting to a particular setting.
\section{Methods}\label{s:methods}
We developed a new, unified framework for implementing turnover. 
We then simulated a deterministic compartmental model of an illustrative STI,
with turnover as per the framework,
to conduct out experiments.
\subsection{A unified framework for implementing turnover}
\label{ss:framework}
\begin{figure}
  \centerline{\includegraphics[width=0.5\linewidth]{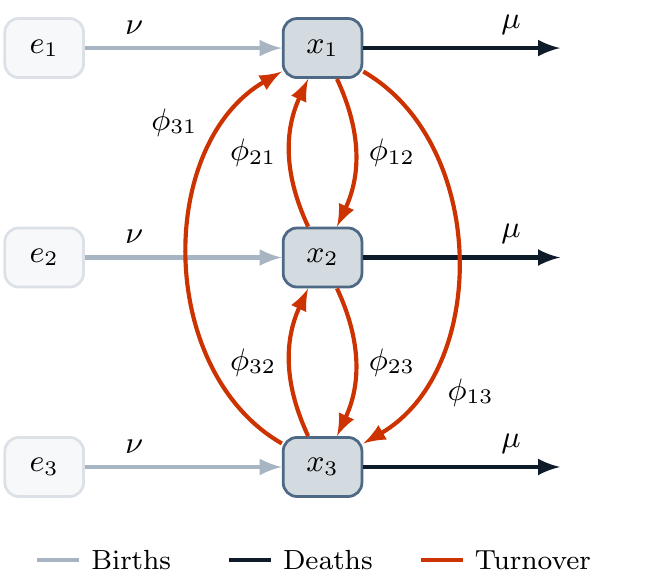}}
  \caption{System of $G = 3$ risk groups and turnover between them.}
  \footnotesize
  $x_i$: number of individuals in risk group~$i$;
  $e_i$: number of individuals available to enter risk group~$i$;
  $\nu$: rate of population entry;
  $\mu$: rate of population exit;
  $\phi_{ij}$: rate of turnover from group~$i$ to group~$j$.
  \label{fig:system}
\end{figure}
We developed a framework for implementing turnover,  
as depicted in Figure~\ref{fig:system}
and detailed in Appendix~\ref{a:framework}. 
In the framework, the simulated population is divided into $G$ risk groups.
The number of individuals in group $i \in [1, \dots, G]$ is denoted $x_i$,
and the relative size of each group is denoted $\hat{x}_i = x_i / N$,
where $N$ is the total population size.
Individuals enter the population at a rate $\nu$ and exit at a rate $\mu$ per year.
The distribution of risk groups at entry into the model
is denoted $\hat{e}_i$, which may be different from $\hat{x}_i$.
The total number of individuals entering group~$i$ per year
is therefore given by $\nu \hat{e}_i N$.
Turnover rates are collected in a $G \times G$ matrix $\phi$,
where $\phi_{ij}$ is the proportion of individuals in group~$i$
who move from group~$i$ into group~$j$ each year.
The framework is independent of the disease model,
and thus transition rates $\phi$ do not depend on health states.
\par
The framework assumes that:
1)~the relative sizes of risk groups
$\bm{\hat{x}} = [\hat{x}_1, \dots, \hat{x}_G]$
are known and should remain constant over time; and
2)~the rates of population entry $\nu$ and exit $\mu$
are known, but that they may vary over time.
An approach to estimate $\nu$ and $\mu$ is detailed in Appendix~\ref{aaa:params-nu-mu}.
The framework then provides a method to estimate
the values of the parameters $\bm{\hat{e}}$ and $\phi$,
representing $G$~and~$G(G-1) = G^2$ total unknowns.
In the framework,
$\bm{\hat{e}}$ and $\phi$ are collected in the vector
$\bm{\theta} = \left[\bm{\hat{e}}, \bm{y}\right]$,
where $\bm{y} = \mathrm{vec}_{i \ne j}(\phi)$.
To uniquely determine the elements of $\bm{\theta}$,
a set of linear constraints are constructed.
Each constraint $k$ takes the form
$b_k = A_k \bm{\theta}$,
where $b_k$ is a constant and $A_k$ is a vector with the same length as $\bm{\theta}$.
The values of $\bm{\theta}$ are then obtained by solving:
\begin{equation}\label{eq:system-matrix}
\bm{b} = A \thinspace \bm{\theta}
\end{equation}
using existing algorithms for solving linear systems~\citep{LAPACK}.
\par
The framework defines four types of constraints, which are based on assumptions,
that can used to solve for the values of
$\bm{\hat{e}}$~and~$\phi$ via~$\bm{\theta}$.
The frameworks is flexible with respect to
selecting and combining these constraints,
guided by the availability of data.
However, exactly $G^2$ non-redundant constraints must be specified
to produce a unique solution,
such that exactly one value of $\bm{\theta}$ satisfies all constraints.
Table~\ref{tab:constraints} summarizes
the four types of constraints,
with their underlying assumptions,
and the types of data that can be used in each case.
Additional details, including
constraint equations, examples, and considerations for combining constraints,
are in Appendix~\ref{aaa:params-turnover}.
\begin{table}
  \caption{Summary of constraint types for defining risk group turnover}
  \label{tab:constraints}
  \centerline{
\footnotesize
\setlength{\tabcolsep}{3pt}
\begin{tabular}{llcl}
  \toprule
  Constraint & Assumption & Parameters & Types of data sources for parameterization \\
  \midrule
  1.~Constant group size
  & \cellbox{0.3\linewidth}{
    the relative population sizes of groups are known or assumed,
    and assumed to not change over time}
  & $\hat{x}_i$
  & \cellbox{0.36\linewidth}{
    demographic health surveys \citep{DHS},
    key population mapping and enumeration \citep{Abdul-Quader2014}}\\
  2.~Specified elements
  & \cellbox{0.3\linewidth}{
    the relative numbers of people entering into each group
    upon entry into the model or after leaving another group
    are known or assumed}
  & $\hat{e}_i$, $\phi_{ij}$
  & \cellbox{0.36\linewidth}{
    demographic health surveys \citep{DHS},
    key population surveys \citep{Baral2014}}\\
  3.~Group duration
  & \cellbox{0.3\linewidth}{
    the average durations of individuals in each group
    are known or assumed}
  & $\delta_i$
  & \cellbox{0.36\linewidth}{
    cohort studies of sexual behaviour over time \citep{Fergus2007},
    key population surveys \citep{Watts2010,Baral2014}}\\
  4.~Turnover rate ratios
  & \cellbox{0.3\linewidth}{
    ratios between different rates of turnover are known or assumed}
  & $\phi_{ij}$
  & \cellbox{0.36\linewidth}{
    demographic health surveys \citep{DHS},
    key population surveys \citep{Baral2014}}\\
  \bottomrule
\end{tabular}}\null
\footnotesize
$\phi_{ij}$:~rate of turnover from group~$i$ to group~$j$;
$\hat{x}_i$:~proportion of individuals in risk group~$i$;
$\hat{e}_i$:~proportion of individuals entering into risk group~$i$;
$\delta_i$:~average duration spent in risk group~$i$.
\end{table}
\subsection{Transmission model}\label{ss:model-sim}
We developed a deterministic, compartmental model of an illustrative
sexually transmitted infection with 3 risk groups.
We did not simulate a specific pathogen, but rather constructed a biological system
that included susceptible, infectious, and treated (or recovered/immune) health states.
The transmission model therefore was mechanistically representative of
sexually transmitted infections like
HIV, where effective antiretroviral treatment represents a health state where
individuals are no longer susceptible nor infectious~\citep{Maartens2014}, or
hepatitis B virus, where a large proportion of individuals who clear their acute infection
develop life-long protective immunity~\citep{Ganem2004}.
\par
\begin{figure}
  \centerline{\includegraphics[width=0.4\linewidth]{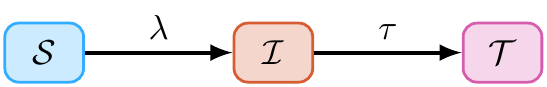}}
  \caption{Modelled health states:
    $\mathcal{S}$: susceptible;
    $\mathcal{I}$: infected;
    $\mathcal{T}$: treated;
    and transitions:
    $\lambda$: force of infection;
    $\tau$: treatment.}
  \label{fig:health-states}
\end{figure}

The model is represented by a set of coupled ordinary differential equations
(Appendix~\ref{aa:eqs-model}) and includes
three health states:
susceptible~$\mathcal{S}$, infectious~$\mathcal{I}$, and treated~$\mathcal{T}$
(Figure~\ref{fig:health-states}),
and $G = 3$ levels of risk:
high~$H$, medium~$M$, and low~$L$.
Risk strata are defined by different number of partners per year,
so that individuals in risk group~$i$ are assumed to
form partnerships at a rate $C_{i}$ per year.
The probability of partnership formation $\rho_{ik}$ between individuals in group~$i$
and individuals in risk group~$k$ is assumed to be
proportionate to the total number of available partnerships within each group%
~\cite{Garnett1994}:
\begin{equation}
\rho_{ik} = \frac
{C_k x_k}
{\sum_{\mathrm{k}}C_{\mathrm{k}} x_{\mathrm{k}}}
\label{eq:rho}
\end{equation}
\par
The biological probability of transmission is defined as $\beta$ per partnership.
Individuals transition from the
susceptible $\mathcal{S}$ to infectious $\mathcal{I}$ health state
via a force of infection $\lambda_i$ per year, per susceptible in risk group~$i$:
\begin{equation}
\lambda_{i} =
C_{i} \sum_k \rho_{ik} \thinspace  \beta \thinspace \frac{\mathcal{I}_k}{x_k}
\label{eq:foi}
\end{equation}
Individuals are assumed to transition from the
infectious $\mathcal{I}$ to treated $\mathcal{T}$ health state
at a rate $\tau$ per year, reflecting diagnosis and treatment.
The treatment rate does not vary by risk group.
Individuals in the treated $\mathcal{T}$ health state are neither infectious nor susceptible,
and individuals cannot become re-infected.
\subsubsection{Implementing turnover within the transmission model}
\label{sss:turnover-implemented}
As described in Section~\ref{ss:framework}, individuals
enter the model at a rate $\nu$,
exit the model at a rate $\mu$,
and transition from risk group~$i$ to group~$j$ at a rate $\phi_{ij}$,
 health state.
The turnover rates $\phi$ and
distribution of individuals entering the model by risk group $\bm{\hat{e}}$
were computed using the methods outlined in
Appendix~\ref{aaa:params-turnover}, based on the following three assumptions.
First, we assumed that
the proportion of individuals entering each risk group $\bm{\hat{e}}$
was equal to the proportion of individuals across risk groups in the model $\bm{\hat{x}}$.
Second, we assumed that
the average duration of time spent in each risk group $\bm{\delta}$ was known.
Third, we assumed that the absolute number of individuals
moving between two risk groups in either direction was balanced,
meaning that if 10 individuals moved from group~$i$ to group~$j$,
then another 10 individuals moved from group~$j$ to group~$i$.
These three assumptions were selected
because they reflect the common assumptions underlying turnover in prior models
\citep{Zhang2012,Henry2015}
and also to avoid any dominant direction of turnover.
That is, we wanted to study
the influence of movement between risk groups in general,
as compared to no movement, and at various rates of movement,
rather than movement predominantly from some groups to some other groups.
The system of equations formulated from the above assumptions and constraints
is given in Appendix~\ref{aa:eqs-turnover}.
To satisfy all three assumptions, there was only one possible value
for each element in $\phi$~and~$\bm{\hat{e}}$.
That is, by specifying these three assumptions,
we generated a unique set of $\phi$~and~$\bm{\hat{e}}$.
\par
Under the above three assumptions,
we still needed to specify the particular values of the parameters
$\bm{\hat{x}}$,~$\bm{\delta}$,~$\nu$,~and~$\mu$.
Such parameter values could be derived from data as described in Appendix~\ref{aaa:params-turnover}.
However, in all our experiments, we used the illustrative values summarized in
Table~\ref{tab:params}.
After resolving the system of equations Eq.~(\ref{eq:system-matrix}) using these values,
$\bm{\hat{e}}$ was equal to $\bm{\hat{x}}$ (assumed), and $\phi$ was:
\begin{equation}
\label{eq:phi-values}
\phi = \left[\begin{array}{ccc}
* & 0.0833 & 0.0867\\
0.0208 & * & 0.0158\\
0.0058 & 0.0042 & *
\end{array}\right]

\end{equation}
\begin{table}
  \centering
  \caption{Default model parameters for experiments}
  \label{tab:params}
  \centerline{
\footnotesize
\begin{tabular}{clc}
	\toprule
	    Symbol     & Description                                                     &        Default value         \\
	\midrule
	 $\bm{\beta}$  & transmission probability per partnership                        &            $0.03$            \\
	    $\tau$     & rate of treatment initiation among infected                     &            $0.1$             \\
	    $N_0$      & initial population size                                         &            $1000$            \\
	\midrule
	$\bm{\hat{x}}$ & proportion of system individuals by risk group                  & $[ 0.05 \es 0.20 \es 0.75 ]$ \\
	$\bm{\hat{e}}$ & proportion of entering individuals risk by risk group           & $[ 0.05 \es 0.20 \es 0.75 ]$ \\
	$\bm{\delta}$  & average duration spent in each risk group                       &    $[ 5 \es 15 \es 25 ]$     \\
	     $C$       & number of partners per year by individuals in each risk group   &     $[ 25 \es 5 \es 1 ]$     \\
	    $\nu$      & rate of population entry                                        &            $0.05$            \\
	    $\mu$      & rate of population exit                                         &            $0.03$            \\
	\bottomrule
\end{tabular}}\null
\footnotesize
All rates have units $\mathrm{year}^{-1}$; durations are in $\mathrm{years}$;
parameters stratified by risk group are written [high, medium, low] risk.
\end{table}
We then simulated epidemics using $\phi$ above
and the parameters shown in Table~\ref{tab:params}.
The transmission model was initialized with $N_0 = 1000$ individuals
who were distributed across risk groups according to $\bm{\hat{x}}$.
We seeded the epidemic with
one infectious individual in each risk group at $t = 0$ in an otherwise 
fully susceptible populatuon.
We numerically solved the system of ordinary differential equations
(Appendix~\ref{aa:eqs-model}) in Python
using Euler's method with a time step of $dt = 0.1$ years.
Code for all aspects of the project is available at:
\href{https://github.com/mishra-lab/turnover}{\small\texttt{https://github.com/mishra-lab/turnover}}.
\subsection{Experiments}
\label{ss:exp}
We designed three experiments to examine the influence of turnover on simulated epidemics.
We analyzed all outcomes at equilibrium,
defined as steady state at $t = 500$ years
with $<1\%$ change in incidence per year.
\subsubsection{Experiment~1: Mechanisms by which turnover influences equilibrium prevalence}
\label{sss:exp-prevalence}
We designed Experiment~1 to explore the mechanisms by which turnover influences
the equilibrium STI prevalence of infection, and
the ratio of prevalence between risk groups (prevalence ratios).
We defined prevalence as $\hat{\mathcal{I}}_i = \dfrac{\mathcal{I}_i}{x_i}$.
Similar to previous studies \citep{Zhang2012,Henry2015},
we varied the rates of turnover using a single parameter.
However, because our model had $G = 3$ risk groups,
multiplying a set of base rates $\phi$ by a scalar factor
would change the relative population sizes of risk groups~$\bm{\hat{x}}$.
Instead of a scalar factor, we controlled the rates of turnover using
the duration of time spent in the high risk group~$\delta_H$,
because of the practical interpretation of $\delta_H$
in the context of STI transmission,
such as the duration in formal sex work \citep{Watts2010}.
A shorter $\delta_H$ yielded faster rates of turnover among all groups.
The duration of time spent in the medium risk group $\delta_M$
was then defined as a value between $\delta_H$ and the maximum duration $\mu^{-1}$
which scaled with $\delta_H$ following:
$\delta_M = \delta_H + \kappa \left(\mu^{-1} - \delta_H\right)$, with $\kappa = 0.3$.
The duration of time in the low risk group $\delta_L$
similarly scaled with $\delta_H$,
but due to existing constraints,
specification of $\delta_H$ and $\delta_M$
ensured only one possible value of $\delta_L$.
Thus, each value of $\delta_H$ yielded a unique set of turnover rates~$\phi$
whose elements all scaled inversely with
the duration in the high risk group~$\delta_H$.
\par
\begin{figure}
  \centerline{\includegraphics[width=0.45\linewidth]{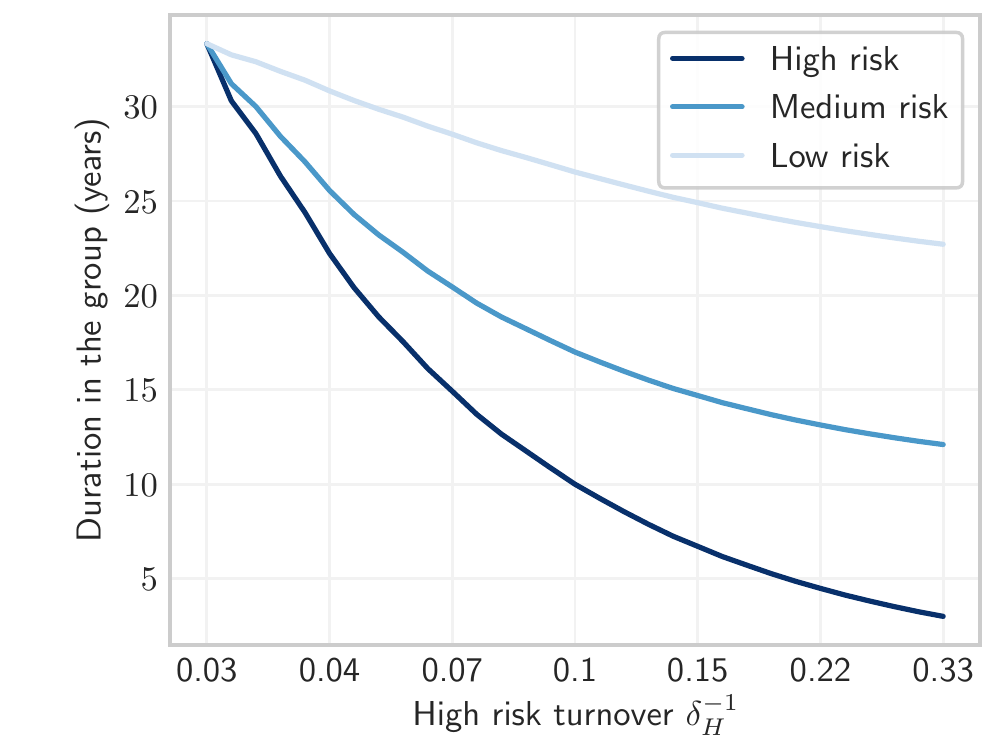}}
  \caption{Average duration of time spent in each risk group versus turnover.}
  \label{fig:dur-group}
  \footnotesize\input{x-axis.tex}
\end{figure}

We varied $\delta_H$ across a range of 33~to~3 years,
reflecting a range from
the full duration of simulated sexual activity $\mu^{-1} \approx 33$ years,
through an average reported duration in sex work as low as 3 years
\citep{Watts2010}.
The resulting duration of time spent in each group
versus turnover in the high risk group $\delta_H^{-1}$
is shown in Figure~\ref{fig:dur-group}.
For each set of turnover rates,
we plotted the equilibrium prevalence in each risk group,
and the prevalence ratios between high/low, high/medium, and medium/low risk groups.
In order to understand the mechanisms by which
turnover influenced prevalence and prevalence ratios (Objective~1),
we additionally plotted the four components which contributed to
gain/loss of infectious individuals in each risk group,
based on Eq.~(\ref{eq:model-I}):
1)~net gain/loss via turnover of infectious individuals,
2)~gain via incident infections,
3)~loss via treatment, and
4)~loss via death.
The influence of turnover on prevalence
was only mediated by components 1~and~2,
since components 3~and~4 were defined as
constant rates which did not change with turnover;
as such, our analysis focused on components 1~and~2.
Finally, to further understand trends in incident infections versus turnover (component~1),
we factored equation Eq.~(\ref{eq:foi}) for incidence $\lambda_i$
into constant and non-constant factors,
and plotted the non-constant factors versus turnover.
\subsubsection{Experiment~2: Inferred risk heterogeneity with vs without turnover}
\label{sss:exp-infer}
We designed Experiment~2 to examine how
the inclusion versus exclusion of turnover influences
the inference of transmission model parameters related to risk heterogeneity,
specifically the numbers of partners per year $C_i$ across risk groups.
The ratio of partner numbers $C_H~/~C_L$
is one way to measure of how different the two risk groups are
with respect to acquisition and transmission risks.
Indeed, ratios of partner numbers are often used when parameterizing 
risk heterogeneity in STI transmission models \citep{Mishra2012}.
\par
First, we fit the transmission model with turnover and without turnover,
to equilibrium infection prevalence across risk groups.
Specifically, we held all other parameters at their default values and
fit the numbers of partners per year in each risk group $C_i$
to reproduce the following:
20\% infection prevalence among the high risk group,
8.75\% among the medium risk group,
3\% among the low risk group,
and 5\% overall.
To identify the set of parameters (i.e.\ partner numbers $C$ in each risk group)
that best reproduced the fitting targets, we minimized
the negative log-likelihood of group-specific and overall prevalence.
Sample sizes of 500, 2000, 7500, and 10,000 were assumed to generate binomial distributions
for the high, medium, low, and overall prevalence targets respectively,
reflecting typical sample sizes in
nationally representative demographic and health surveys \citep{DHS},
multiplied by the relative sizes of risk groups in the model~$\bm{\hat{x}}$.
The minimization was performed using
the SLSQP method~\citep{Kraft1988} from the SciPy Python
\href{https://docs.scipy.org/doc/scipy/reference/generated/scipy.optimize.minimize.html}
{\texttt{minimize}} package.
To address Objective~2, we compared
the fitted (posterior) ratio of partners per year $C_H~/~C_L$
in the model with turnover versus the model without turnover.
\subsubsection{Experiment~3: Influence of turnover on the tPAF of the high risk group}
\label{sss:exp-tpaf}
We designed Experiment~3 to examine how the tPAF of the
high risk group varies when projected
by a model with versus without turnover (Objective~3).
We calculated the tPAF of risk group~$i$ by comparing
the relative difference in cumulative incidence between
a base scenario, and a counterfactual where transmission from group~$i$ is turned off,
starting at the fitted equilibrium.
That is, in the counterfactual scenario,
infected individuals in the high risk group could not transmit the infection.
The tPAF was calculated over different time-horizons (1~to~50 years) as
\citep{Mishra2014}:
\begin{equation}
\textrm{tPAF}_i(t) =
  \frac{\displaystyle\int_{t_{eq}}^{t} I_b(\tau) \, d\tau -
        \displaystyle\int_{t_{eq}}^{t} I_c(\tau) \, d\tau}
       {\displaystyle\int_{t_{eq}}^{t} I_b(\tau) \, d\tau}
\end{equation}
where $t_{eq}$ is the time corresponding to equilibrium,
$I_b(t)$ is the rate of new infections at time $t$ in the base scenario,
and $I_c(t)$ is the rate of new infections at time $t$ in the counterfactual scenario.
We then compared the tPAF generated from the fitted model with turnover
to the tPAF generated from the fitted model without turnover.

\section{Results}\label{s:results}
\subsection{Experiment~1: Mechanisms by which turnover influences equilibrium prevalence}
\label{ss:res-prevalence}
\begin{figure}
  \begingroup\centering
  \begin{subfigure}{0.33\linewidth}
    \includegraphics[width=\linewidth]{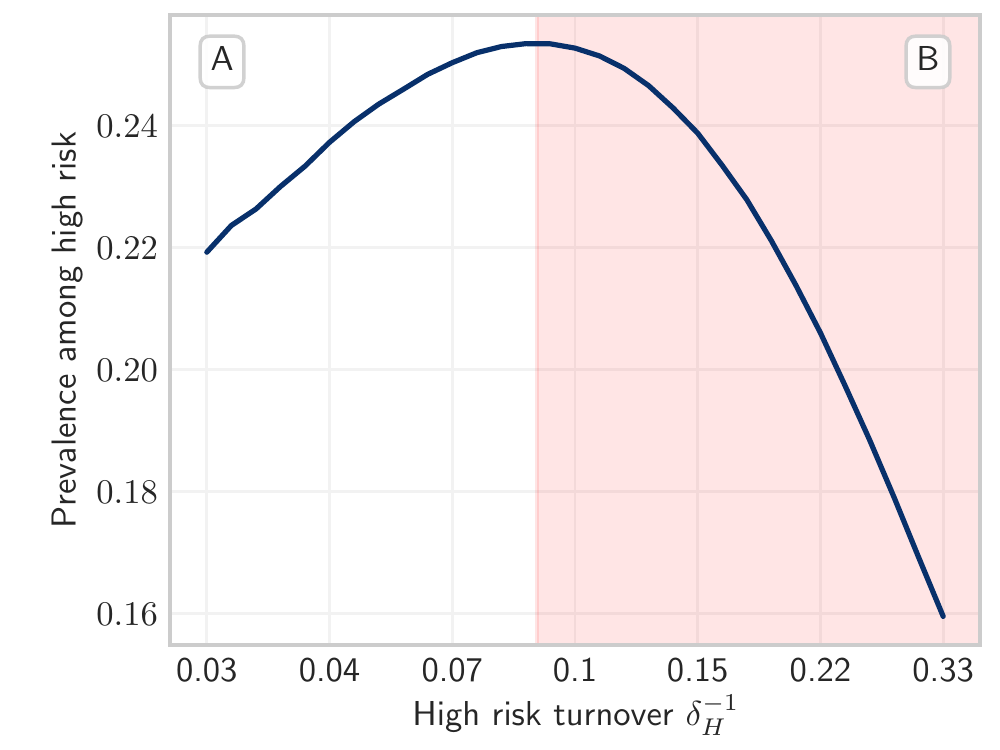}
    \caption{High risk}
    \label{fig:prevalence-high}
  \end{subfigure}
  \begin{subfigure}{0.33\linewidth}
    \includegraphics[width=\linewidth]{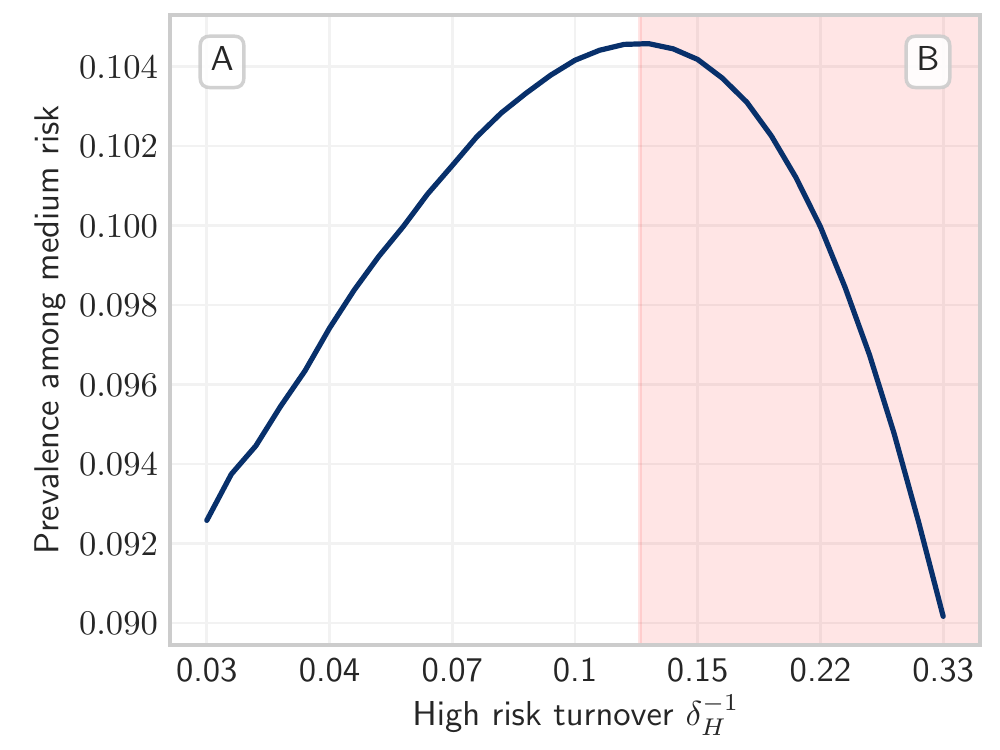}
    \caption{Medium risk}
    \label{fig:prevalence-med}
  \end{subfigure}
  \begin{subfigure}{0.33\linewidth}
    \includegraphics[width=\linewidth]{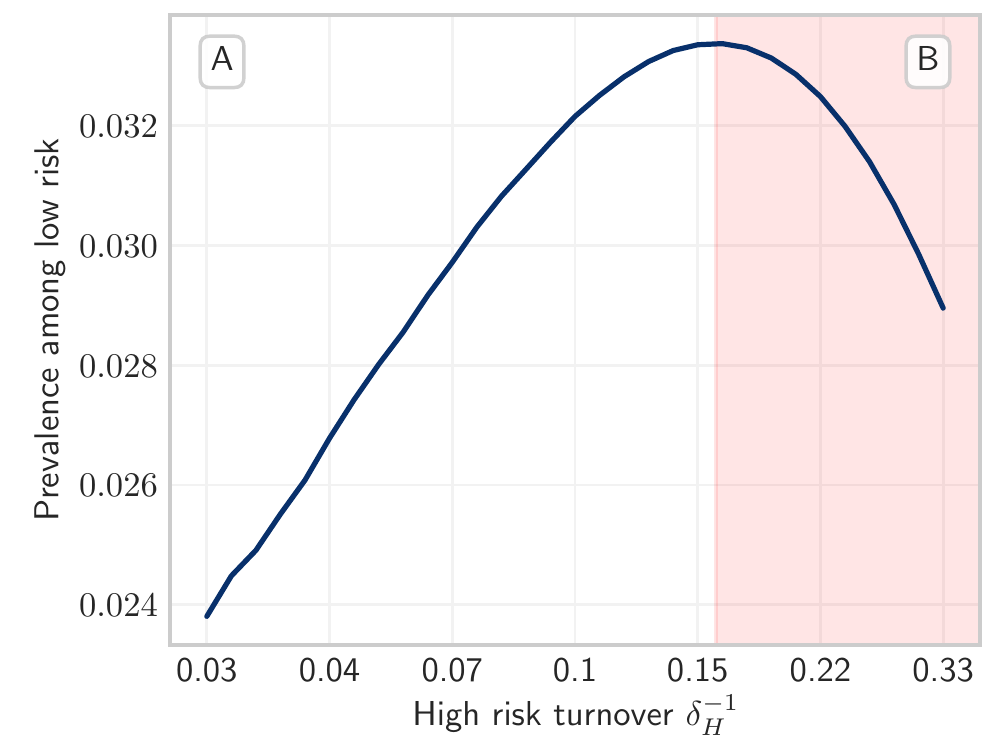}
    \caption{Low risk}
    \label{fig:prevalence-low}
  \end{subfigure}\\
  \endgroup
  \caption{Relationship between equilibrium STI prevalence
    in high, medium, and low risk groups versus turnover rate.
    Regions A~and~B denote where equilibrium prevalence is increasing and decreasing
    with different rates of turnover, respectively.}
  \label{fig:prevalence}
  \footnotesize\input{x-axis.tex}
\end{figure}

Figure~\ref{fig:prevalence} shows the trends in equilibrium STI prevalence among the
high~(\subref{fig:prevalence-high}),
medium~(\subref{fig:prevalence-med}), and
low~(\subref{fig:prevalence-low})
risk groups, at different rates of turnover
which are depicted on the x-axis,
based on duration of time spent in the high risk group.
Figure~\ref{fig:prevalence} reveals an inverted U-shaped relationship
between STI prevalence and turnover in all three risk groups.
That is, equilibrium STI prevalence was higher in systems with slow turnover
versus those with no turnover
(Figure~\ref{fig:prevalence}, region~A).
Equilibrium STI prevalence then peaked at slightly faster turnover
before declining in systems with even faster turnover
(region~B in Figure~\ref{fig:prevalence}).
Comparison of group-specific prevalence in Figure~\ref{fig:prevalence} shows that
the threshold turnover rate at which group-specific prevalence peaked varied by risk group:
prevalence in the high risk group peaked at the lower turnover threshold
(Figure~\ref{fig:prevalence-high}),
while prevalence in low risk group peaked at a higher turnover threshold
(Figure~\ref{fig:prevalence-low}).
To explain the inverted U-shape and different turnover thresholds by group,
we examined the components contributing to prevalence,
first in the high risk group, and then in the low risk group.
\begin{figure}
  \begingroup\centering
  \begin{subfigure}{0.33\linewidth}
    \includegraphics[width=\linewidth]{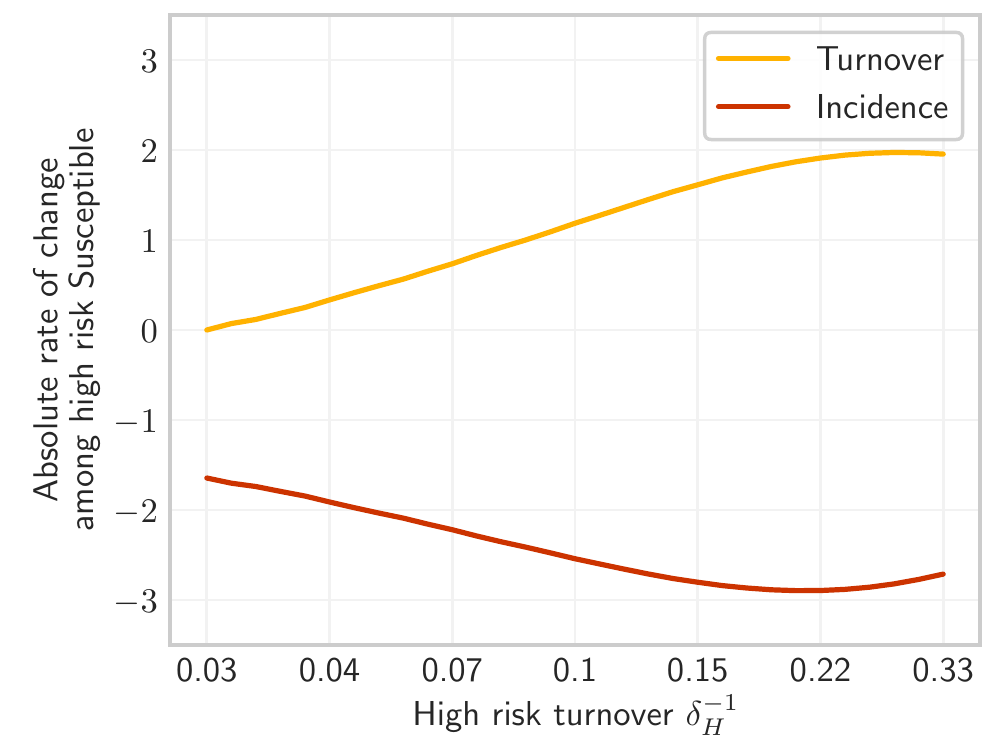}
    \caption{High risk susceptible}
    \label{fig:dX-high-S}
  \end{subfigure}%
  \begin{subfigure}{0.33\linewidth}
    \includegraphics[width=\linewidth]{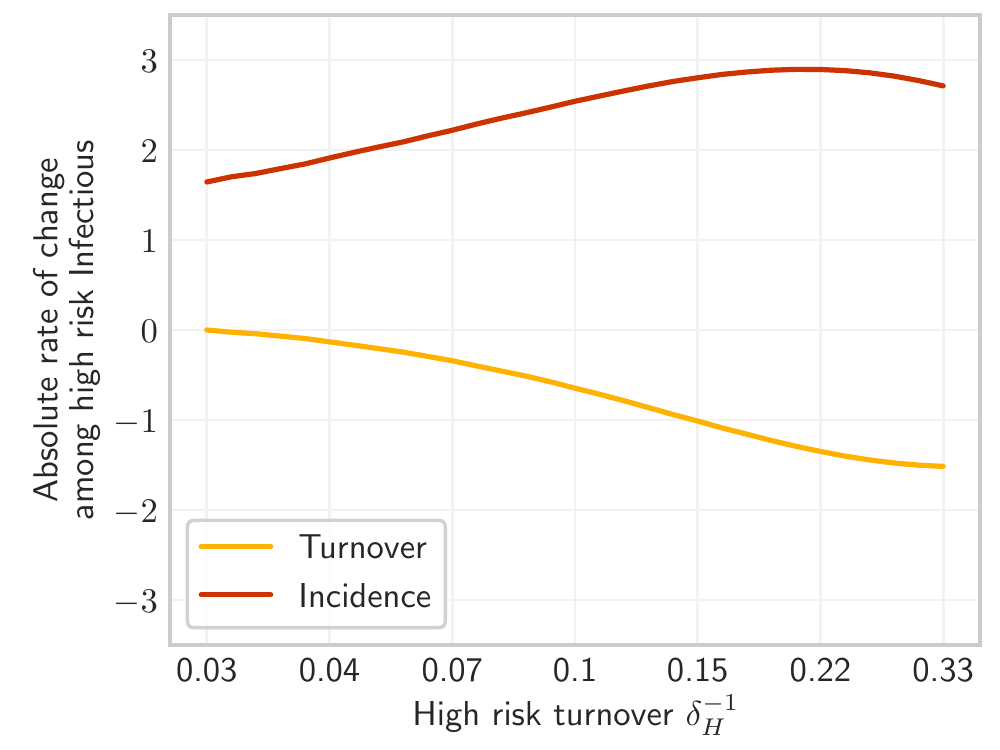}
    \caption{High risk infectious}
    \label{fig:dX-high-I}
  \end{subfigure}%
  \begin{subfigure}{0.33\linewidth}
    \includegraphics[width=\linewidth]{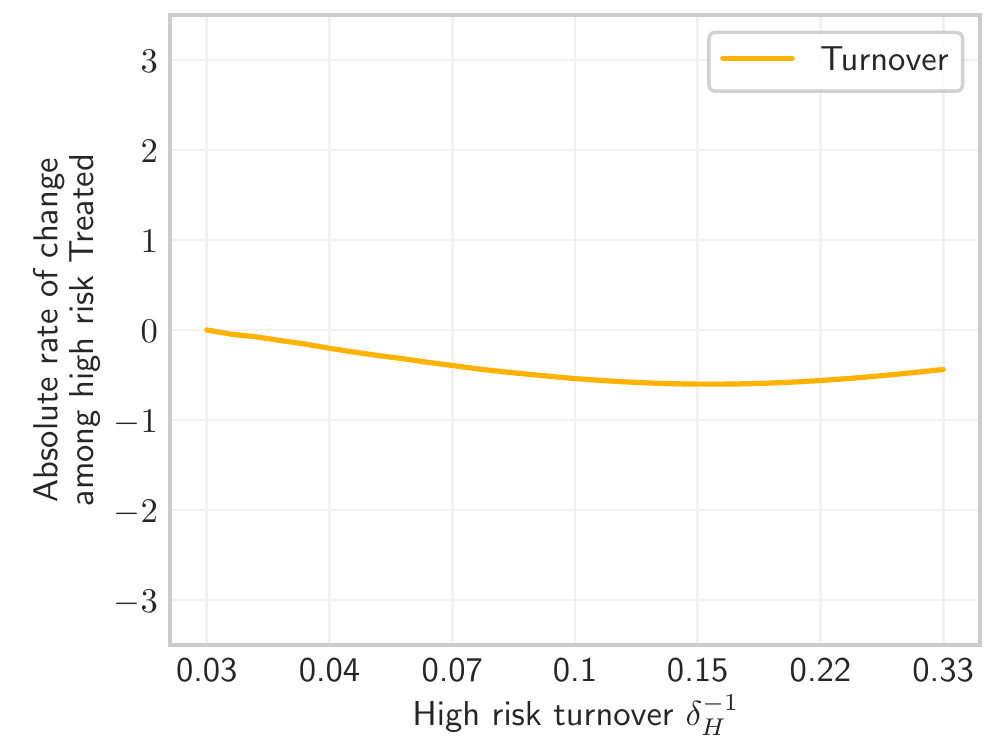}
    \caption{High risk treated}
    \label{fig:dX-high-T}
  \end{subfigure}\\
  \begin{subfigure}{0.33\linewidth}
    \includegraphics[width=\linewidth]{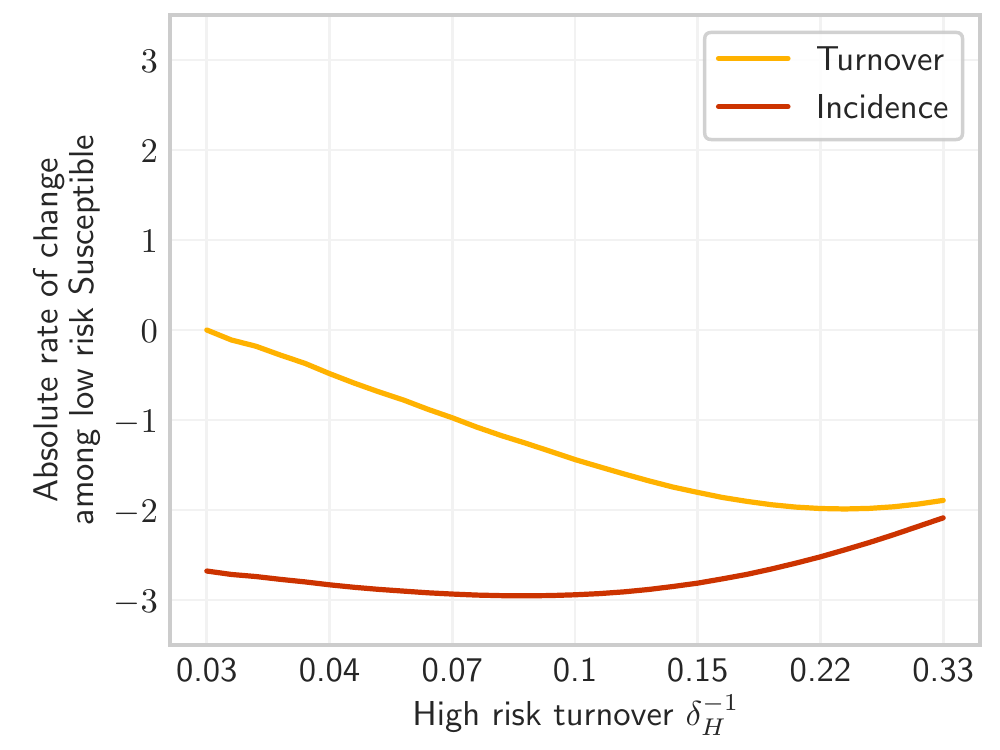}
    \caption{Low risk susceptible}
    \label{fig:dX-low-S}
  \end{subfigure}%
  \begin{subfigure}{0.33\linewidth}
    \includegraphics[width=\linewidth]{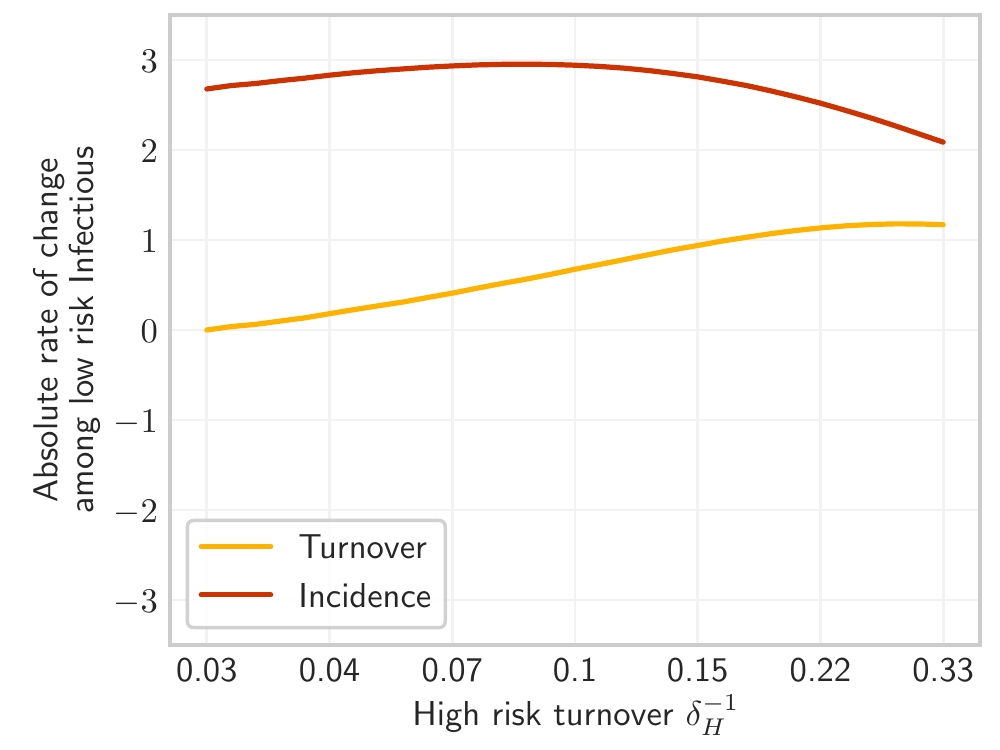}
    \caption{Low risk infectious}
    \label{fig:dX-low-I}
  \end{subfigure}%
  \begin{subfigure}{0.33\linewidth}
    \includegraphics[width=\linewidth]{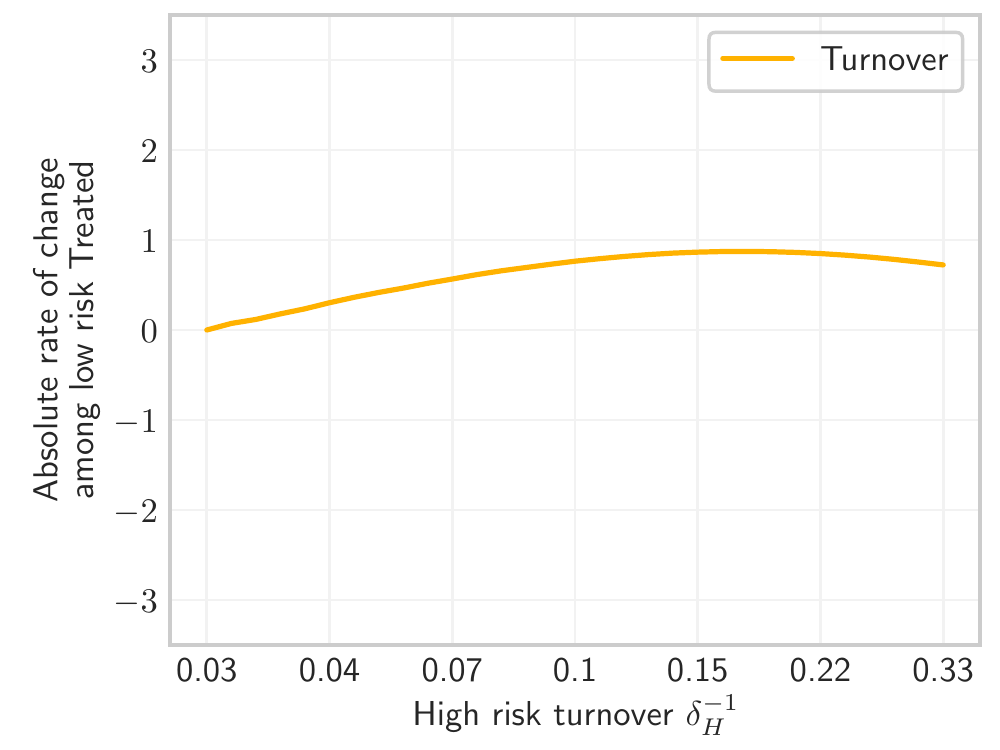}
    \caption{Low risk treated}
    \label{fig:dX-low-T}
  \end{subfigure}\\
  \endgroup
  \caption{Absolute rates of change at equilibrium
    (number of individuals gained/lost per year)
    among individuals in each health state and risk group, due to:
    loss/gain via turnover (yellow), and
    loss/gain via incident infections (red).
    Based on Eq.~(\ref{eq:model}).
    See Figure~\ref{fig:dX-app} for all derivative components.}
  \label{fig:dX}
  \footnotesize\input{x-axis.tex}
\end{figure}
\begin{figure}[!tbp]
  \begingroup\centering
  \includegraphics[width=0.6\linewidth]{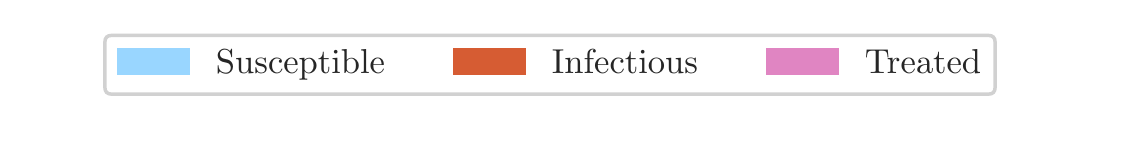}\\
  \includegraphics[width=0.11\linewidth]{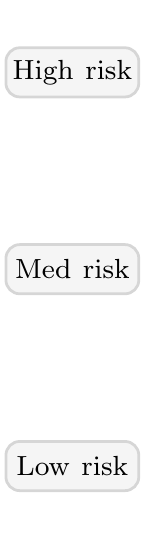}
  \begin{subfigure}[t]{0.22\linewidth}
    \includegraphics[width=\linewidth]{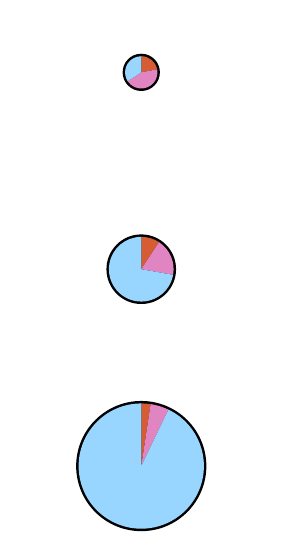}
    \caption{No turnover}
    \centering\footnotesize $\delta_H^{-1} = \input{data/flows-none-phi.tex}$
    \label{fig:flows-none}
  \end{subfigure}%
  \begin{subfigure}[t]{0.22\linewidth}
    \includegraphics[width=\linewidth]{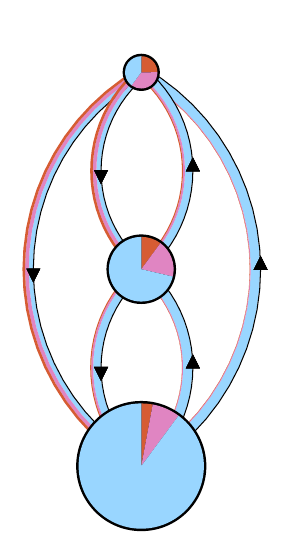}
    \caption{Slow turnover}
    \centering\footnotesize $\delta_H^{-1} = \input{data/flows-low-phi.tex}$
    \label{fig:flows-low}
  \end{subfigure}%
  \begin{subfigure}[t]{0.22\linewidth}
    \includegraphics[width=\linewidth]{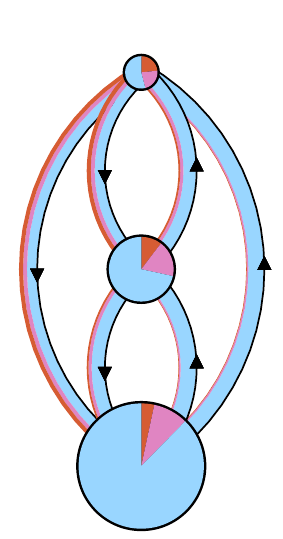}
    \caption{Fast turnover}
    \centering\footnotesize $\delta_H^{-1} = \input{data/flows-high-phi.tex}$
    \label{fig:flows-high}
  \end{subfigure}%
  \begin{subfigure}[t]{0.22\linewidth}
    \includegraphics[width=\linewidth]{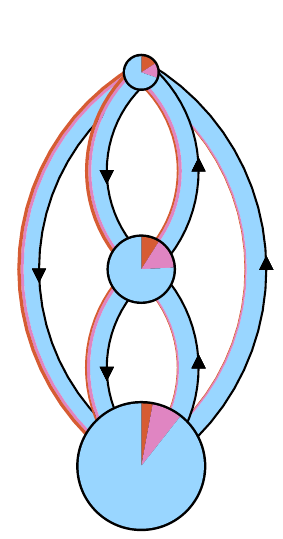}
    \caption{Very fast turnover}
    \centering\footnotesize $\delta_H^{-1} = \input{data/flows-extr-phi.tex}$
    \label{fig:flows-extr}
  \end{subfigure}\\[0.5em]
  $\null\hspace{0.11\linewidth}
  \underbrace{\hspace{0.44\linewidth}}_{\textrm{Region~A}}
  \underbrace{\hspace{0.44\linewidth}}_{\textrm{Region~B}}$
  \endgroup
  \caption{Depiction of health states of individuals in each risk group
    and of individuals moving between risk groups,
    obtained from models at equilibrium
    under four overall rates of turnover.}
  \footnotesize
  Circle sizes are proportional to risk group sizes.
  Circle slices and arrow widths are also proportional to
  the proportion of health states within risk groups and
  among individuals moving between risk groups, respectively.
  However, circle sizes and arrow widths do not have comparable scales.
  Appendix Figure~\ref{fig:1d-health} illustrates
  proportions of health states versus turnover in full.
  \label{fig:flows}
\end{figure}
\par
Figure~\ref{fig:dX} shows the yearly
gain/loss of individuals via turnover, and
gain/loss via incident infections,
in each health state and risk group,
at equilibrium under different rates of turnover.
Figure~\ref{fig:flows} also illustrates
the distribution of health states in each risk group
and among individuals moving between risk groups
under four different rates of turnover.
\subsubsection{Influence of turnover on equilibrium prevalence in the high risk group}
\label{sss:res-prev-high}
As shown in Figure~\ref{fig:flows}, at all four rates of turnover
the proportion of individuals who were in the infectious state (STI prevalence)
was largest in the high risk group.
As infectious individuals left the high risk group via turnover,
they were largely replaced by susceptible individuals from lower risk groups
(Figure~\ref{fig:flows-low} and
Figures~\ref{fig:dX-high-S}~vs~\ref{fig:dX-high-I}, yellow).
The pattern of net outflow of infectious individuals
from the high risk group via turnover
persisted across the range of turnover rates
(Figure~\ref{fig:dX-high-I}, yellow).
This net outflow of infectious individuals via turnover
acted to reduce STI prevalence in the high risk group
(phenomenon~1).
Treated individuals were similarly replaced largely by susceptible individuals
(Figure~\ref{fig:flows-low} and
Figures~\ref{fig:dX-high-S}~vs~\ref{fig:dX-high-T}, yellow).
The net replacement of both infectious and treated individuals with susceptible individuals
in the high risk group acted to reduce herd immunity in that group.
Reduced herd immunity then contributed to a rise in
the number of incident infections in the high risk group,
as the system moved from no turnover to slow turnover
(Figure~\ref{fig:dX-high-I}, red; phenomenon~2).
Incidence was further influenced by a third phenomenon
as systems moved from no turnover to higher rates of turnover:
the net movement of infectious individuals
from high to low risk (Figure~\ref{fig:flows-low})
reduced the average number of partners per year made available by
individuals in the infectious state (Figure~\ref{fig:C-I}).
As shown in Appendix~\ref{aa:eqs-incidence},
modelled incidence in all risk groups was proportional to
the average number of partners per year among infectious individuals
(Figure~\ref{fig:C-I}),
and overall prevalence
(Figure~\ref{fig:prevalence-all}).
Thus, as the average number of partners per year among infectious individuals
fell with faster turnover, incidence decreased
(Figure~\ref{fig:incidence-all}, region~B; phenomenon~3).
\par
Therefore, the inverted U-shaped relationship between turnover rate
and equilibrium STI prevalence in the high risk group was mediated
by the combination of the above three phenomena.
When systems moved from no turnover to slow turnover,
reduction in herd immunity (phenomenon~2) predominated,
leading to increasing equilibrium prevalence with turnover
(Figure~\ref{fig:prevalence-high}, region~A).
When systems were modelled under faster and faster turnover,
outflow of infectious individuals from the group via turnover (phenomenon~1) and
reduction in the average number of partners per year among infectious individuals (phenomenon~3)
predominated, leading to lower equilibrium prevalence at faster rates of turnover
(Figure~\ref{fig:prevalence-high}, region~B).
\begin{figure}
  \begingroup\centering
  \begin{subfigure}[t]{0.33\linewidth}
    \includegraphics[width=\linewidth]{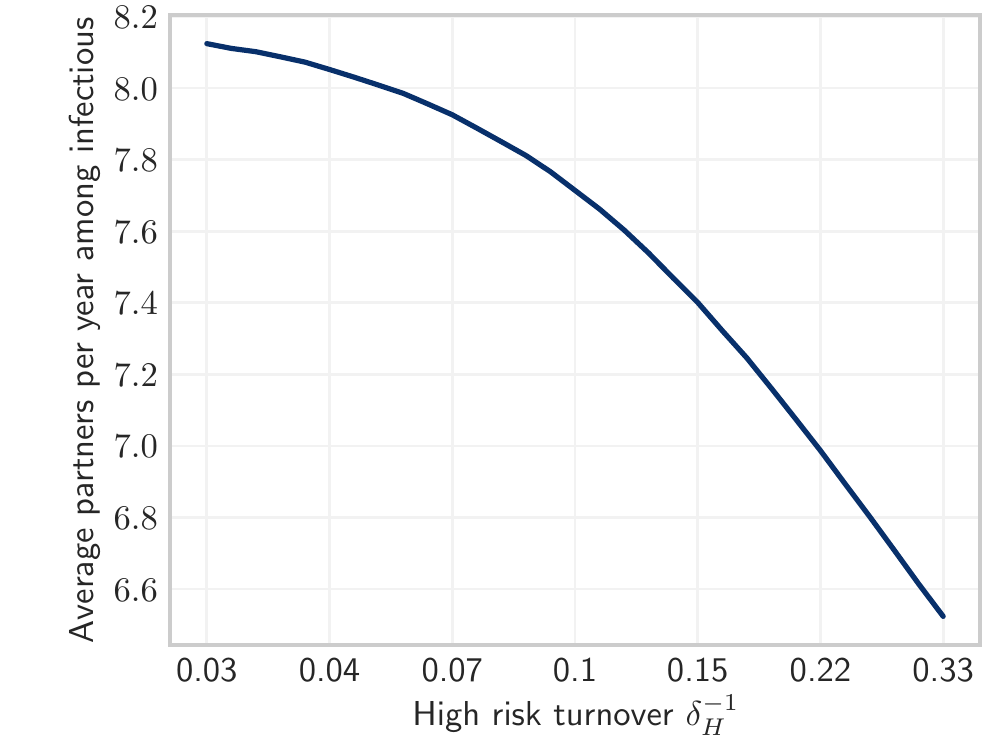}
    \caption{Average number of partners per year among infectious individuals}
    \label{fig:C-I}
  \end{subfigure}%
  \begin{subfigure}[t]{0.33\linewidth}
    \includegraphics[width=\linewidth]{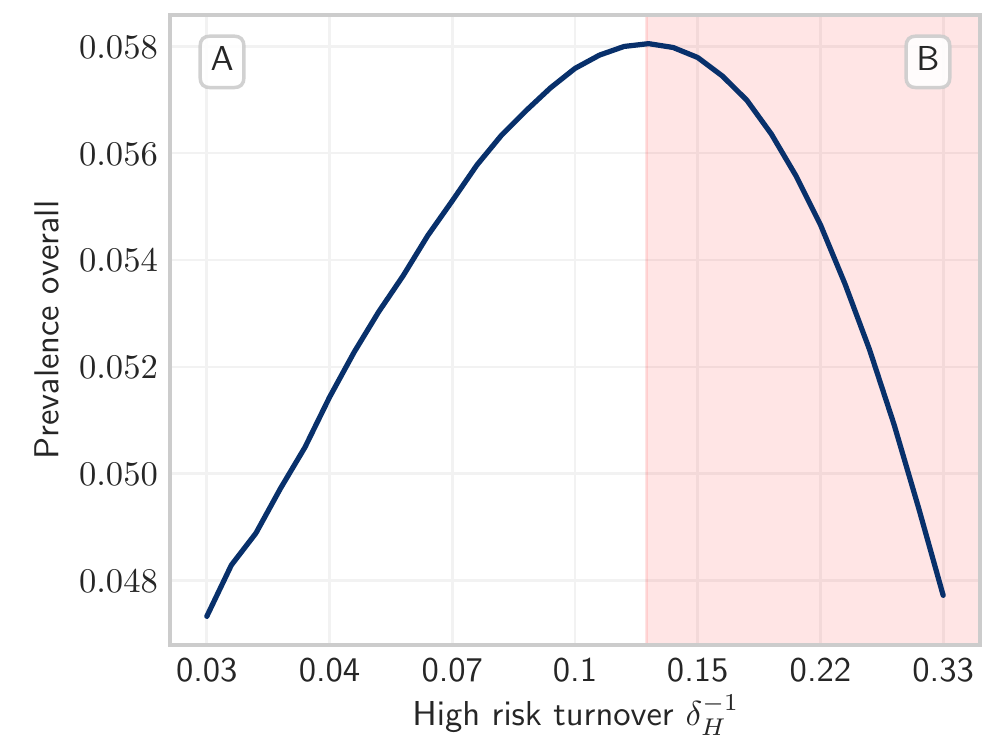}
    \caption{Overall prevalence}
    \label{fig:prevalence-all}
  \end{subfigure}%
  \begin{subfigure}[t]{0.33\linewidth}
    \includegraphics[width=\linewidth]{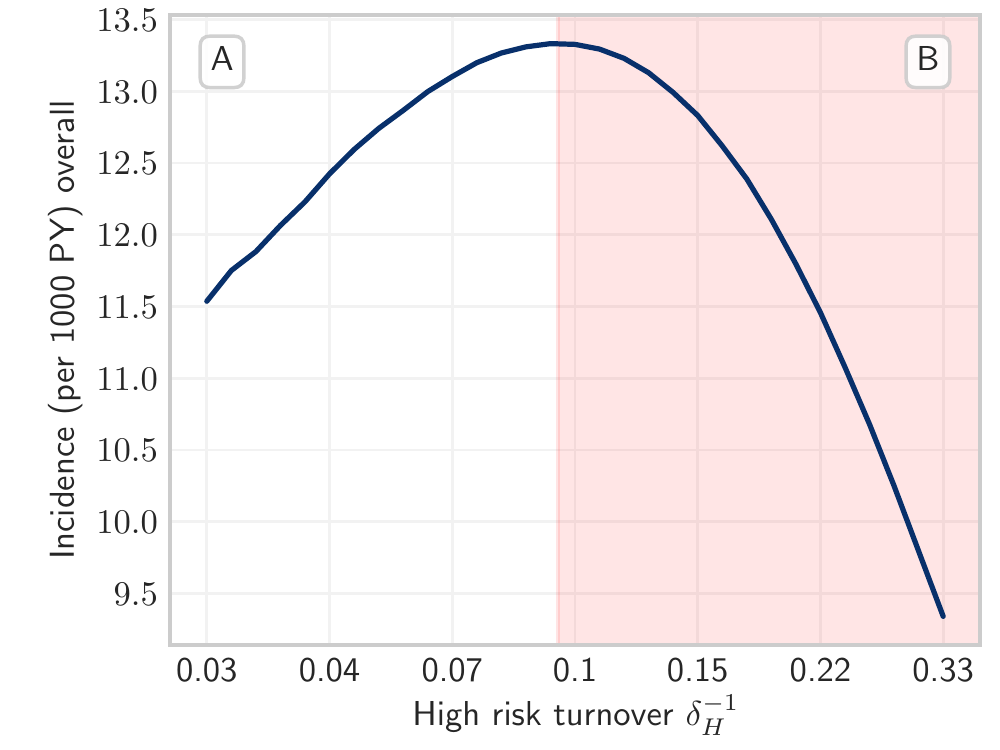}
    \caption{Overall incidence}
    \label{fig:incidence-all}
  \end{subfigure}%
  \\\endgroup
  \caption{Overall incidence and the non-constant factors of incidence versus turnover.
    The product of factors (\subref{fig:C-I}) and (\subref{fig:prevalence-all})
    is proportional to (\subref{fig:incidence-all}) overall incidence.
    See Appendix~\ref{aa:eqs-incidence} for proof.}
  \label{fig:incidence-factors}
  \footnotesize\input{x-axis.tex}
\end{figure}
\subsubsection{Influence of turnover on equilibrium prevalence in the low risk group}
\label{sss:res-prev-low}
As shown in Figure~\ref{fig:flows}, at equilibrium, the low risk group was
composed mainly of susceptible individuals.
Moving from a system no turnover to one with slow turnover
lead to a net inflow of infectious and treated individuals
(Figures~\ref{fig:dX-low-I}~and~\ref{fig:dX-low-T}, yellow),
and a net removal of susceptible individuals
(Figure~\ref{fig:dX-low-S}, yellow).
The net inflow of infectious individuals
(Figure~\ref{fig:dX-low-I}, yellow) contributed to
higher equilibrium prevalence in the low risk group
when the system moved from no turnover to slow turnover
(phenomenon~1).
The inflow of infectious and treated individuals
only slightly reduced the already large proportion who were susceptible
in the low risk group.
Thus, there was little increase in herd immunity within the low risk group
as turnover increased
(phenomenon~2).
However, incident infections still rose in the low risk group
as the system moved from no turnover to slow turnover
(Figure~\ref{fig:dX-low-I}, red)
due to higher incidence in the total population (Figure~\ref{fig:incidence-all})
which was largely driven by
reduced herd immunity in the high risk group
(see Section~\ref{sss:res-prev-high}; phenomenon~2).
Under faster rates of turnover,
incident infections declined in the low risk group
(Figure~\ref{fig:dX-low-I}, red)
due to lower incidence in the total population
(Figure~\ref{fig:incidence-all})
which was driven by
decreasing number of partners per year among infectious individuals
(Figure~\ref{fig:C-I}; phenomenon~3),
as described in Section~\ref{sss:res-prev-high}.
\par
Therefore, as in the high risk group,
the inverted U-shaped relationship between turnover rate
and equilibrium STI prevalence in the low risk group was mediated
by the combination of the above three phenomena.
Moving from no turnover to slow turnover,
the net inflow of infectious individuals (phenomenon~1)
and reduced herd immunity in the high risk group (phenomenon~2)
predominated, leading to higher equilibrium prevalence
(Figure~\ref{fig:prevalence-low}, region~A).
At higher rates of turnover,
a decreasing overall incidence due to
a reduction in the number of partners among infectious individuals (phenomenon~3)
predominated, leading to declining equilibrium prevalence
(Figure~\ref{fig:prevalence-low}, region~B).
\par
In sum, there were three phenomena that
drove shifts in equilibrium STI prevalence across risk groups
at variable rates of turnover:
1)~net flows of infectious individuals from high risk groups into low risk groups;
2)~changes to herd immunity, especially within the high risk group; and
3)~changes to the number of partnerships available with infectious individuals.
\subsubsection{Influence of turnover on STI prevalence ratio between high and low risk groups}
\begin{figure}
  \centerline{\includegraphics[width=0.5\linewidth]{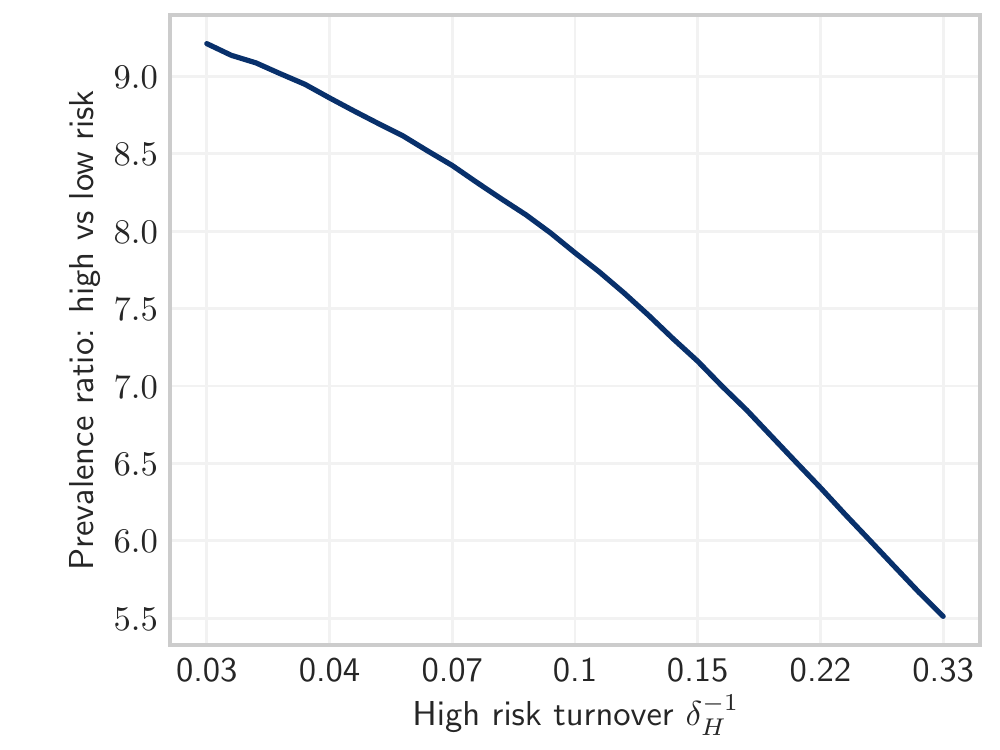}}
  \caption{Equilibrium prevalence ratios between high and low risk groups
    under different rates of turnover.}
  \label{fig:ratio-prevalence-high-low}
  \footnotesize\input{x-axis.tex}
\end{figure}

As discussed in Sections~\ref{sss:res-prev-high}~and~\ref{sss:res-prev-low},
turnover caused
a net outflow of infectious individuals from the high risk group
(Figure~\ref{fig:dX-high-I}, yellow) and
a net inflow of infectious individuals into the low risk group
(Figure~\ref{fig:dX-low-I}, yellow).
In contrast, the influence of turnover on
the rate of incident infections followed a more similar pattern
in both the high and low risk groups
(Figures~\ref{fig:dX-high-I}~and~\ref{fig:dX-low-I}, red).
Therefore, differences in the influence of turnover on prevalence between risk groups
were driven by net movement of infectious individuals from high to low risk,
causing prevalence in the high and low risk groups
to come closer together with faster turnover.
As shown in Figure~\ref{fig:ratio-prevalence-high-low},
the ratio of equilibrium STI prevalence in the high versus low risk groups
was thus reduced under faster turnover rates.
For example, the prevalence ratio
between high and low risk groups was:
$6.7$
in the model under high turnover ($\delta_H = 5$~years) versus
$9.2$
in the model without turnover ($\delta_H = 33$~years)
(Table~\ref{tab:fitting}).
Finally, the propensity for equilibrium STI
prevalence to decrease in the high risk group and for
prevalence to increase in the low risk group with faster turnover
(due to net movement of infectious individuals from high to low risk)
also explains why prevalence peaked at
slower turnover in the high risk group
(Figure~\ref{fig:prevalence-high}) and
faster turnover in the low risk group
(Figure~\ref{fig:prevalence-low}).
\subsection{Experiment~2: Inferred risk heterogeneity with versus without turnover}
\label{ss:res-infer}
\begin{figure}[H]
  \begingroup\centering
  \begin{subfigure}{0.4\linewidth}
    \includegraphics[width=\linewidth]{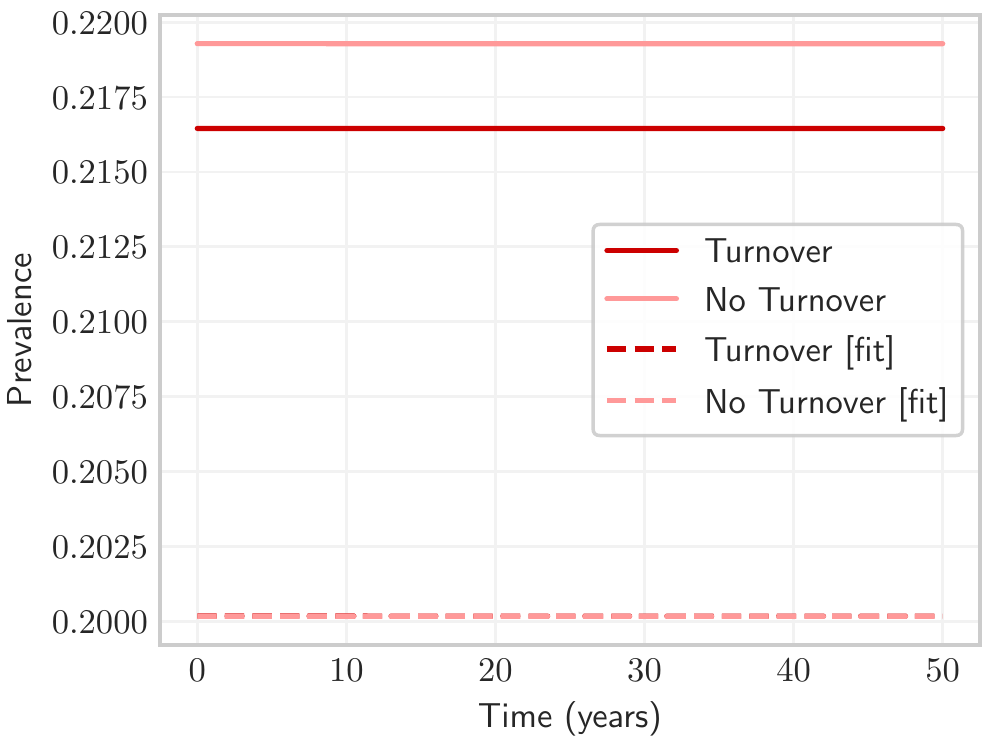}
    \caption{High risk}
    \label{fig:tpaf-prevalence-high}
  \end{subfigure}
  \begin{subfigure}{0.4\linewidth}
    \centering\includegraphics[width=\linewidth]{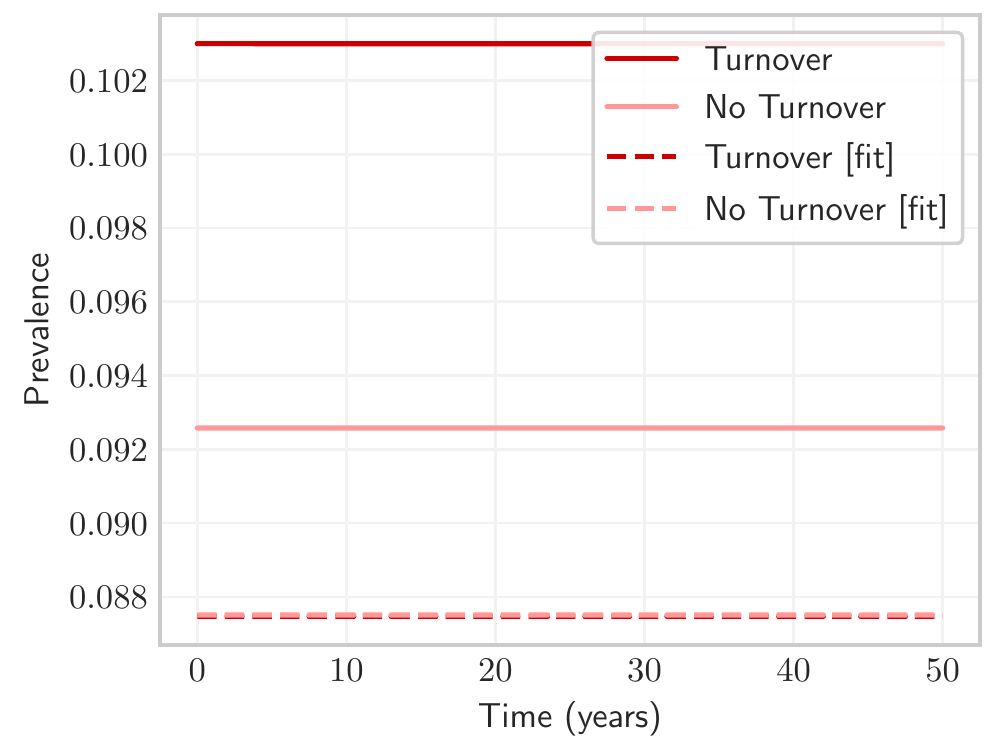}
    \caption{Medium risk}
    \label{fig:tpaf-prevalence-med}
  \end{subfigure}
  \begin{subfigure}{0.4\linewidth}
    \includegraphics[width=\linewidth]{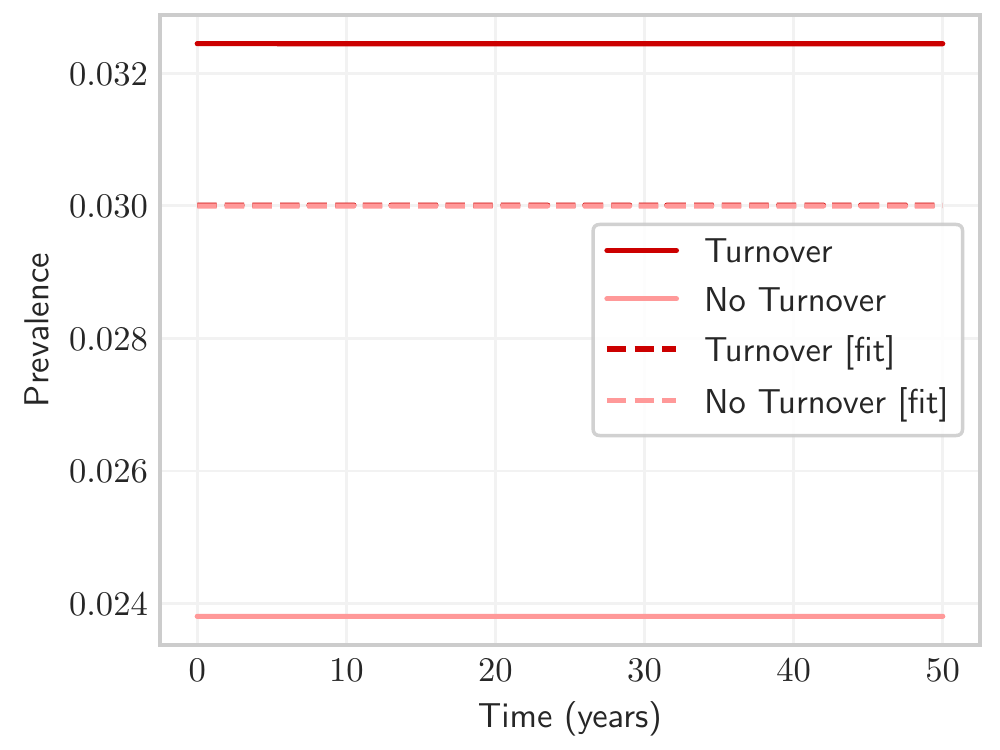}
    \caption{Low risk}
    \label{fig:tpaf-prevalence-low}
  \end{subfigure}
  \begin{subfigure}{0.4\linewidth}
    \includegraphics[width=\linewidth]{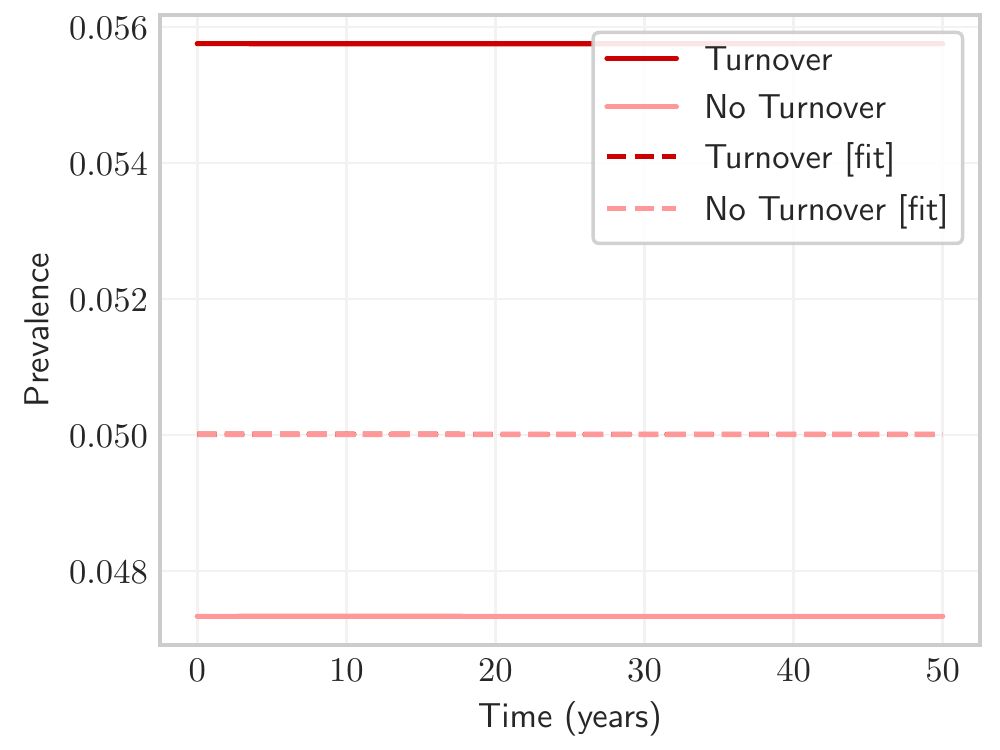}
    \caption{Overall}
    \label{fig:tpaf-prevalence-all}
  \end{subfigure}
  \\\endgroup
  \caption{Equilibrium STI prevalence
    among high, medium, and low risk groups as well as overall,
    with and without turnover,
    and with and without fitted $C_i$ to group-specific prevalence.}
  \label{fig:tpaf-prevalence}
\end{figure}

After model fitting,
our two STI transmission models
(one with turnover and one without turnover)
reproduced the target equilibrium STI prevalence values of 20\%,~8.75\%,~3\%,~and~5\%
in the high, medium, low risk groups, and total population, respectively
(Table~\ref{tab:fitting}; Figure~\ref{fig:tpaf-prevalence}).
When fitting the model with turnover to these group-specific prevalence targets,
the fitted numbers of partners per year $C_i$ (the only non-fixed parameter)
had to compensate for the reduction in
STI prevalence ratio between high and low risk groups
(Figure~\ref{fig:ratio-prevalence-high-low}).
As a result, the ratio of fitted partner numbers
between high and low risk groups ($C_H~/~C_L$)
had to be higher in the model with turnover compared to the model without turnover:
$23.9$~vs~%
$15.2$
(Table~\ref{tab:fitting}).
That is, the inferred level of risk heterogeneity was higher
in the model with turnover than in the model without turnover.
\begin{table}
  \caption{Equilibrium partnership formation rates and prevalence
    among the high and low risk groups
    predicted by the models with and without turnover,
    before and after model fitting.}
  \label{tab:fitting}
  \centerline{
\begin{tabular}{rcccccc}
  \toprule
  & \multicolumn{3}{c}{Number of Partners} & \multicolumn{3}{c}{Prevalence} \\
  \cmidrule(lr){2-4}\cmidrule(lr){5-7}
  Context & High & Low & High/Low & High & Low & High/Low \\\midrule
  Turnover &
  25.0
  & 1.0
  & 25.0
  & 21.6\%
  & 3.2\%
  & \textbf{6.7}\\
  No Turnover &
  25.0
  & 1.0
  & 25.0
  & 21.9\%
  & 2.4\%
  & \textbf{9.2}\\
  Turnover [fit] &
  24.3
  & 1.0
  & \textbf{23.9}
  & 20.0\%
  & 3.0\%
  & 6.7\\
  No Turnover [fit] &
    23.5
  & 1.5
  & \textbf{15.2}
  & 20.0\%
  & 3.0\%
  & 6.7\\
  \bottomrule
\end{tabular}}
\end{table}
\subsection{Experiment~3: Influence of turnover on the tPAF of the high risk group}
\label{ss:res-tpaf}
\begin{figure}[!tbp]
  \centerline{\includegraphics[width=0.5\linewidth]{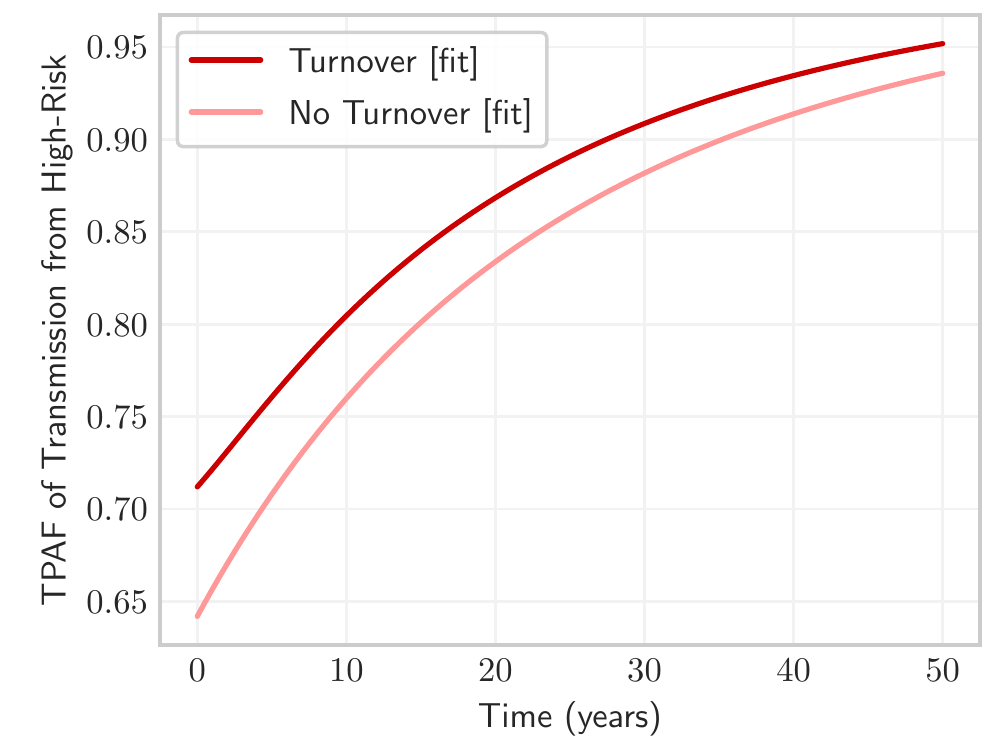}}
  \caption{Transmission population attributable fraction (tPAF)
    of the high risk group in models with and without turnover,
    after fitting the number of partners per year to group-specific prevalence.}
  \label{fig:tpaf-fit}
\end{figure}

Finally, we compared the tPAF of the high risk group
projected by the fitted model with turnover and the fitted model without turnover
(Figure~\ref{fig:tpaf-fit}).
The tPAF projected by both models
increased over longer and longer time horizons,
indicating that unmet prevention and treatment needs of the high risk group
were central to epidemic persistence in both fitted models.
The model with turnover projected a larger tPAF at all
time-horizons compared with the tPAF projected by the model without turnover.
The larger tPAF projected by the model with turnover
stemmed from more risk heterogeneity (Table~\ref{tab:fitting})
which led to more onward transmission from the unmet
prevention and treatment needs of the high risk group.
\section{Discussion}\label{s:discussion}
Using a mechanistic modelling analysis,
we found that turnover could be important when
projecting the tPAF of high risk groups to the overall epidemic.
Mechanistic insights include disentangling
three key phenomena by which turnover
alters equilibrium STI prevalence within risk groups,
and thereby the level of inferred risk heterogeneity between groups via model fitting.
Methodological contributions include a  framework for modelling turnover
which uses a flexible combination of data-driven constraints.
Taken together, our explanatory insights and framework
have mechanistic, public health, and methodological relevance for
the parameterization and use of epidemic models
to project intervention priorities for high~risk~groups. 
\paragraph{Influence of turnover on prevalence}
Building on prior work by \citet{Stigum1994,Zhang2012,Henry2015}
which similarly found an inverted U-shaped relationship between
turnover and overall equilibrium STI prevalence,
we identified three key phenomena that generated this relationship.
These turnover-driven phenomena were:
1)~a net flow of infectious individuals from higher risk groups to lower risk groups;
2)~reduced herd immunity in the higher risk groups due to net gain of susceptible individuals; and
3)~reduced incidence overall due to fewer partner numbers among infectious individuals.
The above three phenomena contributed to the pattern
of declining prevalence ratio between the highest and lowest risk groups
for increasing rates of turnover.
A decline in prevalence ratio due to turnover implies a reduction in risk heterogeneity.
Since risk heterogeneity is associated with epidemic emergence and persistence \citep{May1988}
-- i.e. the basic reproductive number --
our findings are thus consistent with \citet{Henry2015},
who demonstrated that turnover reduces the basic reproductive number by reducing heterogeneity.
Indeed, epidemiological and transmission modelling studies have shown that prevalence ratios
are an important marker of risk heterogeneity, and in turn
the impact of interventions focused on high risk groups \citep{Baral2012,Mishra2012}.
\paragraph{Implications for interventions}
Our comparison of fitted models with and without turnover showed that
if turnover exists in a given setting but is ignored in a model,
the inferred heterogeneity in risk would be lower than in reality,
while reproducing the same STI prevalence in each risk group.
As a result, the projected tPAF of high risk groups
could be systematically underestimated by models that ignore turnover.
Although we examined a single parameter to capture risk
(number of partners per year),
the findings would hold for any combination of factors
that alter the risk per susceptible individual (force of infection), including
biological transmission probabilities and rates of partner change \citep{Anderson1991}.
The public health implications of models ignoring turnover,
and thereby underestimating risk heterogeneity and the tPAF of high risk groups,
is that resources could potentially be misguided away from high risk groups.
For example, epidemic models which fail to include or accurately capture
turnover may underestimate the importance of addressing the unmet
needs of key populations at disproportionate risk of HIV and other STIs, such as
gay men and other men who have sex with men, transgender women, people who use drugs, and sex workers.
In many HIV epidemic models of regions with high HIV prevalence, such as in Southern Africa,
key populations have historically been subsumed into the overall modelled population;
which meant, by design, less risk heterogeneity \citep{Eaton2012,Cori2014,Mishra2016}.
Our findings suggest that even when key populations are included, it is important to
further capture within-person changes in risks over time (such as duration in sex work).
Underestimating risk heterogeneity could also underestimate the resources
required to achieve local epidemic control, as suggested by \citet{Henry2015,Hontelez2013}.
Important next steps surrounding the potential bias in tPAF projections
attributable to inclusion/exclusion of turnover include
quantifying the magnitude of bias,
and characterizing the epidemiologic conditions under which the bias
would be meaningfully large in the context of public health programmes.
\paragraph{Turnover framework}
We developed a unified framework
to parameterize risk group turnover
using available epidemiologic data and/or assumptions.
There are four potential benefits of using the framework to model turnover.
First, the framework defines how specific epidemiologic data and assumptions
could be used as constraints to help define rates of turnover.
Second, the framework allows flexibility in which constraints can be chosen and combined,
so that the constraints best reflect locally available data and/or plausible assumptions.
In fact, the framework can adequately reproduce
several prior implementations of turnover
in various epidemic models \citep{Stigum1994,Eaton2014,Henry2015}.
Third, this flexible approach also allows the framework to scale
to any number of risk groups.
Finally, the framework avoids the need for a burn-in period
to establish a demographic steady-state before introducing infection,
which was required in some previous models \citep{Boily2015}.
\par
As noted above, one benefit of the unified framework for modelling turnover
is clarifying data priorities for parameterizing turnover.
Absolute or relative population size estimates across risk groups
may be obtained from population-based sexual behaviour surveys \citep{DHS},
and from mapping and enumeration of marginalized persons
such as sex workers \citep{Abdul-Quader2014}.
The proportion of individuals who enter into each risk group
may be available through sexual behaviour surveys:
for example, among individuals who became sexually active for the first time in the past year,
the proportion who also engaged in multiple partnerships within the past year.
The average duration of time spent within each risk group, such as
the duration in sex work, may be drawn from
cross-sectional survey questions such as
``for how many years have you been a sex worker?''\ %
albeit with the recognition that such data are censured \citep{Watts2010}.
Longitudinal, or cohort studies that track
self-reported sexual behaviour over time can also provide
estimates of duration of time spent within a given risk strata \citep{Fergus2007},
or provide direct estimates of transition rates between risk strata.
\paragraph{Limitations}
Our framework for modelling turnover was developed
specifically to answer mechanistic questions about the tPAF;
as such, there are two key limitations of the framework in its current form.
First, the framework did not stratify the population by sex or age.
In the context of real-world STI epidemics,
the relative size of risk groups may differ by both sex and age,
such as the often smaller number of females and/or males who sell sex,
versus the larger number of males who pay for sex (clients of female or male sex workers).
Second, the framework does not account for
infection-attributable mortality, such as HIV-attributable mortality.
However, modelling studies have shown that HIV-attributable mortality can reduce the
relative size of higher risk groups who bear a disproportionate burden of HIV,
which in turn can cause an HIV epidemic to decline \citep{Boily1997}.
As such, many models of HIV transmission that include
very small ($<3\%$ of the population) high risk groups, such as female sex workers,
often do not constrain the relative size
of the sub-group populations to be stable over time \citep{Pickles2013}.
By ignoring infection-attributable mortality,
the proposed framework would similarly allow risk groups to change relative size
in response to disproportionate infection-attributable mortality.
Future modifications of the proposed framework include methods to optionally re-balance
infection-attributable mortality, and relevant age-sex stratifications so that
the framework can be applied more broadly to pathogen-specific epidemics.
\par
Our analyses of turnover and tPAF also have several limitations.
First, we did not capture the possibility that some individuals may become
re-susceptible to infection after treatment
-- an important feature of many STIs such as syphilis and gonorrhoea \citep{Fenton2008}.
As shown by \citet{Fenton2008} and \citet{Pourbohloul2003},
the re-supply of susceptible individuals following STI treatment
could fuel an epidemic, and so the influence of turnover on
STI prevalence and tPAF may be different.
Second, our analyses were restricted to equilibrium STI prevalence.
The influence of turnover on prevalence and tPAF
may vary within different phases of an epidemic
-- growth, mature, declining \citep{Wasserheit1996}.
Finally, our analyses reflected an illustrative STI epidemic
in a population with illustrative risk strata.
Important next steps in the examination of
the extent to which turnover influences the tPAF include
pathogen- and population-specific modelling
-- such as the comparisons of model structures by \citet{Hontelez2013,Johnson2016} --
and at different epidemic phases.
\paragraph{Conclusion}
In conclusion, turnover may influence prevalence of infection, and
thus influence inference on risk heterogeneity
when fitting risk-stratified epidemic models.
If models do not capture turnover,
the projected contribution of high risk groups, and thus,
the potential impact of prioritizing interventions to meet their needs, could be underestimated.
To aid the next generation of epidemic models
used to estimate the tPAF of high risk groups -- including key populations --
data collection efforts to parameterize risk group turnover should be prioritized.

\clearpage
\section*{Acknowledgements}
We would like to thank
Kristy Yiu (Unity Health Toronto) for logistical support,
the Siyaphambili research team for helpful discussions,
and Carly Comins (Johns Hopkins University)
for facilitating the modelling discussions with the wider study team.
SM is supported by an Ontario HIV Treatment Network and
Canadian Institutes of Health Research New Investigator Award.
\section*{Contributions}
JK and SM conceptualized the study and drafted the manuscript;
JK designed the experiments with input from LW, HM, and SM.
JK developed the unified framework and conducted the modelling, experiments,
and analyses; conducted the literature review, and drafted the first version of the manuscript. 
LW, HM, SB, and SS led substantial structural
revisions to the manuscript, including assessment of epidemiological constraints and assumptions;
and provided critical discussion surrounding implications of findings.
All authors contributed to interpretation of the results and manuscript revision.
\section*{Funding}
The study was supported by
the National Institutes of Health, Grant number: NR016650;
the Center for AIDS Research, Johns Hopkins University
  through the National Institutes of Health, Grant number: P30AI094189.
\section*{Conflicts of Interest}
Declarations of interest: none.

\clearpage
\section*{References}
\bibliography{ms.bib}

\begin{thebibliography}{45}
\providecommand{\natexlab}[1]{#1}
\providecommand{\url}[1]{\texttt{#1}}
\expandafter\ifx\csname urlstyle\endcsname\relax
  \providecommand{\doi}[1]{doi: #1}\else
  \providecommand{\doi}{doi: \begingroup \urlstyle{rm}\Url}\fi

\bibitem[Abdul-Quader et~al.(2014)Abdul-Quader, Baughman, and
  Hladik]{Abdul-Quader2014}
Abu~S Abdul-Quader, Andrew~L Baughman, and Wolfgang Hladik.
\newblock {Estimating the size of key populations: Current status and future
  possibilities}.
\newblock 9\penalty0 (2):\penalty0 107--114, mar 2014.
\newblock \doi{10.1097/COH.0000000000000041}.

\bibitem[Anderson and May(1991)]{Anderson1991}
Roy~M Anderson and Robert~M May.
\newblock {Infectious diseases of humans: dynamics and control}.
\newblock \emph{Infectious diseases of humans: dynamics and control.}, 1991.

\bibitem[Baral et~al.(2012)Baral, Beyrer, Muessig, Poteat, Wirtz, Decker,
  Sherman, and Kerrigan]{Baral2012}
Stefan Baral, Chris Beyrer, Kathryn Muessig, Tonia Poteat, Andrea~L Wirtz,
  Michele~R Decker, Susan~G Sherman, and Deanna Kerrigan.
\newblock {Burden of HIV among female sex workers in low-income and
  middle-income countries: A systematic review and meta-analysis}.
\newblock \emph{The Lancet Infectious Diseases}, 12\penalty0 (7):\penalty0
  538--549, jul 2012.
\newblock \doi{10.1016/S1473-3099(12)70066-X}.
\newblock URL
  \url{https://www.sciencedirect.com/science/article/pii/S147330991270066X?via{\%}3Dihub}.

\bibitem[Baral et~al.(2013)Baral, Logie, Grosso, Wirtz, and Beyrer]{Baral2013}
Stefan Baral, Carmen~H Logie, Ashley Grosso, Andrea~L Wirtz, and Chris Beyrer.
\newblock {Modified social ecological model: A tool to guide the assessment of
  the risks and risk contexts of HIV epidemics}.
\newblock \emph{BMC Public Health}, 13\penalty0 (1):\penalty0 482, may 2013.
\newblock \doi{10.1186/1471-2458-13-482}.

\bibitem[Baral et~al.(2014)Baral, Ketende, Green, Chen, Grosso, Sithole,
  Ntshangase, Yam, Kerrigan, Kennedy, and Adams]{Baral2014}
Stefan Baral, Sosthenes Ketende, Jessie~L. Green, Ping-An~An Chen, Ashley
  Grosso, Bhekie Sithole, Cebisile Ntshangase, Eileen Yam, Deanna Kerrigan,
  Caitlin~E. Kennedy, and Darrin Adams.
\newblock {Reconceptualizing the HIV epidemiology and prevention needs of
  female sex workers (FSW) in Swaziland}.
\newblock \emph{PLoS ONE}, 9\penalty0 (12):\penalty0 e115465, dec 2014.
\newblock \doi{10.1371/journal.pone.0115465}.

\bibitem[Boily and M{\^{a}}sse(1997)]{Boily1997}
Marie~Claude Boily and Beno{\^{i}}t M{\^{a}}sse.
\newblock {Mathematical models of disease transmission: A precious tool for the
  study of sexually transmitted diseases}.
\newblock \emph{Canadian Journal of Public Health}, 88\penalty0 (4):\penalty0
  255--265, 1997.
\newblock \doi{10.1007/bf03404793}.

\bibitem[Boily et~al.(2015)Boily, Pickles, Alary, Baral, Blanchard, Moses,
  Vickerman, and Mishra]{Boily2015}
Marie~Claude Boily, Michael Pickles, Michel Alary, Stefan Baral, James
  Blanchard, Stephen Moses, Peter Vickerman, and Sharmistha Mishra.
\newblock {What really is a concentrated HIV epidemic and what does it mean for
  West and Central Africa? Insights from mathematical modeling}.
\newblock \emph{Journal of Acquired Immune Deficiency Syndromes}, 68:\penalty0
  S74--S82, mar 2015.
\newblock \doi{10.1097/QAI.0000000000000437}.

\bibitem[Case et~al.(2012)Case, Ghys, Gouws, Eaton, Borquez, Stover, Cuchi,
  Abu-Raddad, Garnett, and Hallett]{Case2012}
Kelsey Case, Peter Ghys, Eleanor Gouws, Jeffery Eaton, Annick Borquez, John
  Stover, Paloma Cuchi, Laith Abu-Raddad, Geoffrey Garnett, and Timothy
  Hallett.
\newblock {Understanding the modes of tranmission model of new HIV infection
  and its use in prevention planning}.
\newblock \emph{Bulletin of the World Health Organization}, 90\penalty0
  (11):\penalty0 831--838, nov 2012.
\newblock \doi{10.2471/blt.12.102574}.

\bibitem[Cori et~al.(2014)Cori, Ayles, Beyers, Schaap, Floyd, Sabapathy, Eaton,
  Hauck, Smith, Griffith, Moore, Donnell, Vermund, Fidler, Hayes, and
  Fraser]{Cori2014}
Anne Cori, Helen Ayles, Nulda Beyers, Ab~Schaap, Sian Floyd, Kalpana Sabapathy,
  Jeffrey~W. Eaton, Katharina Hauck, Peter Smith, Sam Griffith, Ayana Moore,
  Deborah Donnell, Sten~H. Vermund, Sarah Fidler, Richard Hayes, and Christophe
  Fraser.
\newblock {HPTN 071 (PopART): A cluster-randomized trial of the population
  impact of an HIV combination prevention intervention including universal
  testing and treatment: Mathematical model}.
\newblock \emph{PLoS ONE}, 9\penalty0 (1):\penalty0 e84511, jan 2014.
\newblock \doi{10.1371/journal.pone.0084511}.
\newblock URL \url{https://dx.plos.org/10.1371/journal.pone.0084511}.

\bibitem[DataBank(2019)]{WorldBank}
DataBank.
\newblock {Population estimates and projections}, 2019.
\newblock URL
  \url{https://databank.worldbank.org/source/population-estimates-and-projections}.

\bibitem[Eaton and Hallett(2014)]{Eaton2014}
Jeffrey~W. Eaton and Timothy~B. Hallett.
\newblock {Why the proportion of transmission during early-stage HIV infection
  does not predict the long-term impact of treatment on HIV incidence}.
\newblock \emph{Proceedings of the National Academy of Sciences}, 111\penalty0
  (45):\penalty0 16202--16207, nov 2014.
\newblock \doi{10.1073/pnas.1323007111}.

\bibitem[Eaton et~al.(2012)Eaton, Johnson, Salomon, B{\"{a}}rnighausen,
  Bendavid, Bershteyn, Bloom, Cambiano, Fraser, Hontelez, Humair, Klein, Long,
  Phillips, Pretorius, Stover, Wenger, Williams, and Hallett]{Eaton2012}
Jeffrey~W. Eaton, Leigh~F. Johnson, Joshua~A. Salomon, Till B{\"{a}}rnighausen,
  Eran Bendavid, Anna Bershteyn, David~E. Bloom, Valentina Cambiano, Christophe
  Fraser, Jan~A.C. Hontelez, Salal Humair, Daniel~J. Klein, Elisa~F. Long,
  Andrew~N. Phillips, Carel Pretorius, John Stover, Edward~A. Wenger, Brian~G.
  Williams, and Timothy~B. Hallett.
\newblock {HIV treatment as prevention: Systematic comparison of mathematical
  models of the potential impact of antiretroviral therapy on HIV incidence in
  South Africa}.
\newblock \emph{PLoS Medicine}, 9\penalty0 (7):\penalty0 e1001245, jul 2012.
\newblock \doi{10.1371/journal.pmed.1001245}.
\newblock URL \url{https://dx.plos.org/10.1371/journal.pmed.1001245}.

\bibitem[Fenton et~al.(2008)Fenton, Breban, Vardavas, Okano, Martin, Aral, and
  Blower]{Fenton2008}
Kevin~A. Fenton, Romulus Breban, Raffaele Vardavas, Justin~T. Okano, Tara
  Martin, Sevgi Aral, and Sally Blower.
\newblock {Infectious syphilis in high-income settings in the 21st century}.
\newblock \emph{The Lancet Infectious Diseases}, 8\penalty0 (4):\penalty0
  244--253, apr 2008.
\newblock \doi{10.1016/S1473-3099(08)70065-3}.

\bibitem[Fergus et~al.(2007)Fergus, Zimmerman, and Caldwell]{Fergus2007}
Stevenson Fergus, Marc~A Zimmerman, and Cleopatra~H Caldwell.
\newblock {Growth trajectories of sexual risk behavior in adolescence and young
  adulthood}.
\newblock \emph{American Journal of Public Health}, 97\penalty0 (6):\penalty0
  1096--1101, jun 2007.
\newblock \doi{10.2105/AJPH.2005.074609}.

\bibitem[Ganem and Prince(2004)]{Ganem2004}
Don Ganem and Alfred~M. Prince.
\newblock {Hepatitis B Virus Infection — Natural History and Clinical
  Consequences}.
\newblock \emph{New England Journal of Medicine}, 350\penalty0 (11):\penalty0
  1118--1129, mar 2004.
\newblock \doi{10.1056/NEJMra031087}.

\bibitem[Garnett and Anderson(1994)]{Garnett1994}
Geoffrey~P. Garnett and Roy~M. Anderson.
\newblock {Balancing sexual partnership in an age and activity stratified model
  of HIV transmission in heterosexual populations}.
\newblock \emph{Mathematical Medicine and Biology}, 11\penalty0 (3):\penalty0
  161--192, jan 1994.
\newblock \doi{10.1093/imammb/11.3.161}.

\bibitem[Hanley(2001)]{Hanley2001}
J.~A. Hanley.
\newblock {A heuristic approach to the formulas for population attributable
  fraction}.
\newblock \emph{Journal of Epidemiology and Community Health}, 55\penalty0
  (7):\penalty0 508--514, 2001.
\newblock \doi{10.1136/jech.55.7.508}.

\bibitem[Henry and Koopman(2015)]{Henry2015}
Christopher~J. Henry and James~S. Koopman.
\newblock {Strong influence of behavioral dynamics on the ability of testing
  and treating HIV to stop transmission}.
\newblock \emph{Scientific Reports}, 5\penalty0 (1):\penalty0 9467, aug 2015.
\newblock \doi{10.1038/srep09467}.

\bibitem[Hontelez et~al.(2013)Hontelez, Lurie, B{\"{a}}rnighausen, Bakker,
  Baltussen, Tanser, Hallett, Newell, and de~Vlas]{Hontelez2013}
Jan A.C.~C. Hontelez, Mark~N. Lurie, Till B{\"{a}}rnighausen, Roel Bakker, Rob
  Baltussen, Frank Tanser, Timothy~B. Hallett, Marie~Louise Newell, and Sake~J.
  de~Vlas.
\newblock {Elimination of HIV in South Africa through Expanded Access to
  Antiretroviral Therapy: A Model Comparison Study}.
\newblock \emph{PLoS Medicine}, 10\penalty0 (10):\penalty0 e1001534, oct 2013.
\newblock \doi{10.1371/journal.pmed.1001534}.
\newblock URL \url{http://dx.plos.org/10.1371/journal.pmed.1001534}.

\bibitem[ICAP(2019)]{PHIAproject}
ICAP.
\newblock {PHIA Project}, 2019.
\newblock URL \url{https://phia.icap.columbia.edu}.

\bibitem[Johnson and Geffen(2016)]{Johnson2016}
Leigh~F. Johnson and Nathan Geffen.
\newblock {A Comparison of two mathematical modeling frameworks for evaluating
  sexually transmitted infection epidemiology}.
\newblock \emph{Sexually Transmitted Diseases}, 43\penalty0 (3):\penalty0
  139--146, mar 2016.
\newblock \doi{10.1097/OLQ.0000000000000412}.

\bibitem[Knight et~al.(2019)Knight, Wang, Ma, Schwartz, Baral, and
  Mishra]{Knight2019}
Jesse Knight, Linwei Wang, Huiting Ma, Sheree Schwartz, Stefan Baral, and
  Sharmistha Mishra.
\newblock {The influence of risk group turnover in STI/HIV epidemics:
  mechanistic insights from transmission modeling}.
\newblock In \emph{STI \& HIV 2019 World Congress}, Vancouver, BC, Canada,
  2019.
\newblock URL \url{https://sti.bmj.com/content/95/Suppl_1/A83.3}.

\bibitem[Koopman et~al.(1997)Koopman, Jacquez, Welch, Simon, Foxman, Pollock,
  Barth-Jones, Adams, and Lange]{Koopman1997}
J~S Koopman, J~A Jacquez, G~W Welch, C~P Simon, B~Foxman, S~M Pollock,
  D~Barth-Jones, A~L Adams, and K~Lange.
\newblock {The role of early HIV infection in the spread of HIV through
  populations.}
\newblock \emph{Journal of Acquired Immune Deficiency Syndromes}, 14\penalty0
  (3):\penalty0 249--58, mar 1997.
\newblock URL \url{http://www.ncbi.nlm.nih.gov/pubmed/9117458}.

\bibitem[Kraft(1988)]{Kraft1988}
Dieter Kraft.
\newblock {A software package for sequential quadratic programming}.
\newblock Technical Report DFVLR-FB 88-28, DLR German Aerospace Center —
  Institute for Flight Mechanics, Koln, Germany, 1988.

\bibitem[LAPACK(1992)]{LAPACK}
LAPACK.
\newblock {LAPACK: Linear Algebra PACKage}, 1992.
\newblock URL \url{http://www.netlib.org/lapack}.

\bibitem[Lawson and Hanson(1995)]{Lawson1995}
Charles~L Lawson and Richard~J Hanson.
\newblock \emph{{Solving least squares problems}}, volume~15.
\newblock SIAM, 1995.

\bibitem[Maartens et~al.(2014)Maartens, Celum, and Lewin]{Maartens2014}
Gary Maartens, Connie Celum, and Sharon~R. Lewin.
\newblock {HIV infection: Epidemiology, pathogenesis, treatment, and
  prevention}.
\newblock \emph{The Lancet}, 384\penalty0 (9939):\penalty0 258--271, 2014.
\newblock \doi{10.1016/S0140-6736(14)60164-1}.

\bibitem[Maheu-Giroux et~al.(2017)Maheu-Giroux, Vesga, Diabat{\'{e}}, Alary,
  Baral, Diouf, Abo, and Boily]{Maheu-Giroux2017}
Mathieu Maheu-Giroux, Juan~F Vesga, Souleymane Diabat{\'{e}}, Michel Alary,
  Stefan Baral, Daouda Diouf, Kouam{\'{e}} Abo, and Marie~Claude Boily.
\newblock {Changing Dynamics of HIV Transmission in C{\^{o}}te d'Ivoire:
  Modeling Who Acquired and Transmitted Infections and Estimating the Impact of
  Past HIV Interventions (1976-2015)}.
\newblock \emph{Journal of Acquired Immune Deficiency Syndromes}, 75\penalty0
  (5):\penalty0 517--527, 2017.
\newblock \doi{10.1097/QAI.0000000000001434}.

\bibitem[Malthus(1798)]{Malthus1798}
Thomas~Robert Malthus.
\newblock \emph{{An Essay on the Principle of Population}}.
\newblock 1798.

\bibitem[Marston and King(2006)]{Marston2006}
Cicely Marston and Eleanor King.
\newblock {Factors that shape young people's sexual behaviour: a systematic
  review}.
\newblock \emph{Lancet}, 368\penalty0 (9547):\penalty0 1581--1586, nov 2006.
\newblock \doi{10.1016/S0140-6736(06)69662-1}.

\bibitem[May and Anderson(1988)]{May1988}
R.~M. May and R.~M. Anderson.
\newblock {The transmission dynamics of human immunodeficiency virus (HIV).},
  1988.

\bibitem[Mishra et~al.(2012)Mishra, Steen, Gerbase, Lo, and Boily]{Mishra2012}
Sharmistha Mishra, Richard Steen, Antonio Gerbase, Ying~Ru Lo, and Marie~Claude
  Boily.
\newblock {Impact of High-Risk Sex and Focused Interventions in Heterosexual
  HIV Epidemics: A Systematic Review of Mathematical Models}.
\newblock \emph{PLoS ONE}, 7\penalty0 (11):\penalty0 e50691, nov 2012.
\newblock \doi{10.1371/journal.pone.0050691}.

\bibitem[Mishra et~al.(2014)Mishra, Pickles, Blanchard, Moses, and
  Boily]{Mishra2014}
Sharmistha Mishra, Michael Pickles, James~F Blanchard, Stephen Moses, and
  Marie~Claude Boily.
\newblock {Distinguishing sources of HIV transmission from the distribution of
  newly acquired HIV infections: Why is it important for HIV prevention
  planning?}
\newblock \emph{Sexually Transmitted Infections}, 90\penalty0 (1):\penalty0
  19--25, feb 2014.
\newblock \doi{10.1136/sextrans-2013-051250}.

\bibitem[Mishra et~al.(2016)Mishra, Boily, Schwartz, Beyrer, Blanchard, Moses,
  Castor, Phaswana-Mafuya, Vickerman, Drame, Alary, and Baral]{Mishra2016}
Sharmistha Mishra, Marie-Claude Boily, Sheree Schwartz, Chris Beyrer, James~F.
  Blanchard, Stephen Moses, Delivette Castor, Nancy Phaswana-Mafuya, Peter
  Vickerman, Fatou Drame, Michel Alary, and Stefan~D. Baral.
\newblock {Data and methods to characterize the role of sex work and to inform
  sex work programs in generalized HIV epidemics: evidence to challenge
  assumptions}.
\newblock \emph{Annals of Epidemiology}, 26\penalty0 (8):\penalty0 557--569,
  aug 2016.
\newblock \doi{10.1016/j.annepidem.2016.06.004}.

\bibitem[Mukandavire et~al.(2018)Mukandavire, Walker, Schwartz, Boily, Danon,
  Lyons, Diouf, Liestman, Diouf, Drame, Coly, Muhire, Thiam, Diallo, Kane,
  Ndour, Volz, Mishra, Baral, and Vickerman]{Mukandavire2018}
Christinah Mukandavire, Josephine Walker, Sheree Schwartz, Marie-Claude Boily,
  Leon Danon, Carrie Lyons, Daouda Diouf, Ben Liestman, Nafissatou~Leye Diouf,
  Fatou Drame, Karleen Coly, Remy Serge~Manzi Muhire, Safiatou Thiam, Papa
  Amadou~Niang Diallo, Coumba~Toure Kane, Cheikh Ndour, Erik Volz, Sharmistha
  Mishra, Stefan Baral, and Peter Vickerman.
\newblock {Estimating the contribution of key populations towards the spread of
  HIV in Dakar, Senegal}.
\newblock \emph{Journal of the International AIDS Society}, 21:\penalty0
  e25126, jul 2018.
\newblock \doi{10.1002/jia2.25126}.

\bibitem[Pickles et~al.(2013)Pickles, Boily, Vickerman, Lowndes, Moses,
  Blanchard, Deering, Bradley, Ramesh, Washington, Adhikary, Mainkar,
  Paranjape, and Alary]{Pickles2013}
Michael Pickles, Marie~Claude Boily, Peter Vickerman, Catherine~M Lowndes,
  Stephen Moses, James~F Blanchard, Kathleen~N Deering, Janet Bradley,
  Banadakoppa~M Ramesh, Reynold Washington, Rajatashuvra Adhikary, Mandar
  Mainkar, Ramesh~S Paranjape, and Michel Alary.
\newblock {Assessment of the population-level effectiveness of the Avahan
  HIV-prevention programme in South India: A preplanned, causal-pathway-based
  modelling analysis}.
\newblock \emph{The Lancet Global Health}, 1\penalty0 (5):\penalty0 e289--e299,
  nov 2013.
\newblock \doi{10.1016/S2214-109X(13)70083-4}.

\bibitem[Pourbohloul et~al.(2003)Pourbohloul, Rekart, and
  Brunham]{Pourbohloul2003}
Babak Pourbohloul, Michael~L. Rekart, and Robert~C. Brunham.
\newblock {Impact of mass treatment on syphilis transmission: A mathematical
  modeling approach}.
\newblock \emph{Sexually Transmitted Diseases}, 30\penalty0 (4):\penalty0
  297--305, apr 2003.
\newblock \doi{10.1097/00007435-200304000-00005}.

\bibitem[Pr{\"{u}}ss-Ust{\"{u}}n et~al.(2013)Pr{\"{u}}ss-Ust{\"{u}}n, Wolf,
  Driscoll, Degenhardt, Neira, and Calleja]{Pruss-Ustun2013}
Annette Pr{\"{u}}ss-Ust{\"{u}}n, Jennyfer Wolf, Tim Driscoll, Louisa
  Degenhardt, Maria Neira, and Jesus Maria~Garcia Calleja.
\newblock {HIV Due to Female Sex Work: Regional and Global Estimates}.
\newblock \emph{PLoS ONE}, 8\penalty0 (5):\penalty0 e63476, 2013.
\newblock \doi{10.1371/journal.pone.0063476}.

\bibitem[Shubber et~al.(2014)Shubber, Mishra, Vesga, and Boily]{Shubber2014}
Zara Shubber, Sharmistha Mishra, Juan~F. Vesga, and Marie~Claude Boily.
\newblock {The HIV modes of transmission model: A systematic review of its
  findings and adherence to guidelines}.
\newblock \emph{Journal of the International AIDS Society}, 17\penalty0
  (1):\penalty0 18928, jan 2014.
\newblock \doi{10.7448/IAS.17.1.18928}.

\bibitem[Stigum et~al.(1994)Stigum, Falck, and Magnus]{Stigum1994}
Hein Stigum, W.~Falck, and P.~Magnus.
\newblock {The core group revisited: The effect of partner mixing and migration
  on the spread of gonorrhea, chlamydia, and HIV}.
\newblock \emph{Mathematical Biosciences}, 120\penalty0 (1):\penalty0 1--23,
  mar 1994.
\newblock \doi{10.1016/0025-5564(94)90036-1}.

\bibitem[{The DHS Program}(2019)]{DHS}
{The DHS Program}.
\newblock {Data}, 2019.
\newblock URL \url{https://www.dhsprogram.com}.

\bibitem[Wasserheit and Aral(1996)]{Wasserheit1996}
J.~N. Wasserheit and S.~O. Aral.
\newblock {The Dynamic Topology Of Sexually Transmitted Disease Epidemics:
  Implications For Prevention Strategies}.
\newblock \emph{Journal of Infectious Diseases}, 174\penalty0 (Supplement
  2):\penalty0 S201--S213, oct 1996.
\newblock \doi{10.1109/ICPDS.2016.7756727}.
\newblock URL
  \url{https://academic.oup.com/jid/article-lookup/doi/10.1093/infdis/174.Supplement{\_}2.S201}.

\bibitem[Watts et~al.(2010)Watts, Zimmerman, Foss, Hossain, Cox, and
  Vickerman]{Watts2010}
C.~Watts, C.~Zimmerman, A.~M. Foss, M.~Hossain, A.~Cox, and P.~Vickerman.
\newblock {Remodelling core group theory: the role of sustaining populations in
  HIV transmission}.
\newblock \emph{Sexually Transmitted Infections}, 86\penalty0 (Suppl
  3):\penalty0 iii85--iii92, dec 2010.
\newblock \doi{10.1136/sti.2010.044602}.

\bibitem[Yorke et~al.(1978)Yorke, Hethcote, and Nold]{Yorke1978}
James~A Yorke, Herbert~W Hethcote, and Annett Nold.
\newblock {Dynamics and control of the transmission of gonorrhea}.
\newblock \emph{Sexually Transmitted Diseases}, 5\penalty0 (2):\penalty0
  51--56, 1978.
\newblock \doi{10.1097/00007435-197804000-00003}.

\bibitem[Zhang et~al.(2012)Zhang, Zhong, Romero-Severson, Alam, Henry, Volz,
  and Koopman]{Zhang2012}
Xinyu Zhang, Lin Zhong, Ethan Romero-Severson, Shah~Jamal Alam, Christopher~J
  Henry, Erik~M Volz, and James~S Koopman.
\newblock {Episodic HIV Risk Behavior Can Greatly Amplify HIV Prevalence and
  the Fraction of Transmissions from Acute HIV Infection}.
\newblock \emph{Statistical Communications in Infectious Diseases}, 4\penalty0
  (1), nov 2012.
\newblock \doi{10.1515/1948-4690.1041}.

\end{thebibliography}
\clearpage
\initappendix
\section{Turnover Framework}\label{a:framework}
We introduce a system of parameters and constraints
to describe risk group turnover
in deterministic epidemic models with heterogeneity in risk.%
\footnote{A preliminary version of this framework was used by \citet{Knight2019}.}
We then describe how the system can be used in practical terms,
based on different assumptions and data available for parameterizing turnover in risk.
We conclude by framing previous approaches to this task using the proposed system.
\subsection{Notation}
\label{aa:notation}
Consider a population divided into $G$ risk groups.
We denote the number of individuals in risk group $i \in [1, \dots, G]$ as $x_i$
and the set of all risk groups as $\bm{x} = \{x_1, \dots, x_G\}$.
The total population size is $N = \sum_i {x_i}$,
and the relative population size of each group
is denoted as $\hat{x}_i = x_i / N$.
Individuals enter the population at a rate $\nu$ per year,
and exit at a rate $\mu$ per year.
We model the distribution of risk groups
among individuals entering into the population as $\bm{\hat{e}}$,
which may be different from individuals already in the population~$\bm{\hat{x}}$.%
\footnote{We could equivalently stratify the rate of entry $\nu$ by risk group;
  however, we find that the mathematics in subsequent sections
  are more straightforward using~$\bm{\hat{e}}$.}
Thus, the total number of individuals
entering into population $\bm{x}$ per year is given by $\nu N$,
and the number of individuals
entering into group~$i$ specifically is given by~$\hat{e}_i \nu N$.
\par
\begin{figure}
  \centering
  \includegraphics[width=0.5\linewidth]{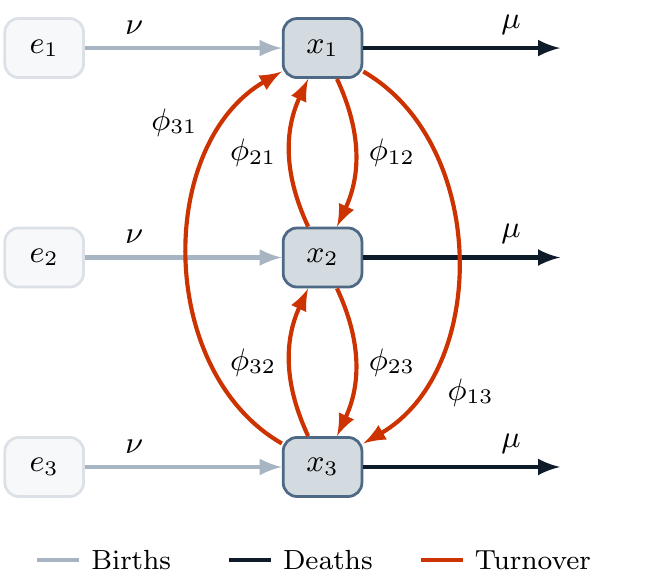}
  \caption{System of $G = 3$ risk groups and turnover between them.}
  \footnotesize\raggedright
  $x_i$: number of individuals in risk group~$i$;
  $e_i$: number of individuals available to enter risk group~$i$;
  $\nu$: rate of population entry;
  $\mu$: rate of population exit;
  $\phi_{ij}$: rate of turnover from group~$i$ to group~$j$.
  \label{fig:system-app}
\end{figure}

Turnover transitions may then occur between any two groups, in either direction.
Therefore we denote the turnover rates as a $G \times G$~matrix~$\phi$.
The element $\phi_{ij}$ corresponds to the proportion of individuals in group $i$
who move from group~$i$ to group~$j$ each year.
An example matrix is given in Eq.~(\ref{eq:phi}),
where we write the diagonal elements as $*$ since they represent
transitions from a group to itself.
\begin{equation}\label{eq:phi}
\phi = \left[\begin{array}{cccc}
           *          & x_1  \rightarrow x_2 & \cdots & x_1 \rightarrow x_G \\[0.5em]
  x_2 \rightarrow x_1 &          *           & \cdots & x_2 \rightarrow x_G \\[0.5em]
        \vdots        &        \vdots        & \ddots &       \vdots        \\[0.5em]
  x_G \rightarrow x_1 & x_G \rightarrow x_2  & \cdots &          *
\end{array}\right]
\end{equation}
Risk groups, transitions, and the associated rates
are also shown for $G = 3$ in Figure~\ref{fig:system-app}.
\subsection{Parameterization}
\label{aa:params}
Next, we present methods to illustrate how
epidemiologic data can be used to parametrize
turnover in epidemic models.
We construct a system like the one above
which reflects the risk group dynamics observed in a specific context.
We assume that the relative sizes of the risk groups in the model ($\bm{\hat{x}}$)
are already known, and should remain constant over time.
Thus, what remains is to estimate the values of the parameters:
$\nu$, $\mu$, $\bm{\hat{e}}$, and $\phi$,
using commonly available sources of data.
\subsubsection{Total Population Size}
\label{aaa:params-nu-mu}
The total population size $N(t)$ is a function of
the rates of population entry $\nu(t)$ and exit $\mu(t)$, given an initial size~$N_0$.
We allow the proportion entering the system to vary by risk group via $\bm{\hat{e}}$,
while the exit rate has the same value for each group.
We assume that there is no disease-attributable death.
Because the values of $\nu$ and $\mu$ are the same for each risk group, 
they can be estimated independent of
$\bm{\hat{x}}$,~$\bm{\hat{e}}$,~and~$\phi$.
\par
The difference between entry and exit rates
defines the rate of population growth:
\begin{equation}\label{eq:growth-G}
\mathcal{G}(t) = \nu(t) - \mu(t) 
\end{equation}
The total population may then be defined using an initial population size $N_0$ as:
\begin{equation}\label{eq:growth-vary}
N(t) = N_0 \exp{\left(\int_{0}^{\,t}{\log\big(1+\mathcal{G}(\tau) \big)d\tau}\right)}
\end{equation}
which, for constant growth, simplifies to the familiar expression~\citep{Malthus1798}:
\begin{equation}\label{eq:growth-const}
N(t) = N_0 {(1 + \mathcal{G})}^{t}
\end{equation}
Census data, such as \citep{WorldBank}, can be used to source
the total population size in a given geographic setting over time $N(t)$,
thus allowing Eqs.~(\ref{eq:growth-vary})~and~(\ref{eq:growth-const})
to be used to estimate~$\mathcal{G}(t)$.
\par
If the population size is assumed to be constant,
then $\mathcal{G}(t) = 0$ and $\nu(t) = \mu(t)$.
If population growth occurs at a stable rate, then
$\mathcal{G}$ is fixed at a constant value
which can be estimated via Eq.~(\ref{eq:growth-const})
using any two values of $N(t)$, separated by a time interval $\tau$:
\begin{equation}\label{eq:growth-backwards}
\mathcal{G}_{\tau} = {\frac{N(t+\tau)}{N(t)}}^{\frac{1}{\tau}} -1
\end{equation}
If the rate of population growth $\mathcal{G}$ varies over time,
then Eq.~(\ref{eq:growth-backwards}) can be reused for consecutive time intervals,
and the complete function $\mathcal{G}(t)$ approximated piecewise by constant values.
The piecewise approximation can be more feasible
than exact solutions using Eq.~(\ref{eq:growth-vary}),
and can reproduce $N(t)$ accurately for small enough intervals $\tau$,
such as one year.
\par
Now, given a value of $\mathcal{G}(t)$,
either $\nu(t)$ must be chosen and $\mu(t)$ calculated using Eq.~(\ref{eq:growth-G}),
or $\mu(t)$ must be chosen, and $\nu(t)$ calculated.
Most modelled systems assume
a constant duration of time that individuals spend in the model $\delta(t)$~\citep{Anderson1991}
which is related to the rate of exit $\mu$ by:
\begin{equation}\label{eq:duration-model}
\delta(t) = \mu^{-1}(t)
\end{equation}
In the context of sexually transmitted infections, the duration of time usually reflects
the average sexual life-course of individuals from age 15~to~50 years,
such that $\delta = 35$ years.
The duration $\delta$ may also vary with time to reflect changes in life expectancy.
The exit rate $\mu(t)$ can then be defined as $\delta^{-t}(t)$
following Eq.~(\ref{eq:duration-model}),
and the entry rate $\nu(t)$ defined as $\mathcal{G}(t) - \mu(t)$
following Eq.~(\ref{eq:growth-G}).
\subsubsection{Turnover}
\label{aaa:params-turnover}
Next, we present methods for resolving
the distribution of individuals entering the risk model $\bm{\hat{e}}(t)$ and
the rates of turnover $\phi(t)$,
assuming that entry and exit rates $\nu(t)$ and $\mu(t)$ are known.
Similar to above, we first formulate the problem as a system of equations.
Then, we explore the data and assumptions required
to solve for the values of parameters in the system.
The $(t)$ notation is omitted throughout this section for clarity,
though time-varying parameters can be estimated by
repeating the necessary calculations for each~$t$.
\par
The number of risk groups $G$ dictates the number of
unknown elements in $\bm{\hat{e}}$ and $\phi$: $G$ and $G(G-1)$, respectively.
We collect these unknowns in the vector
$\bm{\theta} = \left[\bm{\hat{e}}, \bm{y}\right]$,
where $\bm{y} = \mathrm{vec}_{i \ne j}(\phi)$.
For example, for $G = 3$, the vector $\bm{\theta}$ is defined as:
\begin{equation}
\bm{\theta} = \left[
\begin{array}{ccccccccc}
\hat{e}_1 & \hat{e}_2 & \hat{e}_3 & \phi_{12} & \phi_{13} & \phi_{21} & \phi_{23} & \phi_{31} & \phi_{32}
\end{array}\right]
\end{equation}
We then define a linear system of equations
which uniquely determine the elements of $\bm{\theta}$:
\begin{equation}\label{eq:system-matrix-app}
\bm{b} = A \thinspace \bm{\theta}
\end{equation}
where $A$ is a $M \times G^2$ matrix
and $\bm{b}$ is a $M$-length vector.
Specifically, each row in $A$ and $\bm{b}$ defines a constraint:
an assumed mathematical relationship involving one or more elements of
$\bm{\hat{e}}$~and~$\phi$.
For example, a simple constraint could be to assume the value $\hat{e}_2 = 0.20$.
Each of the following four sections introduces a type of constraint, including:
assuming a constant group size,
specifying elements of $\bm{\theta}$ directly,
assuming an average duration in a group,
and specifying a relationship between two individual rates of turnover.
Constraints may be selected and combined together based on
availability of data and plausibility of assumptions.
However, a total of $M = G^2$ constraints must be defined
in order to obtain a ``unique solution'':
exactly one value of $\bm{\theta}$ which satisfies all constraints.
The values of $\bm{\hat{e}}$ and $\phi$
can then be calculated algebraically by solving Eq.~(\ref{eq:system-matrix-app})
with $\bm{\theta} = A^{-1}\bm{b}$,
for which many algorithms exist~\citep{LAPACK}.
\paragraph{1.~Constant group size}
\label{con:const-group}
One epidemiologic feature that epidemic models consider
is whether or not the relative sizes of risk groups are constant over time
\citep{Henry2015,Boily2015}.
Assuming constant group size implies a stable level of heterogeneity over time.
To enforce this assumption,
we define the ``conservation of mass'' equation for group~$i$,
wherein the rate of change of the group
is defined as the sum of flows in\,/\,out of the group:
\begin{equation}\label{eq:mass-balance-1}
\frac{d}{dt}x_i
= \nu N \hat{e}_i + \sum_{j}{\phi_{ji} \thinspace x_j}
- \mu \thinspace x_i - \sum_{j}{\phi_{ij} \thinspace x_i}
\end{equation}
Eq.~(\ref{eq:mass-balance-1}) is written in terms of
absolute population sizes $\bm{x}$,
but can be written as proportions $\bm{\hat{x}}$
by dividing all terms by~$N$.
If we assume that the proportion of each group $\hat{x}_i$ is constant over time,
then the desired rate of change for risk group~$i$
will be equal to the rate of population growth of the risk group, $\mathcal{G} x_i$.
Substituting $\frac{d}{dt}x_i = \mathcal{G} x_i$
into Eq.~(\ref{eq:mass-balance-1}),
and simplifying yields:
\begin{equation}\label{eq:mass-balance-2}
\nu \thinspace x_i
= \nu \thinspace N \hat{e}_i + \sum_{j}{\phi_{ji} \thinspace x_j}
- \sum_{j}{\phi_{ij} \thinspace x_i}
\end{equation}
Factoring the left and right hand sides in terms of $\bm{\hat{e}}$ and $\phi$,
we obtain $G$ unique constraints.
For $G = 3$, this yields the following 3 rows as the basis of $\bm{b}$ and~$A$:
\begin{equation}\label{eq:eg-basis}
\bm{b} = \left[\begin{array}{c}
\nu x_1 \\ \nu x_2 \\ \nu x_3
\end{array}\right];\qquad
A = \left[\begin{array}{ccccccccc}
   \nu  & \cdot & \cdot & -x_1  & -x_1  &  x_2  & \cdot &  x_3  & \cdot \\
  \cdot &  \nu  & \cdot &  x_1  & \cdot & -x_2  & -x_2  & \cdot &  x_3  \\
  \cdot & \cdot &  \nu  & \cdot &  x_1  & \cdot &  x_2  & -x_3  & -x_3
\end{array}\right] 
\end{equation}
These $G$ constraints ensure risk groups do not change size over time.
However, a unique solution requires
an additional $G(G-1)$ constraints.
For $G = 3$, this corresponds to 6 additional constraints.
\paragraph{2.~Specified elements}
\label{con:spec-element}
The simplest type of additional constraint is to
directly specify the values of individual elements in $\bm{\hat{e}}$~or~$\phi$.
Such constraints may be appended to $\bm{b}$~and~$A$
as an additional row $k$ using indicator notation.%
\footnote{Indicator notation, also known as ``one-hot notation'' is used to
  select one element from another vector, based on its position.
  An indicator vector is 1 in the same location as the element of interest,
  and 0 everywhere else.}
That is, with $b_k$ as the specified value $v$,
and $A_k$ as the indicator vector,
with $1$ in the same position as the desired element in~$\bm{\theta}$:
\begin{equation}\label{eq:spec-elem}
b_k = v;\qquad
A_k = [0,\dots,1,\dots,0]
\end{equation}
For example, for $G = 3$, if it is known that 20\% of individuals
enter directly into risk group $2$ upon entry into the model ($\hat{e}_2 = 0.20$),
then $\bm{b}$ and $A$ can be augmented with:
\begin{equation}\label{eq:eg-spec}
b_k = \left[\begin{array}{c} 0.20 \end{array}\right];\qquad
A_k = \left[\begin{array}{ccccccccc}
  \cdot & 1 & \cdot & \cdot & \cdot & \cdot & \cdot & \cdot & \cdot
\end{array}\right] 
\end{equation}
since $\hat{e}_2$ is the second element in~$\bm{\theta}$.
If the data suggest zero turnover from group~$i$ to group~$j$,
then Eq.~(\ref{eq:eg-spec}) can also be used to set $\phi_{ij} = 0$.
\par
The elements of $\bm{\hat{e}}$ must sum to one.
Therefore, specifying all elements in $\bm{\hat{e}}$
will only provide $G-1$ constraints,
as the last element will be either redundant or violate the sum-to-one rule.
As shown in Appendix~\ref{aa:eqs-e-redundant},
the sum-to-one rule is actually implicit in Eq.~(\ref{eq:eg-basis}),
so it is not necessary to supply a constraint like $1 = \sum_{i} \hat{e}_i$.
\paragraph{3.~Group duration}
\label{con:group-dur}
Type~1 constraints assume that the relative population size of each group remains constant.
Another epidemiologic feature that epidemic models considered
is whether or not the duration of time spent within a given risk group remains constant.
For example, in STI transmission models that include formal sex work,
it can be assumed that the duration in formal sex work work remains stable over time,
such as in \citep{Mishra2014,Boily2015}.
The duration $\delta_i$ is defined as the inverse of all rates of exit from the group:
\begin{equation}\label{eq:duration-group}
\delta_i = {\bigg(\mu + \sum_{j}{\phi_{ij}}\bigg)}^{-1}
\end{equation}
Estimates of the duration in a given group can be sourced from
cross-sectional survey data where participants are asked about
how long they have engaged in a particular practice
-- such as sex in exchange for money \citep{Watts2010}.
Data on duration may also be sourced from longitudinal data,
where repeated measures of self-reported sexual behaviour,
or proxy measures of sexual risk data,
are collected \citep{DHS,PHIAproject}.
Data on duration in each risk group can then be used to define $\phi$ by
rearranging Eq.~(\ref{eq:duration-group}) to yield:
${\delta_{i}}^{-1} - \mu = \sum_{j}{\phi_{ij}}$.
For example, if for $G = 3$,
the average duration in group $1$ is known to be $\delta_1 = 5$ years,
then $\bm{b}$ and $A$ can be augmented with another row $k$:
\begin{equation}\label{eq:eg-dur}
b_k = \left[\begin{array}{c}
{5}^{-1} - \mu
\end{array}\right];\qquad
A_k = \left[\begin{array}{ccccccccc}
  \cdot & \cdot & \cdot & 1 & 1 & \cdot & \cdot & \cdot & \cdot
\end{array}\right]
\end{equation}
\par
Similar to specifying all elements of $\bm{\hat{e}}$,
specifying $\delta_i$ may result in conflicts or redundancies with other constraints.
A conflict means it will not be possible to resolve values of $\phi$
which simultaneously satisfy all constraints, while
a redundancy means that adding one constraint does not help resolve
a unique set of values~$\bm{\theta}$.
For example, for $G = 3$,
if Type~2 constraints are used to specify $\phi_{12} = 0.1$ and $\phi_{13} = 0.1$,
and $\mu = 0.05$, then by Eq.~(\ref{eq:duration-group}), we must have
$\delta_1 = 4$.
Specifying any other value for $\delta_1$ will result in a conflict,
while specifying $\delta_1 = 4$ is redundant,
since it is already implied.
There are innumerable situations in which this may occur,
so we do not attempt to describe them all.
Section~\ref{p:solving} describes how to identify
conflicts and redundancies when they are not obvious.
\paragraph{4.~Turnover rate ratios}
\label{con:rel-turnover}
In many cases, it may be difficult to
obtain estimates of a given turnover rate $\phi_{ij}$
for use in Type~2 constraints.
However, it may be possible to estimate
relative relationships between rates of turnover,
such as:
\begin{equation}\label{eq:ratio}
r\,\phi_{ij} = \phi_{i'j'}
\end{equation}
where $r$ is a ratio relating the values of $\phi_{ij}$~and~$\phi_{i'j'}$.
For example, for $G = 3$,
let $T_1$ be the total number of individuals entering group $1$ due to turnover.
If we know that
70\% of $T_1$ originates from group $2$, while
30\% of $T_1$ originates from group $3$,
then $0.7\,T_1 = \phi_{23} \, x_2$ and $0.3\,T_1 = \phi_{13} \, x_1$,
and thus: $\phi_{23} \left(\frac{0.3\,x_2}{0.7\,x_1}\right) = \phi_{13}$.
This constraint can then be appended as another row $k$ in $\bm{b}$ and $A$ like:
\begin{equation}\label{eq:eg-ratio}
b_k = \left[\begin{array}{c}
0
\end{array}\right];\qquad
A_k = \left[\begin{array}{ccccccccc}
  \cdot & \cdot & \cdot & \cdot & \left(\frac{0.3\,x_2}{0.7\,x_1}\right) & \cdot & 1 & \cdot & \cdot
\end{array}\right] 
\end{equation}
The example in Eq.~(\ref{eq:eg-ratio}) is based on
what proportions of individuals entering a risk group~$j$
came from which former risk group~$i$,
but similar constraints may be defined based on
what proportions of individuals exiting a risk group~$i$
enter into which new risk group~$j$.
It can also be assumed that
the absolute number of individuals moving between two risk groups is equal,
in which case the relationship is:
$\phi_{ij} \left(\frac{x_i}{x_j}\right) = \phi_{ji}$.
All constraints of this type will have $b_k = 0$.
\paragraph{Solving the System}
\label{p:solving}
Table~\ref{tab:constraints-app} summarizes the four types of constraints described above.
Given a set of sufficient constraints on $\bm{\theta}$
to ensure exactly one solution, the system of equations Eq.~(\ref{eq:system-matrix-app})
can be solved using $\bm{\theta} = A^{-1}\bm{b}$.
The resulting values of $\bm{\hat{e}}$ and $\phi$ can then be used
in the epidemic model.
\par
However, we may find that we have an insufficient number of constraints, implying that
there are multiple values of the vector $\bm{\theta}$ which satisfy the constraints.
An insufficient number of constraints may be identified
by a ``rank deficiency'' warning
in numerical solvers of Eq.~(\ref{eq:system-matrix-app}) \citep{LAPACK}.
Even if $A$ has $G^2$ rows,
the system may have an insufficient number of constraints
because some constraints are redundant.
In this situation, we can pose the problem as a minimization problem, namely:
\begin{equation}\label{eq:system-optimize}
\bm{\theta}^{*} = {\arg \min}
\thinspace f(\bm{\theta}),
\quad \textrm{subject to:}
\enspace\bm{b} = A\thinspace\bm{\theta};
\enspace\bm{\theta} \ge 0
\end{equation}
where $f$ is a function which penalizes certain values of $\bm{\theta}$.
For example, $f = {\left|\left| \,\cdot\, \right|\right|}_2$
penalizes large values in $\bm{\theta}$,
so that the smallest values of $\bm{\hat{e}}$ and $\phi$
which satisfy the constraints will be resolved.%
\footnote{Numerical solutions to such problems are widely available,
  such as the Non-Negative Lease Squares solver \citep{Lawson1995},
  available in Python:
  \href{https://docs.scipy.org/doc/scipy/reference/generated/scipy.optimize.nnls.html}
  {\texttt{https://docs.scipy.org/doc/scipy/reference/generated/scipy.optimize.nnls.html}}.}
\par
Similarly, we may find that no solution exists for the given constraints,
since two or more constraints are in conflict.
Conflicting constraints may be identified by a non-zero error
in the solution to Eq.~(\ref{eq:system-matrix-app}) \citep{LAPACK}.
In this case, the conflict should be resolved by
changing or removing one of the conflicting constraints.
\begin{table}
  \centering
  \caption{Summary of constraint types for defining risk group turnover}
  \label{tab:constraints-app}
  \begin{tabular}{lccl}
  \toprule
  Name                    &            Eq.            &        E.g.         & Data requirements                                                         \\
  \midrule
  1.~Constant group size  & (\ref{eq:mass-balance-2}) & (\ref{eq:eg-basis}) & all values of $\hat{x}_i$ and $\nu$                                       \\
  2.~Specified elements   &   (\ref{eq:spec-elem})    & (\ref{eq:eg-spec})  & any value of $\hat{e}_i$ or $\phi_{ij}$                                   \\
  3.~Group duration       & (\ref{eq:duration-group}) &  (\ref{eq:eg-dur})  & any value of $\delta_i$                                                   \\
  4.~Turnover rate ratios &     (\ref{eq:ratio})      & (\ref{eq:eg-ratio}) & any relationship between two turnover rates $\phi_{ij}$ and $\phi_{i'j'}$ \\
  \bottomrule
\end{tabular}\\[1em]
\footnotesize\flushleft
$\nu$:~rate of population entry;
$\phi_{ij}$:~rate of turnover from group~$i$ to group~$j$;
$\hat{x}_i$:~proportion of individuals in risk group~$i$;
$\hat{e}_i$:~proportion of individuals entering into risk group~$i$;
$\delta_i$:~average duration spent in risk group~$i$.
\end{table}
\subsection{Previous Approaches}
\label{aa:prev-appr}
Few epidemic models of sexually transmitted infections
with heterogeneity in risk
have simulated turnover among risk groups,
and those models which have simulated turnover
have done so in various ways.
In this section, we review
three prior implementations of turnover and their assumptions.
We then highlight how the approach proposed in Section~\ref{aa:params}
could be used to achieve the same objectives.
\par
\citet{Stigum1994} simulated turnover among $G = 2$ risk groups
in a population with no exogenous entry or exit
($\nu = \mu = 0$ and hence $\bm{\hat{e}}$ is not applicable).
Turnover between the groups was balanced
in order to maintain constant risk group sizes (Type~1 constraint),%
\footnote{Due to its simplicity,
  this constraint is actually an example of both Type~1 and Type~4 constraints.}
while the rate of turnover from high to low
was specified as $\kappa$ (Type~2 constraint).
Thus, the turnover system used by \citet{Stigum1994} can be written
in the proposed framework as:
\begin{equation}\label{eq:sys-Stigum1994}
\left[\begin{array}{c}
    0    \\
  \kappa
\end{array}\right]
=
\left[\begin{array}{cc}
  \hat{x}_1 & -\hat{x}_2 \\
      1     &   \cdot
\end{array}\right]
\left[\begin{array}{c}
  \phi_{12} \\
  \phi_{21}
\end{array}\right]
,\qquad
\hat{e}_1 = \hat{e}_2 = 0
\end{equation}
\par
\citet{Henry2015} also simulated turnover among $G = 2$ risk groups,
but considered exogenous entry and exit, both at a rate $\mu$.
The authors used the notation $f_i$ for our $\hat{x}_i$, and assumed that
the population of individuals entering into the modelled population
had the same distribution of risk groups as the modelled population itself:
$\hat{e}_i = f_i$ (Type~2 constraint).
The authors further maintained constant risk group sizes (Type~1 constraint)
by analytically balancing turnover between the two groups using:
$\phi_{12} = \omega \hat{x}_2 ;\enspace \phi_{21} = \omega \hat{x}_1$,
where $\omega$ is a constant.
However, it can be shown that this analytical approach
is also the solution to the following combination of Type~1 and Type~2 constraints:
\begin{equation}\label{eq:sys-Henry2015}
\left[\begin{array}{c}
      0      \\
  \omega f_2
\end{array}\right]
=
\left[\begin{array}{cc}
  f_1 & -f_2  \\
   1  & \cdot
\end{array}\right]
\left[\begin{array}{c}
  \phi_{12} \\
  \phi_{21}
\end{array}\right]
,\qquad
\hat{e}_i = f_i
\end{equation}
\par
\citet{Eaton2014} simulated turnover among $G = 3$ risk groups,
considering a distribution of risk among
individuals entering into the modelled population $\bm{\hat{e}}$
which was different from $\bm{\hat{x}}$.
Turnover was considered from
high-to-medium, high-to-low, and medium-to-low risk,
all with an equal rate $\psi$;
the reverse transition rates were set to zero
(six total Type~2 constraints).
Given the unidirectional turnover,
risk group sizes were maintained using the values of $\hat{e}_i$,
computed using Type~1 constraints as follows:
\begin{equation}\label{eq:sys-Eaton2014-simple}
\left[\begin{array}{l}
\nu x_1 + 2 x_1 \psi \\
\nu x_2 -   x_1 \psi + x_2 \psi \\
\nu x_3 -   x_1 \psi - x_2 \psi
\end{array}\right]
=
\left[\begin{array}{ccc}
\nu  & \cdot & \cdot \\
\cdot &  \nu  & \cdot \\
\cdot & \cdot &  \nu  \\
\end{array}\right]
\left[\begin{array}{c}
e_1 \\
e_2 \\
e_3
\end{array}\right]
,\qquad
\begin{array}{c}
\phi_{12} = \phi_{13} = \phi_{23} = \psi\\
\phi_{21} = \phi_{31} = \phi_{32} = 0
\end{array}
\end{equation}

\par
In sum, the framework for modelling turnover presented in this section
aims to generalize all previous implementations.
In so doing, we hope to clarify the requisite assumptions,
dependencies on epidemiologic data,
and relationships between previous approaches.

\clearpage
\section{Supplemental Equations}\label{a:eqs}
\begin{table}[H]
  \centering
  \caption{Notation}
  \label{tab:notation}
  \small\renewcommand{\arraystretch}{0.8}
  \begin{tabular}{cl}
    \toprule
        Symbol      & Definition                                                           \\
    \midrule
          $i$       & risk group index                                                     \\
          $j$       & risk group index for ``other'' group in turnover                     \\
          $k$       & risk group index for ``other'' group in incidence                    \\
          $t$       & time                                                                 \\
    $\mathcal{S}_i$ & number of susceptible individuals in risk group~$i$                  \\
    $\mathcal{I}_i$ & number of infectious individuals in risk group~$i$                   \\
    $\mathcal{T}_i$ & number of treated individuals in risk group~$i$                      \\
          $N$       & total population size                                                \\
         $\nu$      & rate of population entry                                             \\
         $\mu$      & rate of population exit                                              \\
      $\phi_{ij}$   & rate of turnover from group~$i$ to group~$j$                         \\
      $\lambda_i$   & force of infection among susceptibles in risk group~$i$              \\
        $\tau$      & rate of treatment initiation among infected                          \\
      $\hat{x}_i$   & proportion of individuals in risk group~$i$                          \\
      $\hat{e}_i$   & proportion of individuals entering into risk group~$i$               \\
      $\delta_i$    & average duration spent in risk group~$i$                             \\
         $C_i$      & number of partners per year among individuals in risk group~$i$      \\
        $\beta$     & probability of transmission per partnership                          \\
      $\rho_{ik}$   & probability of partnership formation between risk groups $i$ and $k$ \\
    \bottomrule
  \end{tabular}
\end{table}

\subsection{Model Equations}\label{aa:eqs-model}
\newcommand{\bracelabel}[2]{%
  \underbrace{\smash{\phantom{#1}}}_{\textrm{#2}}
}
\begin{subequations}
\label{eq:model}
\begin{alignat}{9}
\frac{d}{dt}\mathcal{S}_i(t) &=
      + \sum_j \phi_{ji} \mathcal{S}_j(t)
&&    - \sum_j \phi_{ij} \mathcal{S}_i(t)
&&    - \mu \mathcal{S}_i(t)
&&    + \nu \hat{e}_i N(t)
&&    - \lambda_i(t) \mathcal{S}_i(t)
&&
\label{eq:model-S}\\
\frac{d}{dt}\mathcal{I}_i(t) &=
      + \sum_j \phi_{ji} \mathcal{I}_j(t)
&&    - \sum_j \phi_{ij} \mathcal{I}_i(t)
&&    - \mu \mathcal{I}_i(t)
&&
&&    + \lambda_i(t) \mathcal{S}_i(t)
&&    - \tau \mathcal{I}_i(t)
\label{eq:model-I}\\
\frac{d}{dt}\mathcal{T}_i(t) &=
      + \sum_j \phi_{ji} \mathcal{T}_j(t)
&&    - \sum_j \phi_{ij} \mathcal{T}_i(t)
&&    - \mu \mathcal{T}_i(t)
&& 
&&
&&    + \tau \mathcal{I}_i(t)
\label{eq:model-T}\\[-1em]
&\hphantom{=}
      \bracelabel{+ \sum_j \phi_{ji} \mathcal{S}_j(t)}{turnover into}
&&    \bracelabel{+ \sum_j \phi_{ij} \mathcal{S}_i(t)}{turnover from}
&&    \bracelabel{+ \mu \mathcal{S}_i(t)}{death}
&&    \bracelabel{+ \nu \hat{e}_i N(t)}{birth}
&&    \bracelabel{+ \lambda_i(t) \mathcal{S}_i(t)}{incidence}
&&    \bracelabel{+ \tau \mathcal{I}_i(t)}{treatment}
\nonumber
\end{alignat}
\end{subequations}
\\ 
\subsection{Complete Example Turnover System}\label{aa:eqs-turnover}
\begin{equation}\label{eq:turnover-system-g=3}
\footnotesize
\begin{array}{r}
  \left.\begin{array}{c}\\ \textrm{constant group size}  \\ \\\end{array}\right\{\\
  \left.\begin{array}{c}\\ \textrm{specified $e$}        \\ \\\end{array}\right\{\\
  \left.\begin{array}{c}\\ \textrm{group duration}       \\ \\\end{array}\right\{\\
  \left.\begin{array}{c}\\ \textrm{turnover rate ratios} \\ \\\end{array}\right\{\\
\end{array}
\left[\begin{array}{c}
         \nu x_1        \\
         \nu x_2        \\
         \nu x_3        \\
          e^*_1         \\
          e^*_2         \\
          e^*_3         \\
  {\delta_1}^{-1} - \mu \\
  {\delta_2}^{-1} - \mu \\
  {\delta_3}^{-1} - \mu \\
            0           \\
            0           \\
            0
\end{array}\right]
=
\left[\begin{array}{ccccccccc}
   \nu  & \cdot & \cdot & -x_1  & -x_1  &  x_2  & \cdot &  x_3  & \cdot \\
  \cdot &  \nu  & \cdot &  x_1  & \cdot & -x_2  & -x_2  & \cdot &  x_3  \\
  \cdot & \cdot &  \nu  & \cdot &  x_1  & \cdot &  x_2  & -x_3  & -x_3  \\
    1   & \cdot & \cdot & \cdot & \cdot & \cdot & \cdot & \cdot & \cdot \\
  \cdot &   1   & \cdot & \cdot & \cdot & \cdot & \cdot & \cdot & \cdot \\
  \cdot & \cdot &   1   & \cdot & \cdot & \cdot & \cdot & \cdot & \cdot \\
  \cdot & \cdot & \cdot &   1   &   1   & \cdot & \cdot & \cdot & \cdot \\
  \cdot & \cdot & \cdot & \cdot & \cdot &   1   &   1   & \cdot & \cdot \\
  \cdot & \cdot & \cdot & \cdot & \cdot & \cdot & \cdot &   1   &   1   \\
  \cdot & \cdot & \cdot &  x_1  & \cdot & -x_2  & \cdot & \cdot & \cdot \\
  \cdot & \cdot & \cdot & \cdot &  x_1  & \cdot & \cdot & -x_3  & \cdot \\
  \cdot & \cdot & \cdot & \cdot & \cdot & \cdot &  x_2  & \cdot & -x_3
\end{array}\right]
\left[\begin{array}{c}
     e_1     \\
     e_2     \\
     e_3     \\
  \phi_{12} \\
  \phi_{13} \\
  \phi_{21} \\
  \phi_{23} \\
  \phi_{31} \\
  \phi_{32} \\
\end{array}\right]
\end{equation}
\\

\subsection{Redundancy in specifying all elements of $\hat{e}$}\label{aa:eqs-e-redundant}
Whenever it is assumed that risk groups do not change size,
$G$ rows of the form shown in Eq.~(\ref{eq:eg-basis})
are added to $\bm{b}$ and $A$:
\begin{equation}\tag{\ref{eq:eg-basis}}
\bm{b} = \left[\begin{array}{c}
\nu x_1 \\ \nu x_2 \\ \nu x_3
\end{array}\right];\qquad
A = \left[\begin{array}{ccccccccc}
 \nu  & \cdot & \cdot & -x_1  & -x_1  &  x_2  & \cdot &  x_3  & \cdot \\
\cdot &  \nu  & \cdot &  x_1  & \cdot & -x_2  & -x_2  & \cdot &  x_3  \\
\cdot & \cdot &  \nu  & \cdot &  x_1  & \cdot &  x_2  & -x_3  & -x_3  \\
\end{array}\right]
\end{equation}
After multiplying by $\bm{\theta}$, these $G$ rows can be row-reduced by summing to obtain:
\begin{equation}
\begin{aligned}
\left[ \nu x_1 + \nu x_2 + \nu x_3 \right] &= 
\left[ \nu e_1 + \nu e_2 + \nu e_3
+ 0\,\phi_{12} + 0\,\phi_{13} + 0\,\phi_{21} + 0\,\phi_{23} + 0\,\phi_{31} + 0\,\phi_{32}
\right]\\
\nu \left[ x_1 + x_2 + x_3 \right] &= 
\nu \left[ e_1 + e_2 + e_3 \right]
\end{aligned}
\end{equation}
which therefore implies that $\sum_{i} x_i = \sum_{i} e_i$,
or equivalently $\sum_{i} \hat{x}_i = \sum_{i} \hat{e}_i = 1$.
Thus, it is redundant to specify all $G$ elements of $\hat{e}$,
as the final element will be dictated by constant group size constraints.
\subsection{Factors of Incidence}\label{aa:eqs-incidence}
Substituting the proportional mixing definition of $\rho_{ik}$ into
the incidence equation, Eq.~(\ref{eq:foi}), we have:
\begin{align}
\lambda_i
  &= C_i \sum_k \rho_{ik} \beta \thinspace \frac{\mathcal{I}_k}{\mathcal{X}_k}
    \nonumber\\
  &= C_i \beta \sum_k
    \frac{C_k \mathcal{X}_k}{\sum_{\mathrm{k}}C_{\mathrm{k}}\mathcal{X}_{\mathrm{k}}}
    \frac{\mathcal{I}_k}{\mathcal{X}_k}
    \nonumber\\
  &= C_i \beta \underbrace{
      \frac{\sum_k C_k \mathcal{I}_k}{\sum_{\mathrm{k}}C_{\mathrm{k}}\mathcal{X}_{\mathrm{k}}}
    }_{f}
\end{align}
We can factor the term $f$ as:
\begin{align}
  f
  &= \frac{\sum_k C_k \mathcal{I}_k}{\sum_{\mathrm{k}}C_{\mathrm{k}}\mathcal{X}_{\mathrm{k}}}
     \nonumber\\
  &= \frac{\sum_k C_k \mathcal{I}_k}{\sum_k \mathcal{I}_k}
       \cdot
     \frac{\sum_k \mathcal{I}_k}{\sum_k \mathcal{X}_k}
       \cdot
     \frac{\sum_k \mathcal{X}_k}{\sum_k C_k \mathcal{X}_k}
\intertext{which we recognize as the following terms:}
  &= \hat{C}_{\mathcal{I}} \cdot \hat{\mathcal{I}} \cdot \hat{C}^{-1}
\end{align}
Namely,
\begin{enumerate}
  \item $\hat{C}_{\mathcal{I}}$ is the average number of partners among infectious individuals
  \item $\hat{\mathcal{I}}$ is the proportion of the population who are infectious (overall prevalence)
  \item $\hat{C}$ is the average number of partners among all individuals (constant)
\end{enumerate}
Therefore, only two non-constant factors control incidence per susceptible:
1) the average number of partners among infectious individuals $\hat{C}_{\mathcal{I}}$, and
2) overall prevalence $\hat{\mathcal{I}}$.
The product of these factors $\hat{C}_{\mathcal{I}}\,\hat{\mathcal{I}}$,
scaled by $\beta \, C_i / \hat{C}$,
then gives $\lambda_i$.
In fact, the incidence in each group individually is proportional to
incidence overall, as $C_i$ is only factor depending on $i$.

\clearpage
\section{Supplemental Results}\label{a:results}
\subsection{Equilibrium health states and rates of transition}
\begin{figure}[H]
  \begingroup\centering
  \begin{subfigure}{0.4\linewidth}
    \includegraphics[width=\linewidth]{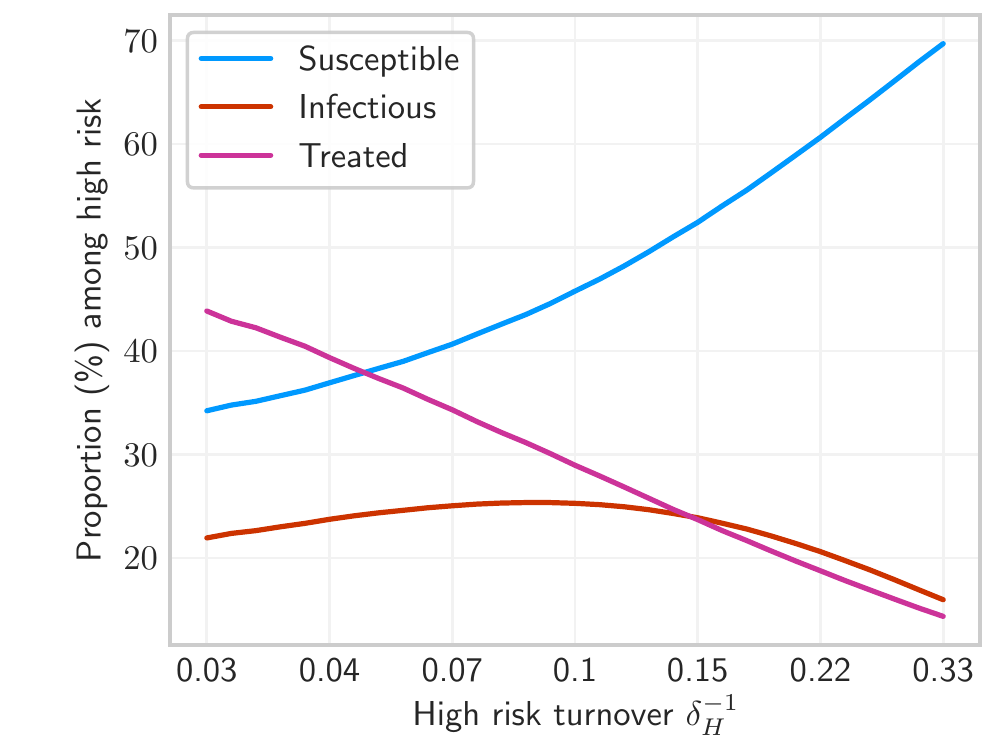}
    \caption{High risk}
    \label{fig:1d-health-high}
  \end{subfigure}
  \begin{subfigure}{0.4\linewidth}
    \includegraphics[width=\linewidth]{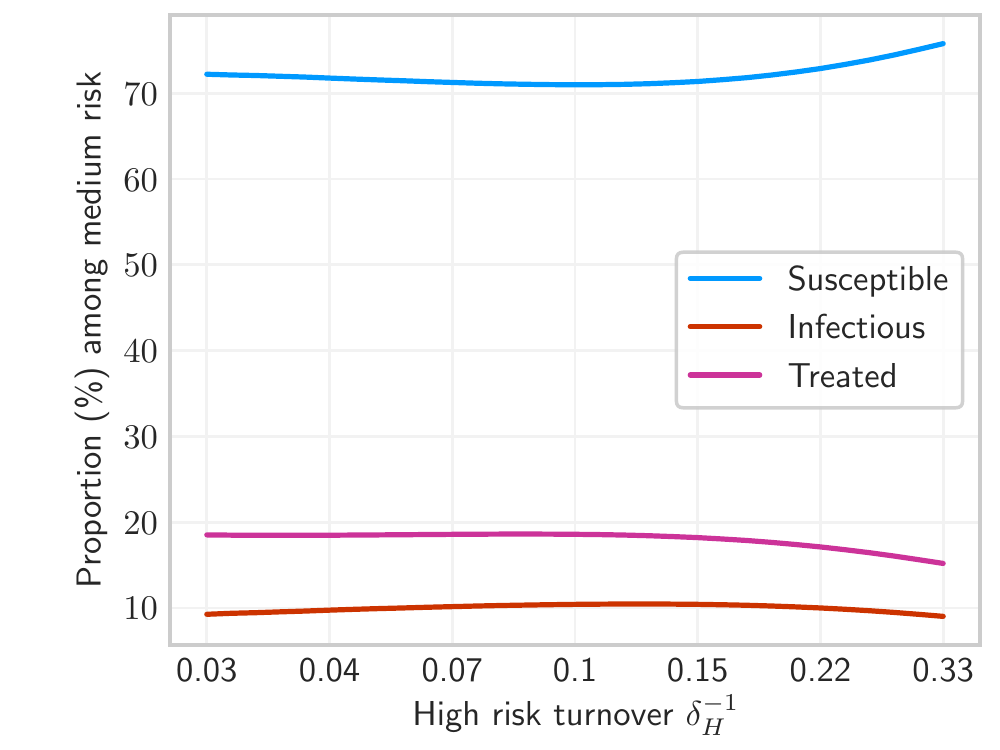}
    \caption{Medium risk}
    \label{fig:1d-health-med}
  \end{subfigure}
  \begin{subfigure}{0.4\linewidth}
    \includegraphics[width=\linewidth]{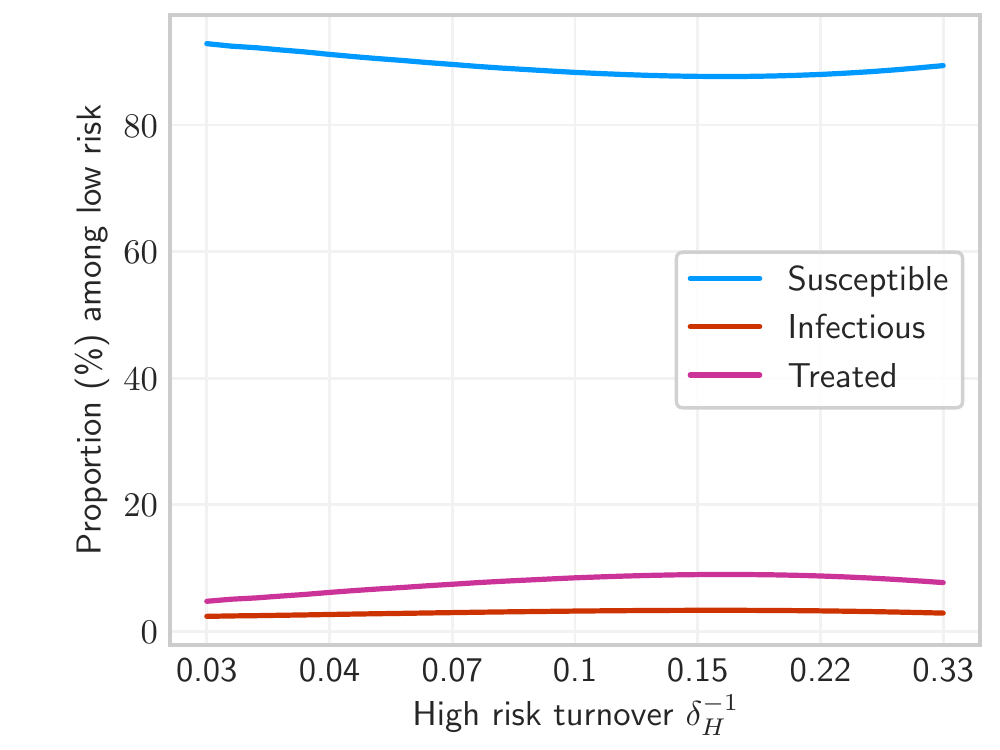}
    \caption{Low risk}
    \label{fig:1d-health-low}
  \end{subfigure}
  \begin{subfigure}{0.4\linewidth}
    \includegraphics[width=\linewidth]{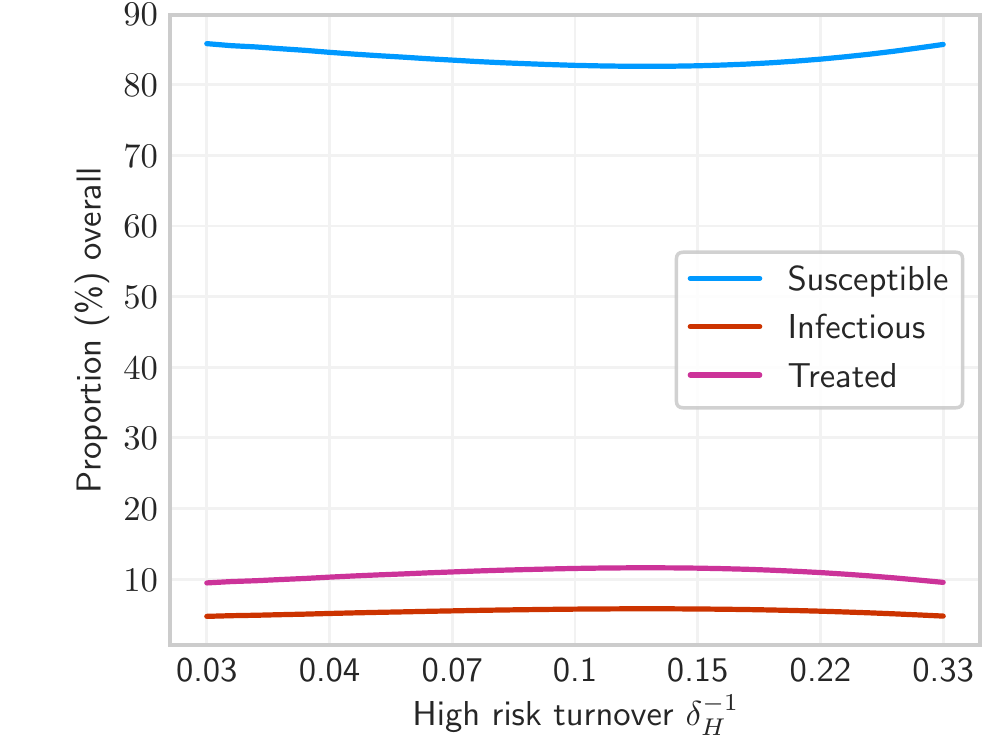}
    \caption{Overall}
    \label{fig:1d-health-all}
  \end{subfigure}
  \\\endgroup
  \caption{Equilibrium health state proportions under different rates of turnover.}
  \label{fig:1d-health}
  \footnotesize\input{x-axis.tex}
\end{figure}
\begin{figure}[H]
  \begingroup\centering
  \includegraphics[width=\linewidth]{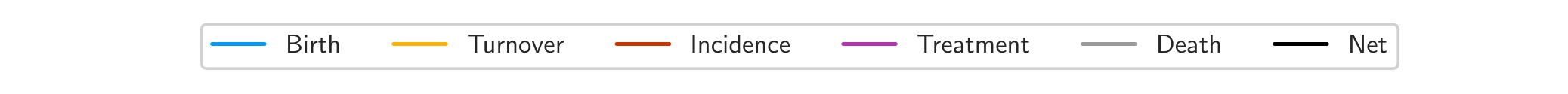}
  \begin{subfigure}{0.33\linewidth}
    \includegraphics[width=\linewidth]{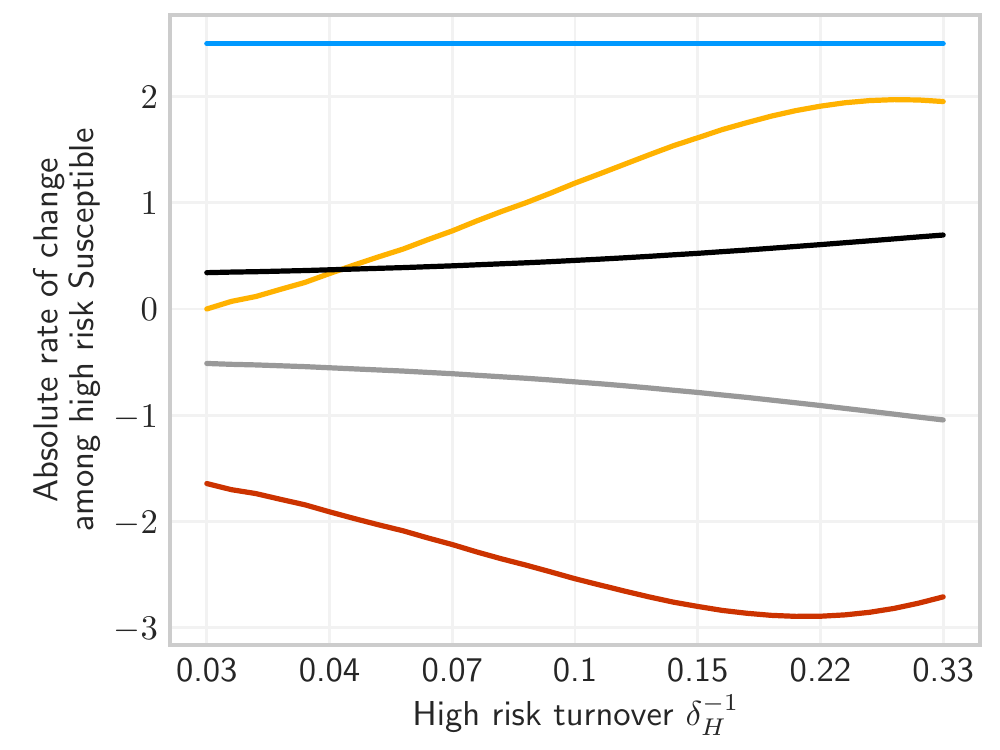}
    \caption{High risk susceptible}\label{fig:dX-app-high-S}
  \end{subfigure}%
  \begin{subfigure}{0.33\linewidth}
    \includegraphics[width=\linewidth]{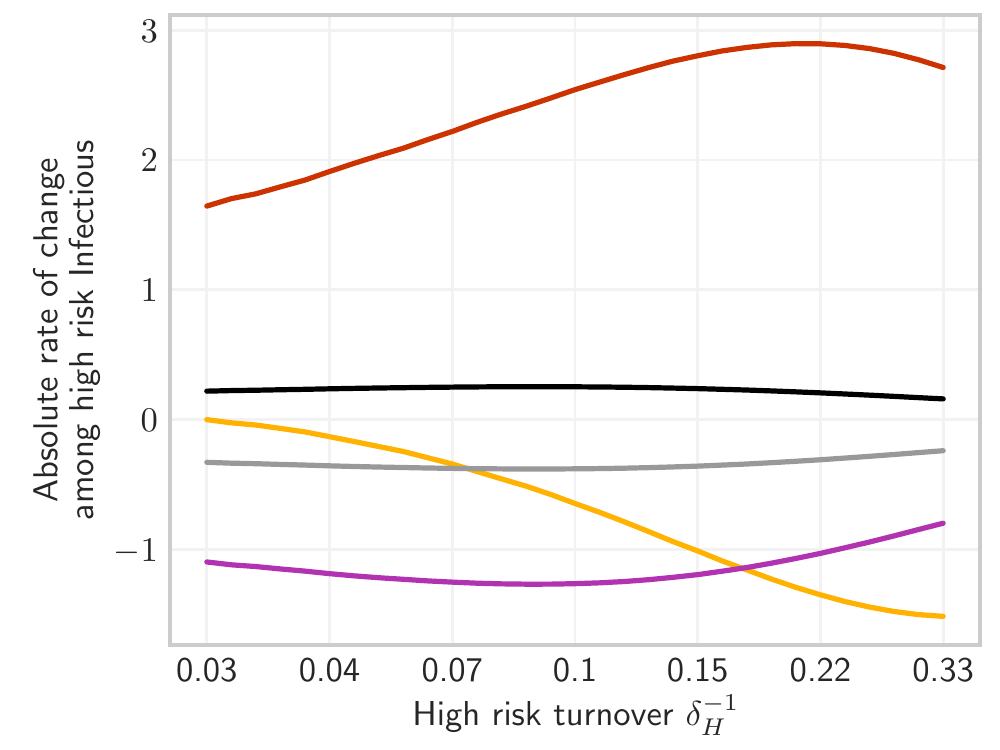}
    \caption{High risk infectious}\label{fig:dX-app-high-I}
  \end{subfigure}%
  \begin{subfigure}{0.33\linewidth}
    \includegraphics[width=\linewidth]{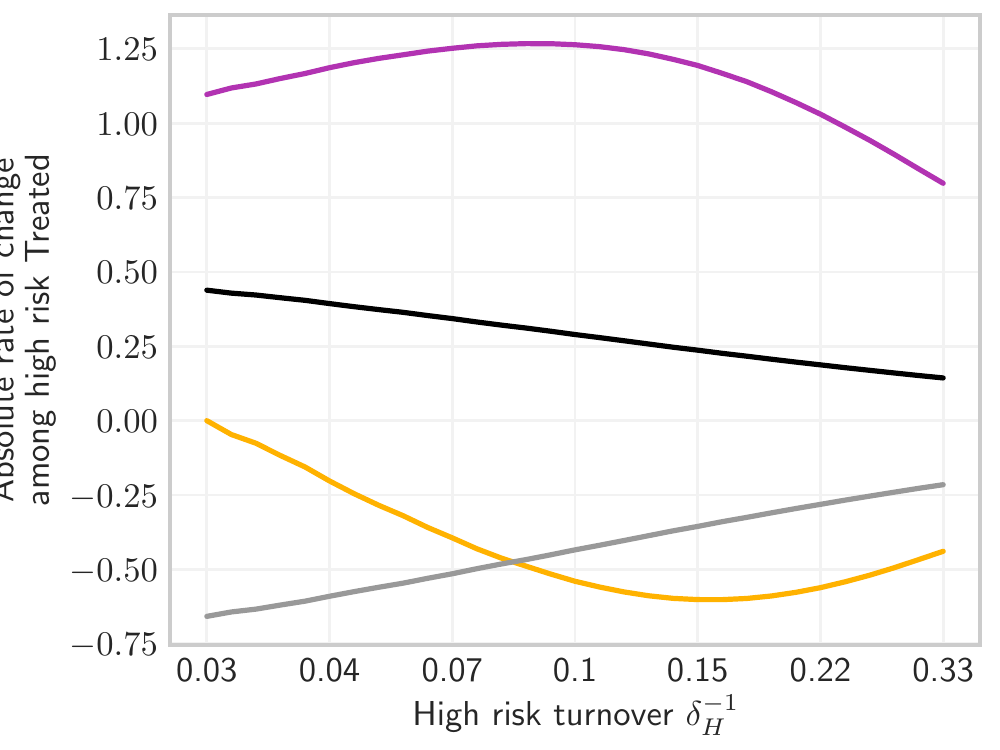}
    \caption{High risk treated}\label{fig:dX-app-high-T}
  \end{subfigure}\\
  \begin{subfigure}{0.33\linewidth}
    \includegraphics[width=\linewidth]{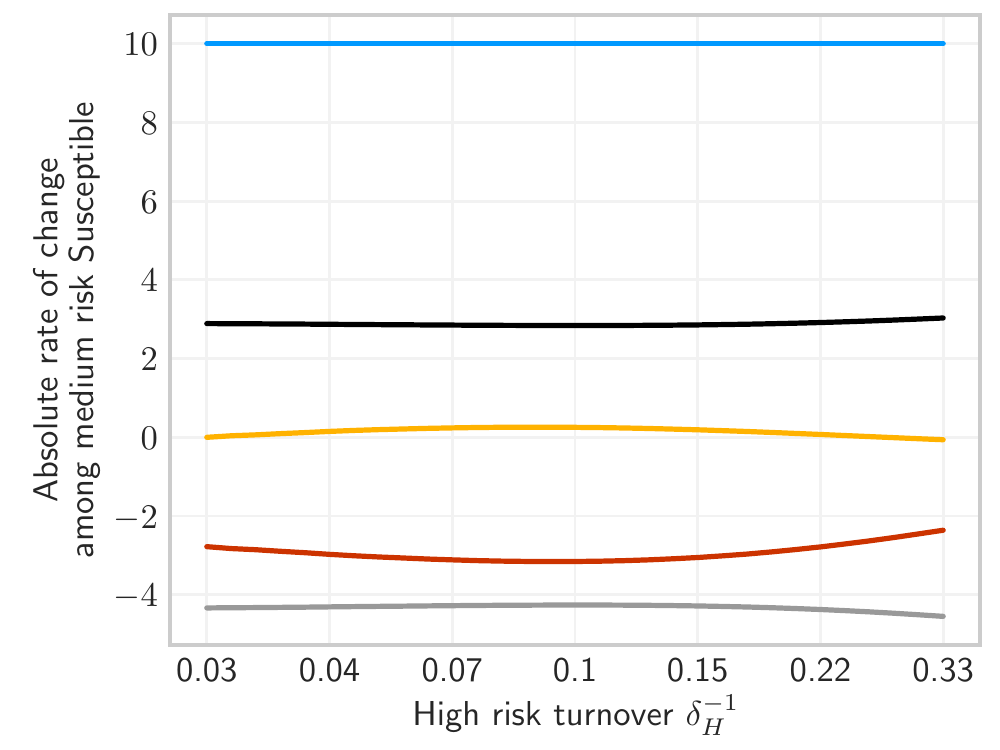}
    \caption{Medium risk susceptible}\label{fig:dX-app-med-S}
  \end{subfigure}%
  \begin{subfigure}{0.33\linewidth}
    \includegraphics[width=\linewidth]{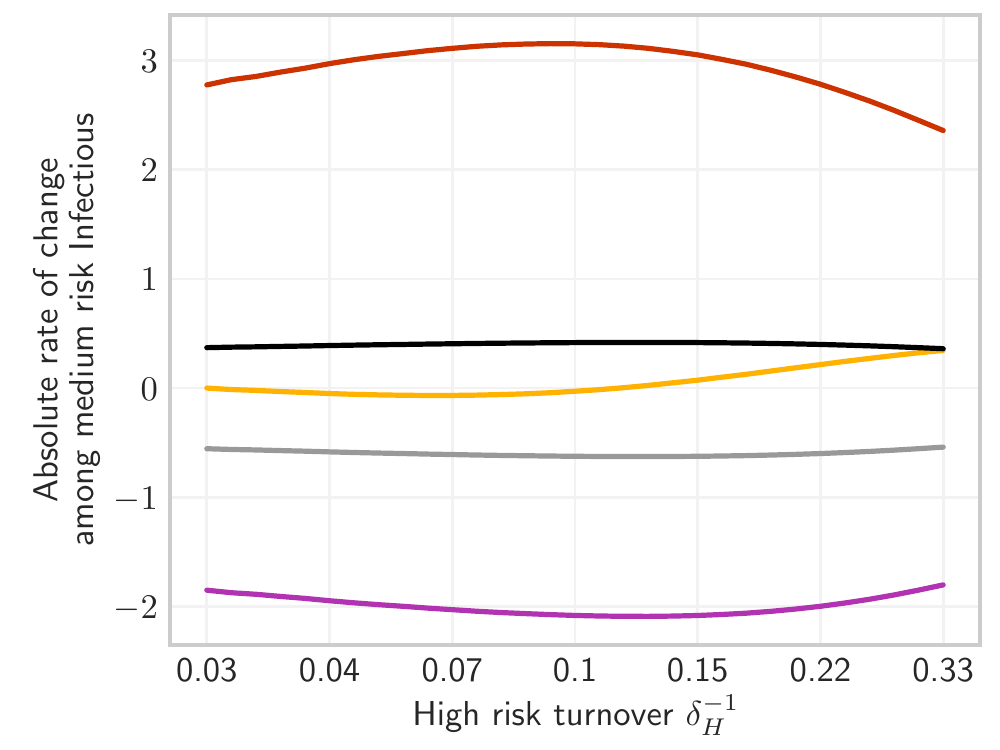}
    \caption{Medium risk infectious}\label{fig:dX-app-med-I}
  \end{subfigure}%
  \begin{subfigure}{0.33\linewidth}
    \includegraphics[width=\linewidth]{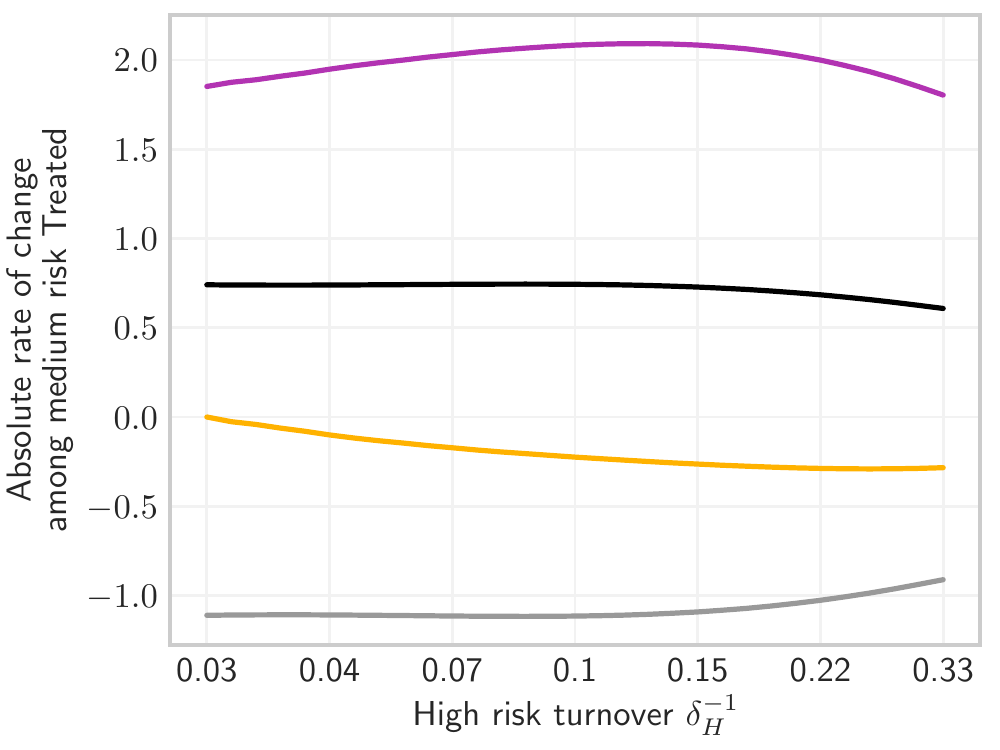}
    \caption{Medium risk treated}\label{fig:dX-app-med-T}
  \end{subfigure}\\
  \begin{subfigure}{0.33\linewidth}
    \includegraphics[width=\linewidth]{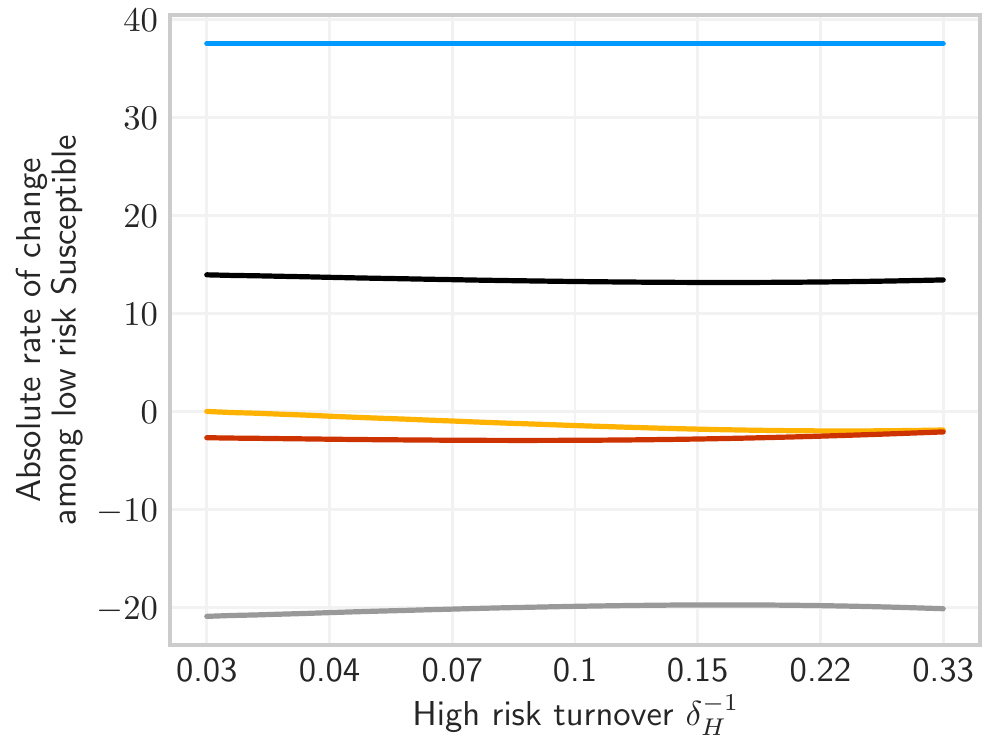}
    \caption{Low risk susceptible}\label{fig:dX-app-low-S}
  \end{subfigure}%
  \begin{subfigure}{0.33\linewidth}
    \includegraphics[width=\linewidth]{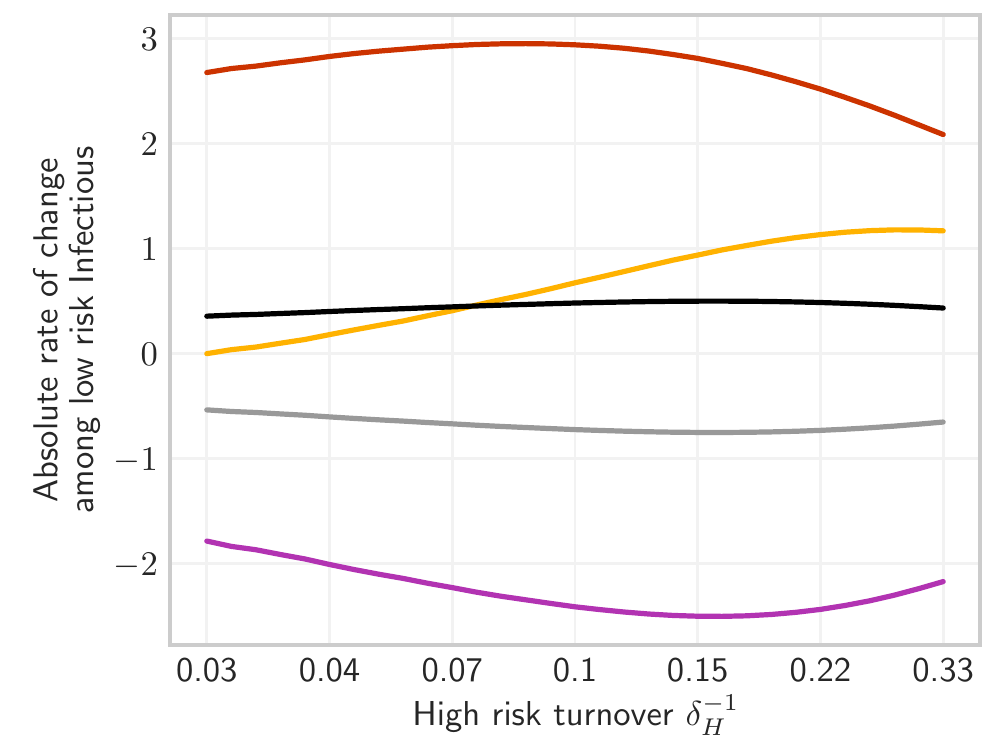}
    \caption{Low risk infectious}\label{fig:dX-app-low-I}
  \end{subfigure}%
  \begin{subfigure}{0.33\linewidth}
    \includegraphics[width=\linewidth]{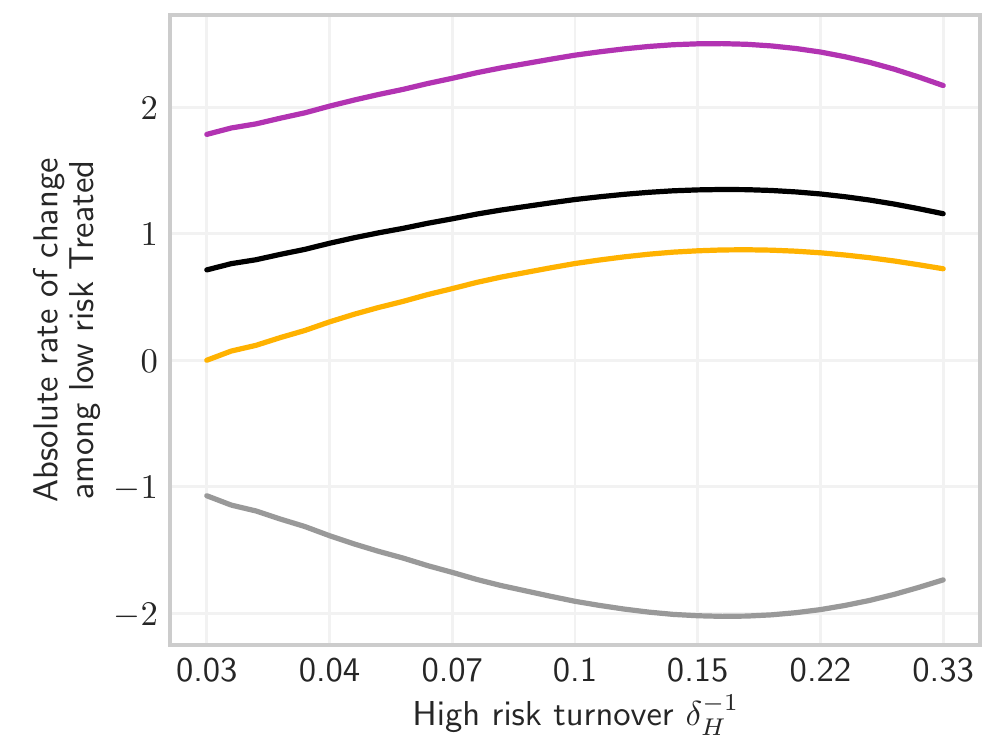}
    \caption{Low risk treated}\label{fig:dX-app-low-T}
  \end{subfigure}\\
  \endgroup
  \caption{Absolute rates of change at equilibrium
    (number of individuals gained/lost per year)
    among individuals in each health state and risk group,
    broken down by type of change:
    gain via births,
    loss/gain via incident infections,
    loss/gain via treatment,
    loss/gain via turnover,
    loss via death,
    and net change.
    Based on Eq.~(\ref{eq:model}).}
  \label{fig:dX-app}
  \footnotesize\input{x-axis.tex}
  Rates of change do not sum to zero due to population growth.
\end{figure}
\begin{figure}[H]
  \centerline{\includegraphics[width=0.5\linewidth]{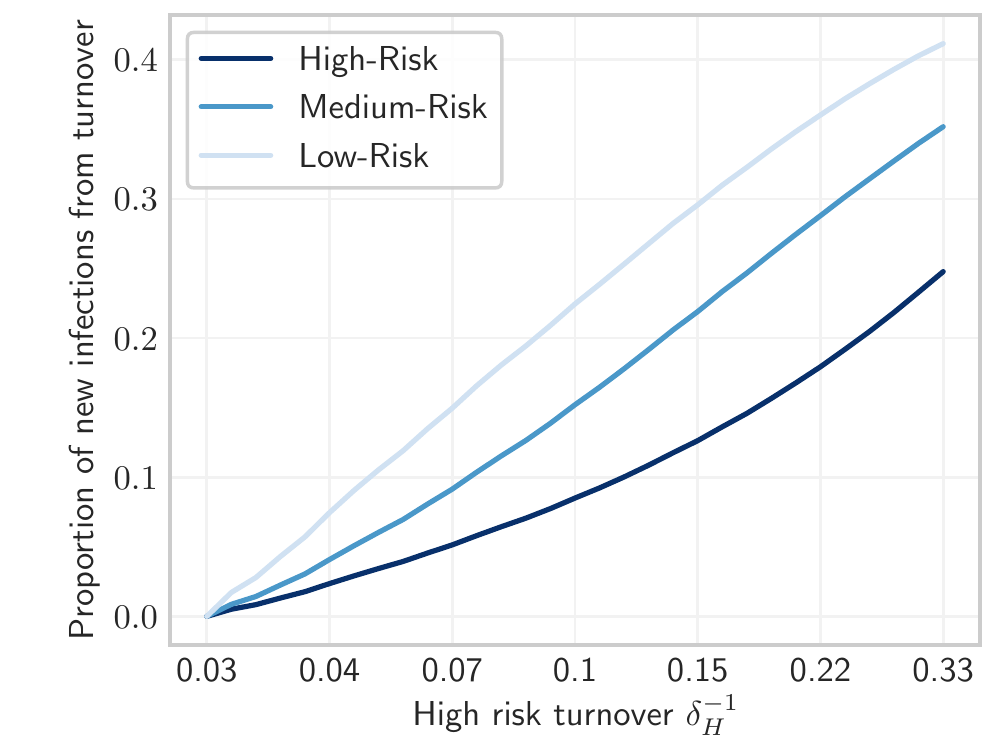}}
  \caption{Proportion of new infectious individuals in each risk group
    which are from turnover of infectious individuals,
    as opposed to incident infection of susceptible individuals in the risk group.}
  \label{fig:new-inf-phi-vs-lambda}
\end{figure}
\subsection{Equilibrium Prevalence Ratios}
\begin{figure}[H]
  \begingroup\centering
  \begin{subfigure}{0.33\linewidth}
    \includegraphics[width=\linewidth]{{1d-ratio-prevalence-high-low-tau=0.1}.pdf}
    \caption{High vs Low risk}
    \label{fig:1d-ratio-prevalence-high-low}
  \end{subfigure}%
  \begin{subfigure}{0.33\linewidth}
    \includegraphics[width=\linewidth]{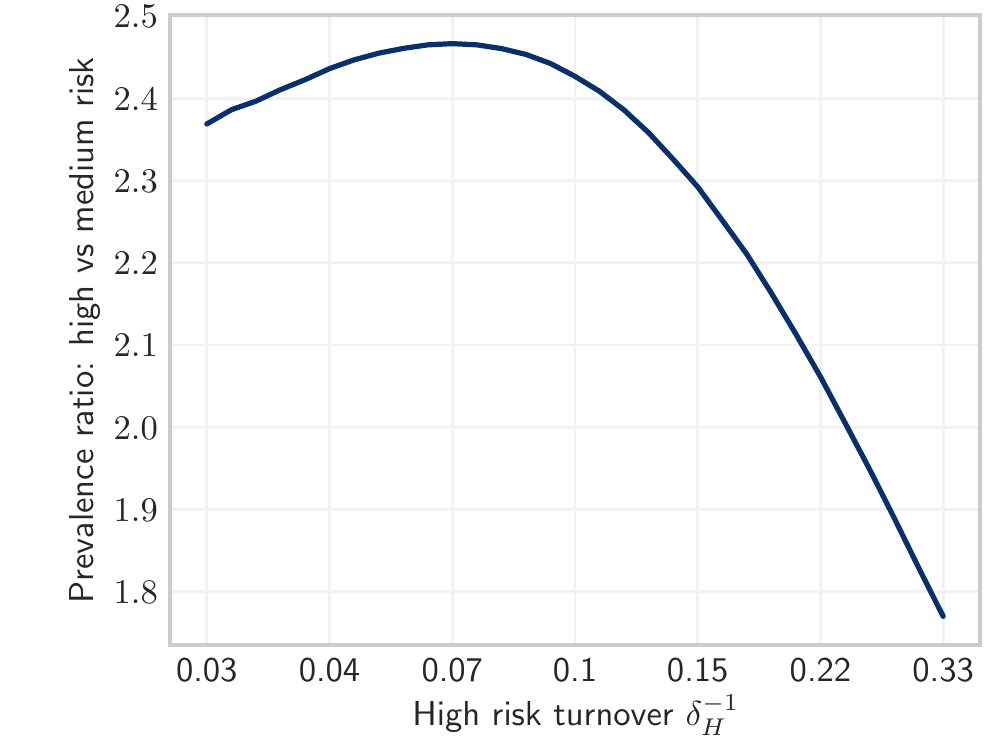}
    \caption{High vs Medium Risk}
    \label{fig:1d-ratio-prevalence-high-med}
  \end{subfigure}%
  \begin{subfigure}{0.33\linewidth}
    \includegraphics[width=\linewidth]{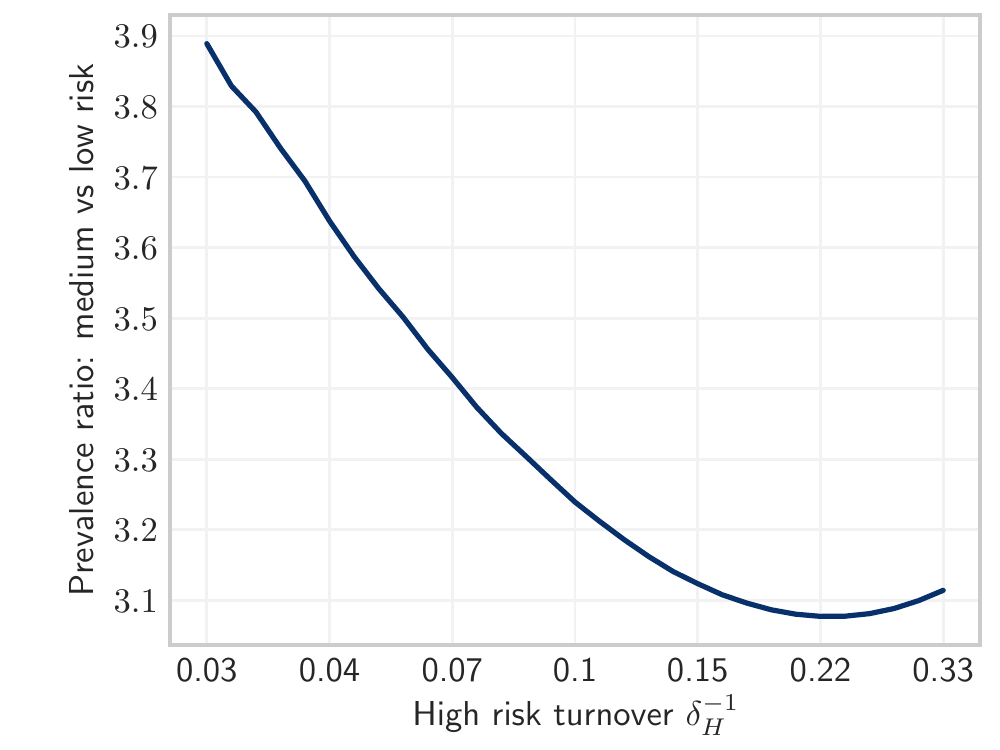}
    \caption{Medium vs Low Risk}
    \label{fig:1d-ratio-prevalence-med-low}
  \end{subfigure}%
  \\\endgroup
  \caption{Equilibrium prevalence ratios between risk groups
    under different rates of turnover.}
  \label{fig:1d-ratio-prevalence}
  \footnotesize\input{x-axis.tex}
\end{figure}
\subsection{Equilibrium Incidence}
\begin{figure}[H]
  \begingroup\centering
  \begin{subfigure}{0.33\linewidth}
    \includegraphics[width=\linewidth]{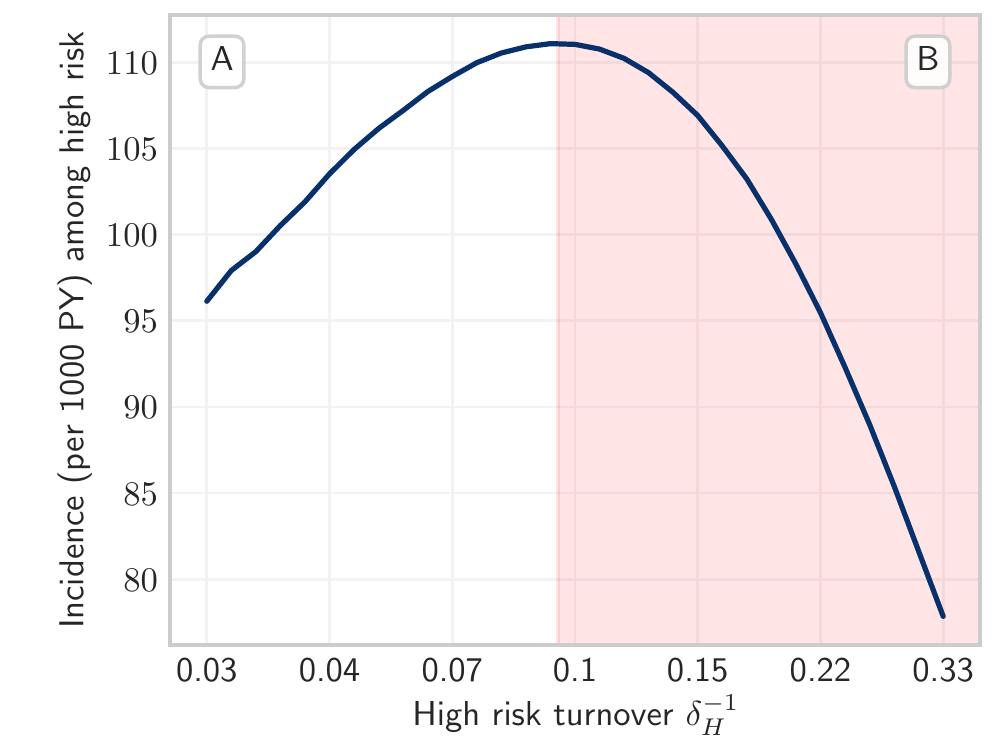}
    \caption{High risk}
    \label{fig:1d-incidence-high}
  \end{subfigure}%
  \begin{subfigure}{0.33\linewidth}
    \includegraphics[width=\linewidth]{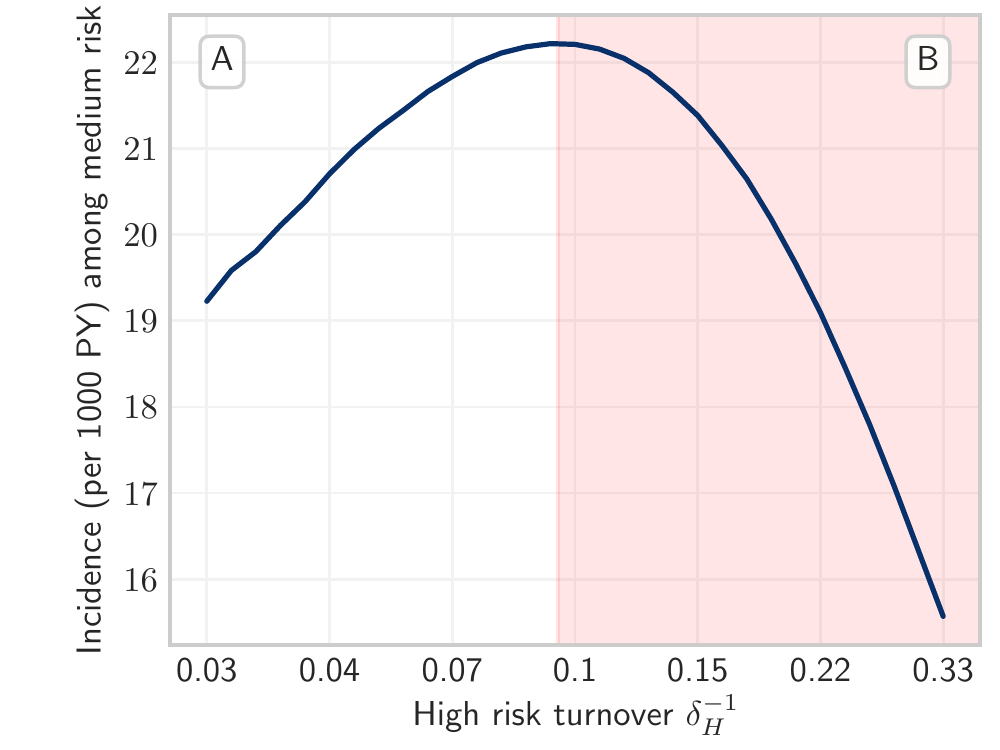}
    \caption{Medium risk}
    \label{fig:1d-incidence-med}
  \end{subfigure}%
  \begin{subfigure}{0.33\linewidth}
    \includegraphics[width=\linewidth]{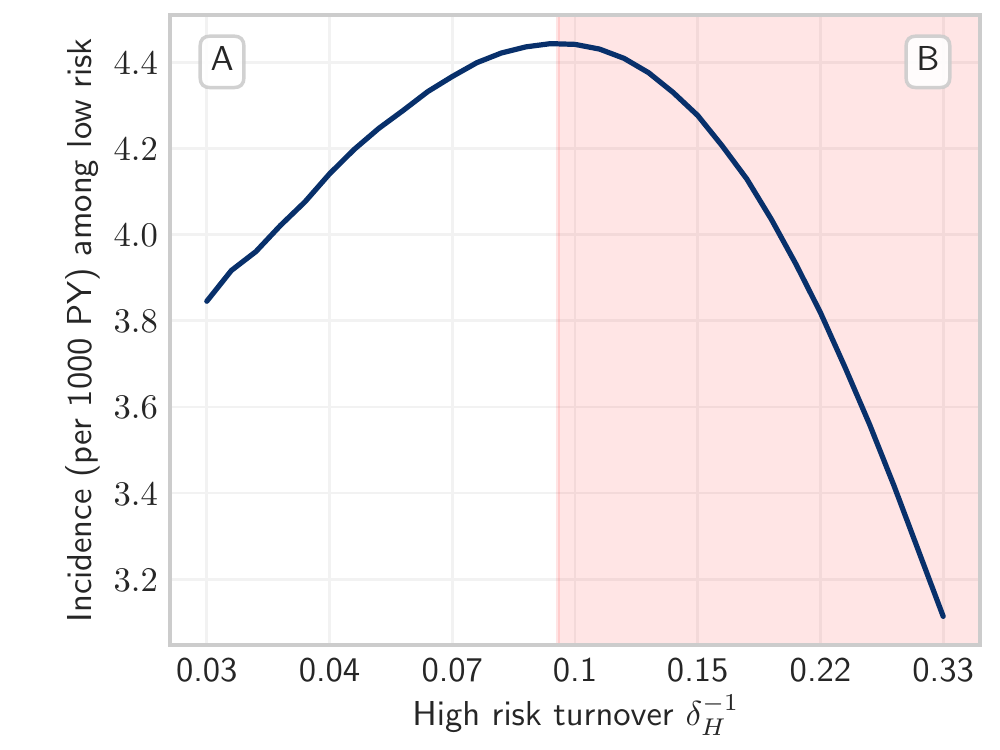}
    \caption{Low risk}
    \label{fig:1d-incidence-low}
  \end{subfigure}%
  \\\endgroup
  \caption{Equilibrium incidence among high, medium, and low risk groups
    under different rates of turnover.}
  \label{fig:1d-incidence}
  \footnotesize\input{x-axis.tex}
  Incidence in each risk group is proportional to overall incidence
  with $C_i$ as a scale factor.
\end{figure}
\begin{figure}[H]
  \begingroup\centering
  \begin{subfigure}{0.33\linewidth}
    \includegraphics[width=\linewidth]{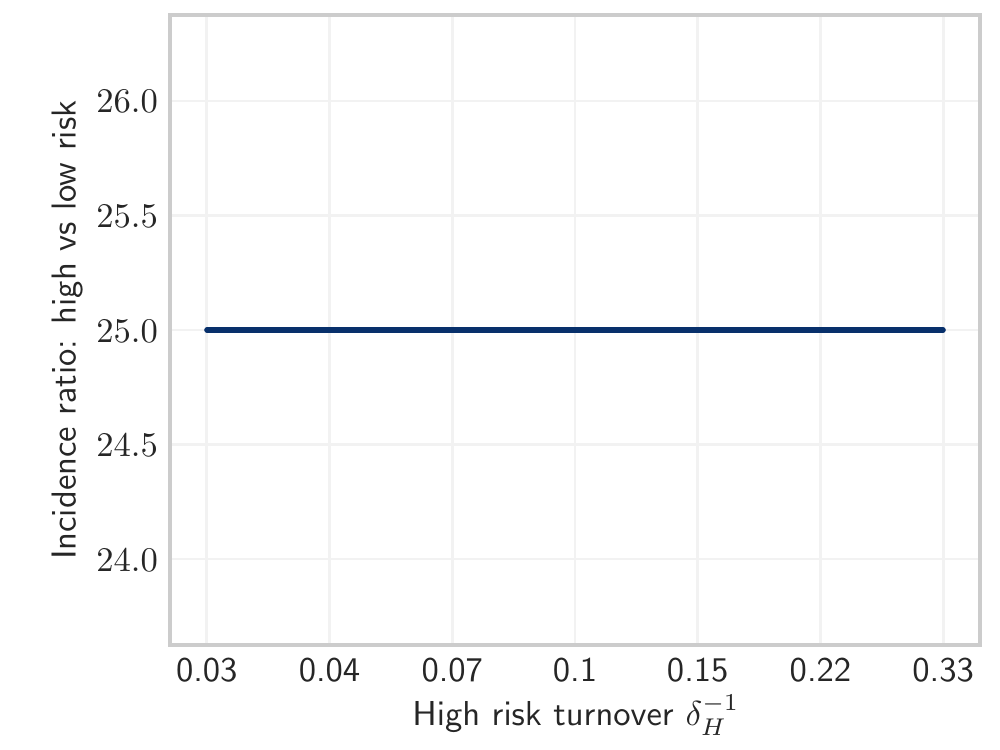}
    \caption{High vs low risk}
    \label{fig:1d-ratio-incidence-high-low}
  \end{subfigure}%
  \begin{subfigure}{0.33\linewidth}
    \includegraphics[width=\linewidth]{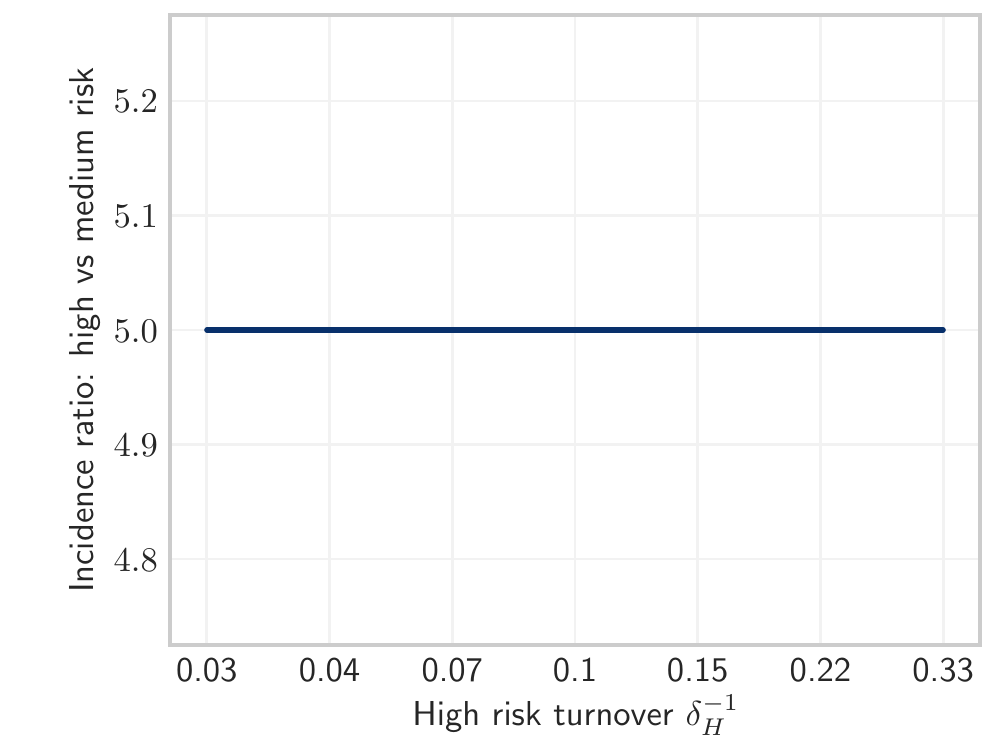}
    \caption{High vs medium risk}
    \label{fig:1d-ratio-incidence-high-med}
  \end{subfigure}%
  \begin{subfigure}{0.33\linewidth}
    \includegraphics[width=\linewidth]{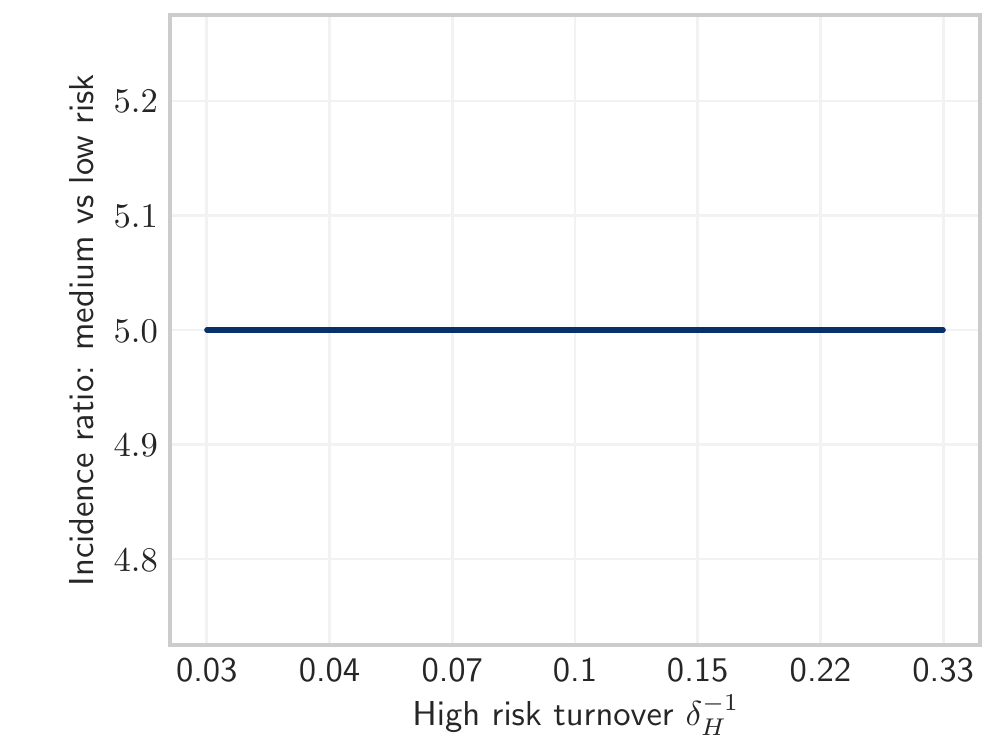}
    \caption{Medium vs low risk}
    \label{fig:1d-ratio-incidence-med-low}
  \end{subfigure}
  \\\endgroup
  \caption{Equilibrium incidence ratios between risk groups
    under different rates of turnover.
    Incidence ratios do not depend on turnover.}
  \label{fig:1d-ratio-incidence}
  \footnotesize\input{x-axis.tex}
\end{figure}
\subsection{Equilibrium prevalence and number of partners before and after model fitting}
\begin{figure}[H]
  \begingroup\centering
  \begin{subfigure}{0.4\linewidth}
    \includegraphics[width=\linewidth]{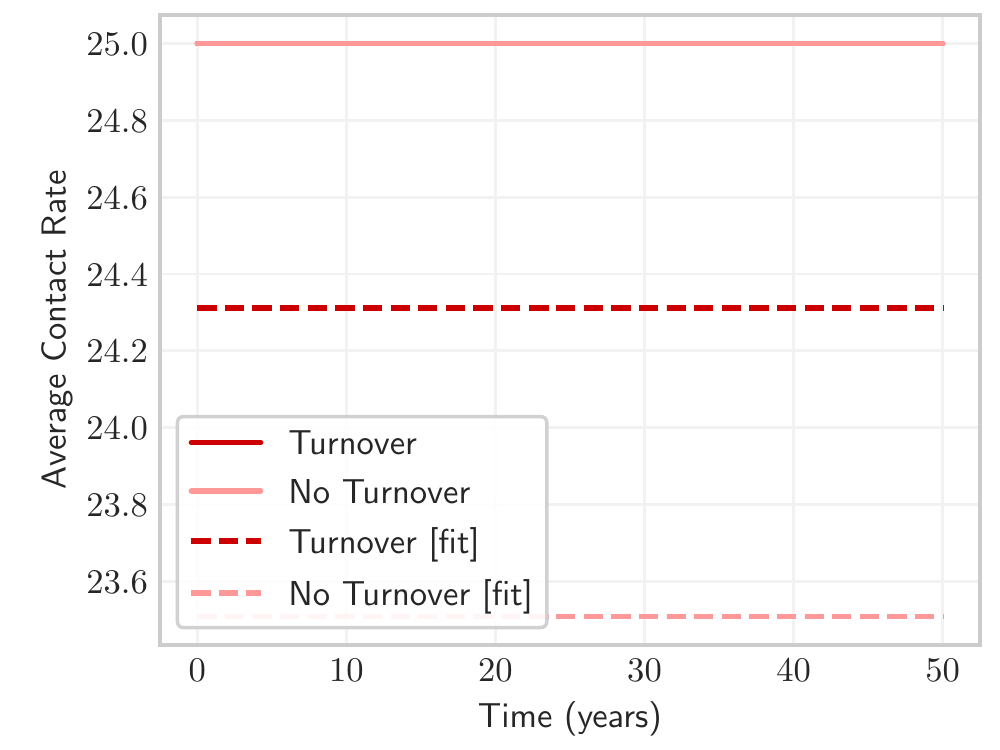}
    \caption{High risk}
    \label{fig:tpaf-C-high}
  \end{subfigure}
  \begin{subfigure}{0.4\linewidth}
    \includegraphics[width=\linewidth]{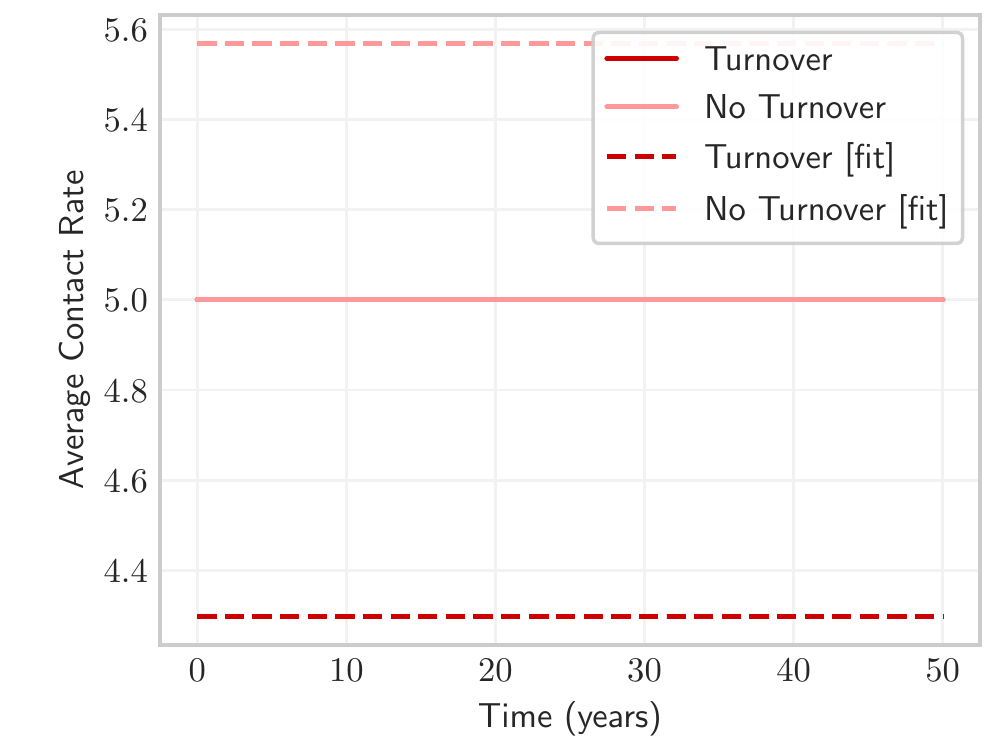}
    \caption{Medium risk}
    \label{fig:tpaf-C-med}
  \end{subfigure}
  \begin{subfigure}{0.4\linewidth}
    \includegraphics[width=\linewidth]{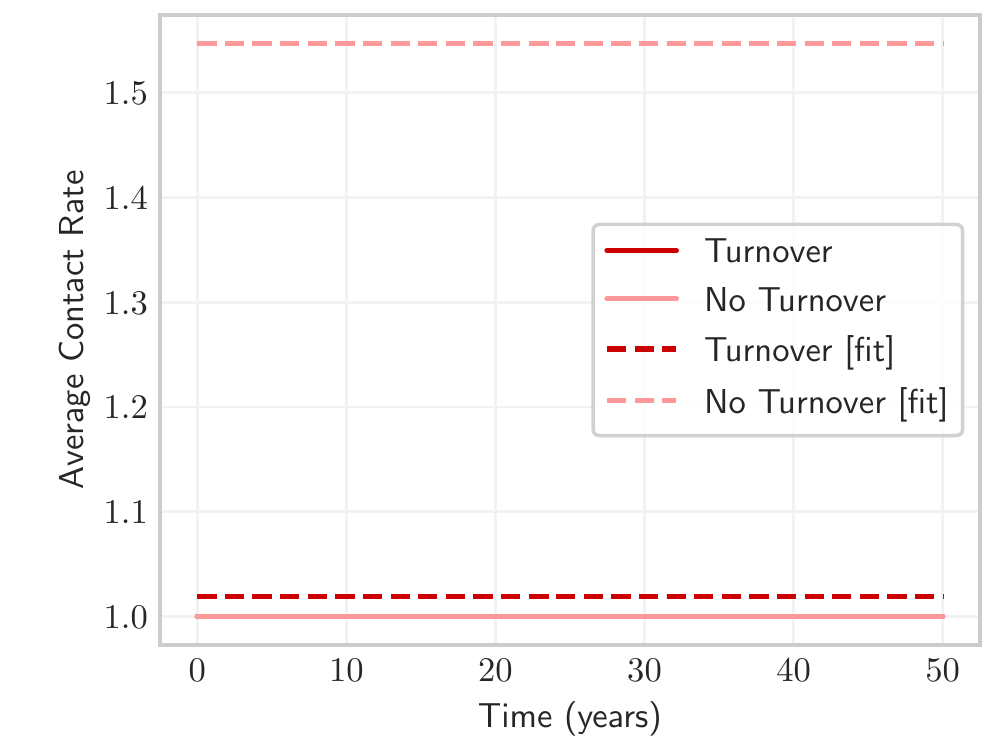}
    \caption{Low risk}
    \label{fig:tpaf-C-low}
  \end{subfigure}
  \begin{subfigure}{0.4\linewidth}
    \includegraphics[width=\linewidth]{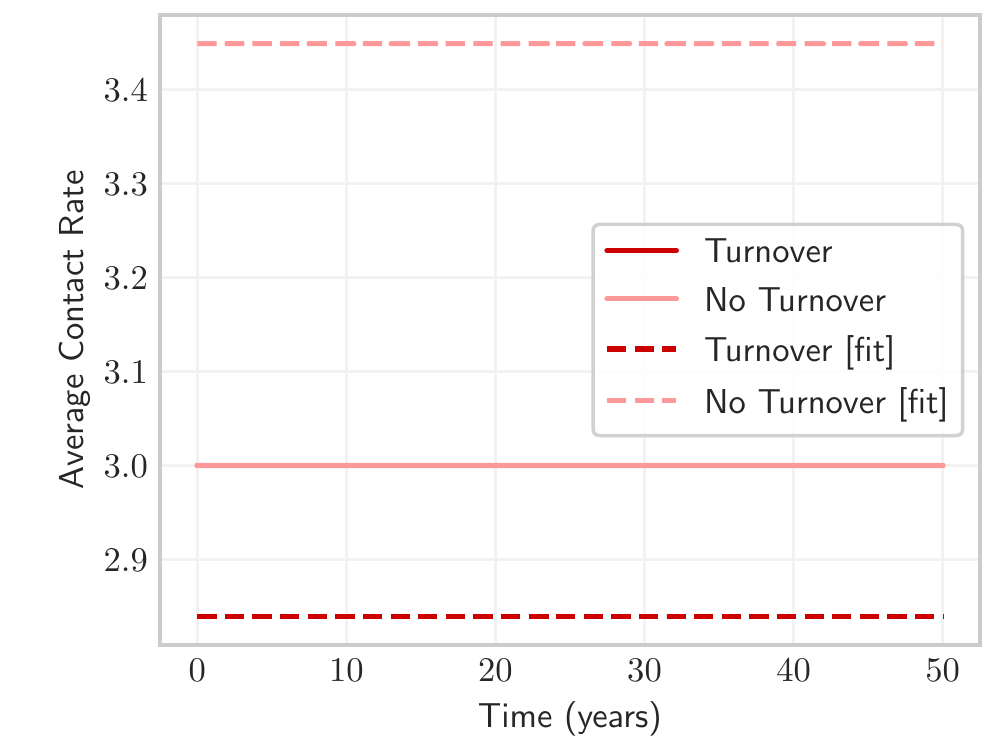}
    \caption{Overall}
    \label{fig:tpaf-C-all}
  \end{subfigure}
  \\\endgroup
  \caption{Numbers of partners $C_i$
    among high, medium, and low risk groups as well as overall,
    with and without turnover,
    and with and without model fitting to group-specific prevalence.}
  \label{fig:tpaf-C}
\end{figure}
\subsection{Influence of turnover on the tPAF of the highest risk group before model fitting}
\begin{figure}[H]
  \centerline{\includegraphics[width=0.5\linewidth]{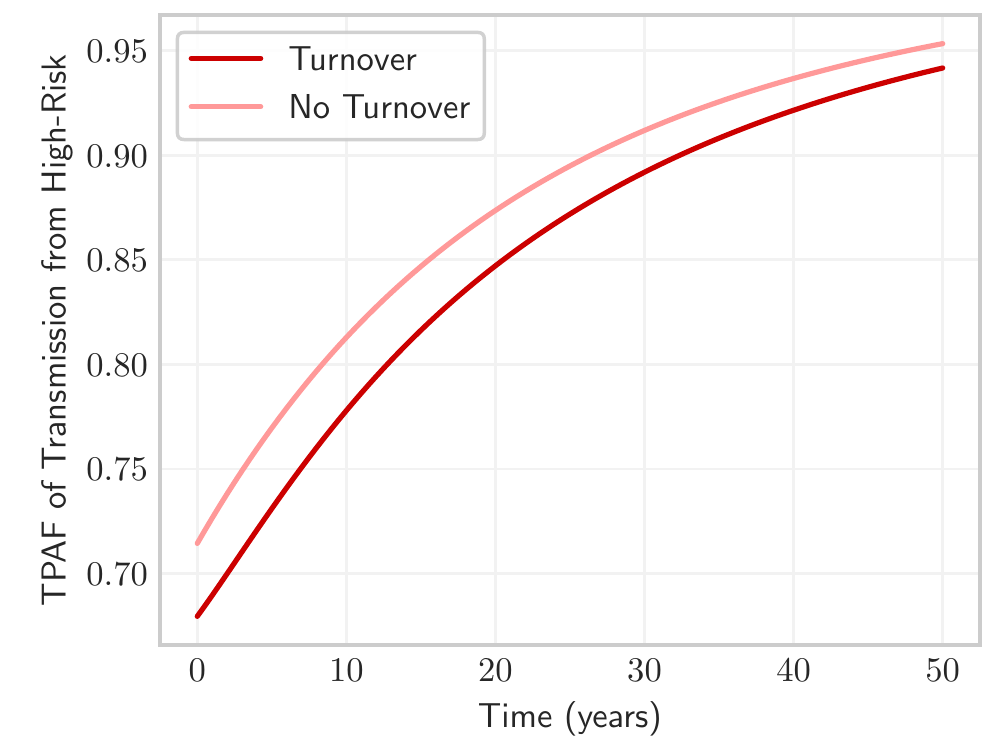}}
  \caption{Transmission population attributable fraction (tPAF)
    of the high risk group in models with and without turnover,
    before model fitting.}
  \label{fig:tpaf-raw}
\end{figure}
\subsection{Effect of treatment rate on the influence of turnover on equilibrium prevalence}
\begin{figure}[H]
  \begin{minipage}{0.5\linewidth}
    \begingroup\centering
    \begin{subfigure}{0.8\linewidth}
      \includegraphics[width=\linewidth]{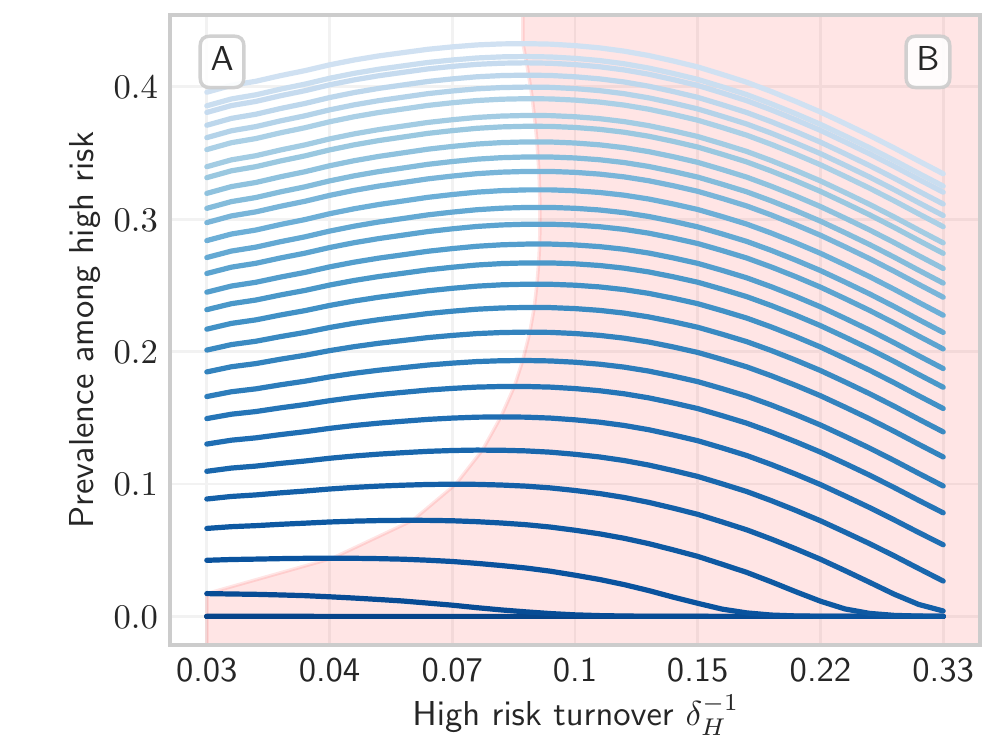}
      \caption{Prevalence among high risk}
      \label{fig:2d-prevalence-high}
    \end{subfigure}\\
    \begin{subfigure}{0.8\linewidth}
      \includegraphics[width=\linewidth]{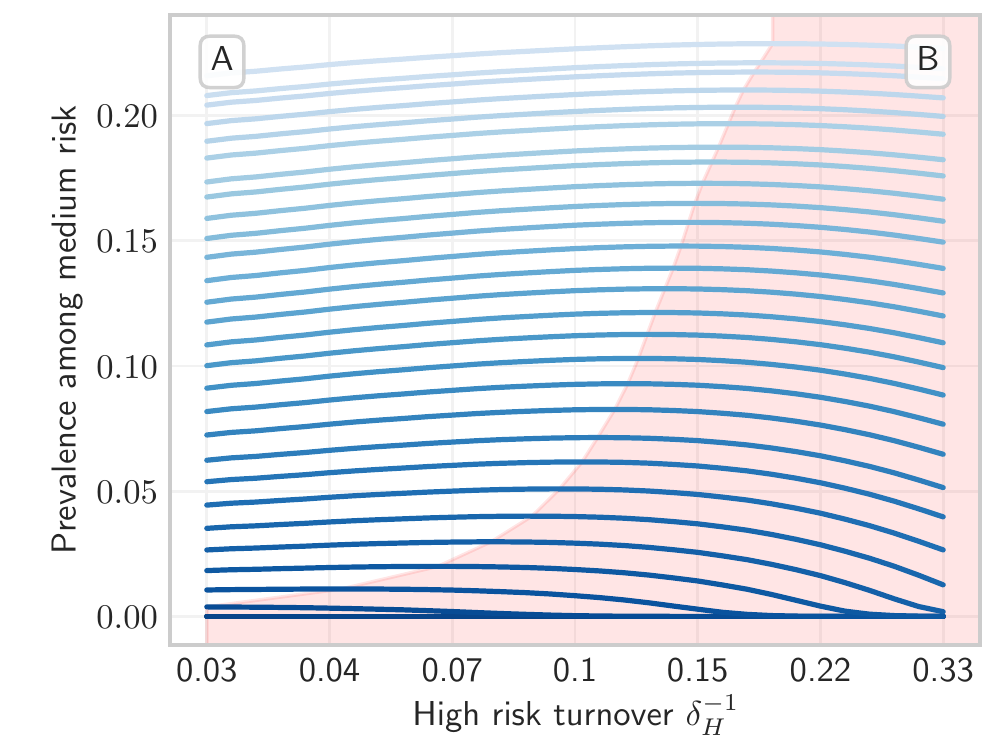}
      \caption{Prevalence among medium risk}
      \label{fig:2d-prevalence-med}
    \end{subfigure}\\
    \begin{subfigure}{0.8\linewidth}
      \includegraphics[width=\linewidth]{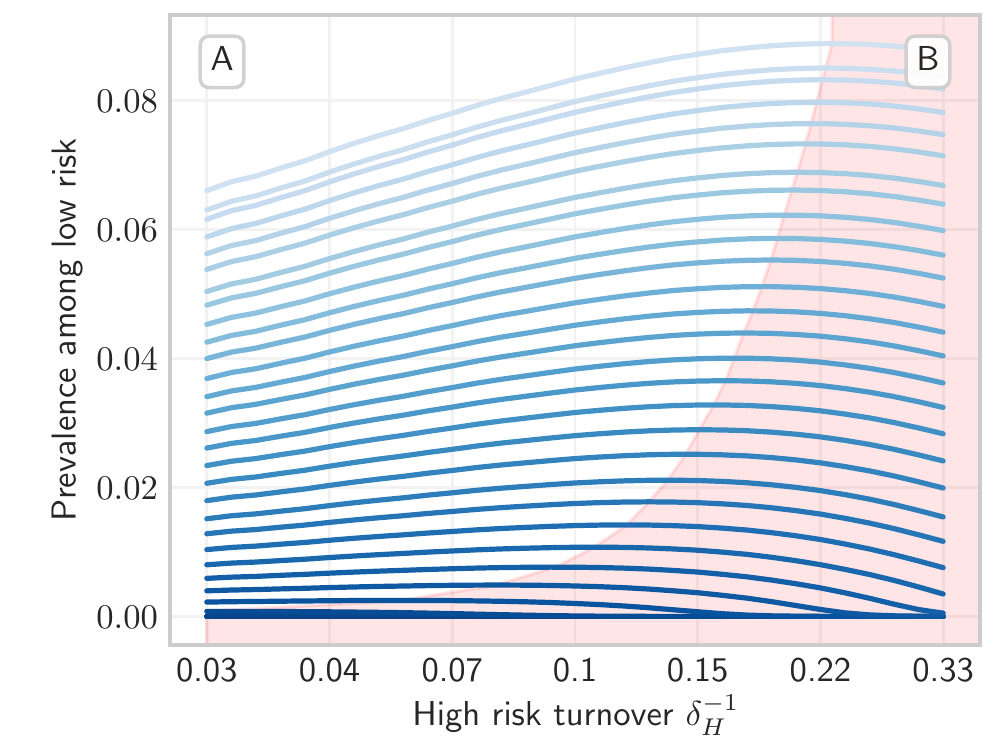}
      \caption{Prevalence among low risk}
      \label{fig:2d-prevalence-low}
    \end{subfigure}\\
    \endgroup
  \end{minipage}%
  \begin{minipage}{0.5\linewidth}
    \begingroup\centering
    \begin{subfigure}[t]{0.8\linewidth}
      \includegraphics[width=\linewidth]{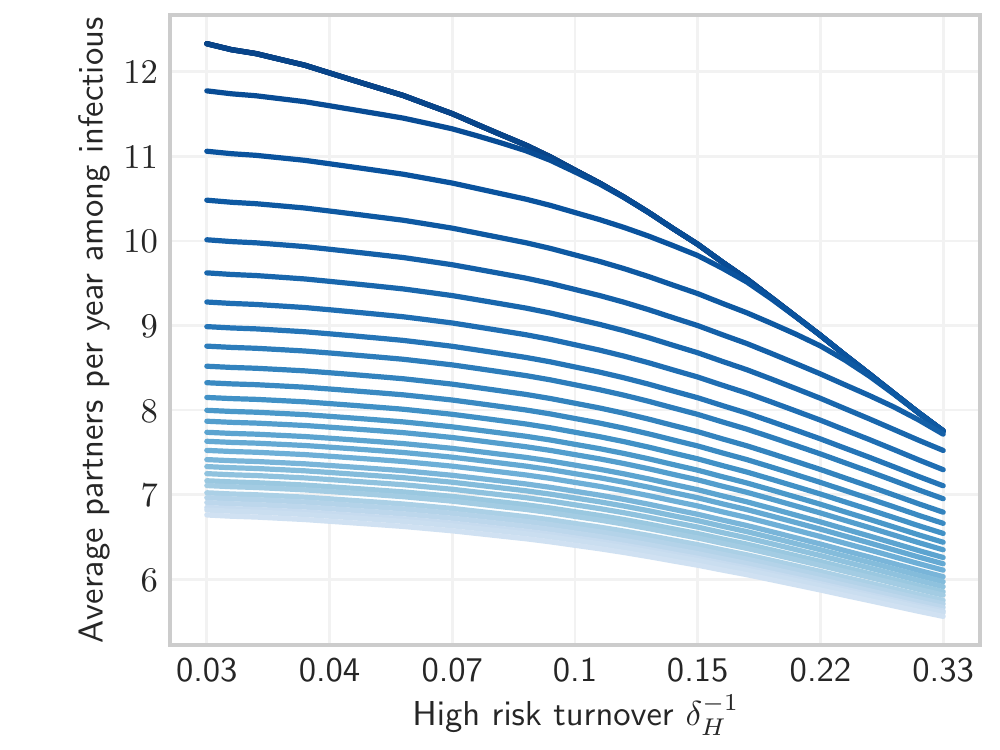}
      \caption{Average $C$ among infectious individuals $\hat{C}_{\mathcal{I}}$}
      \label{fig:2d-C-I}
    \end{subfigure}\\
    \begin{subfigure}[t]{0.8\linewidth}
      \includegraphics[width=\linewidth]{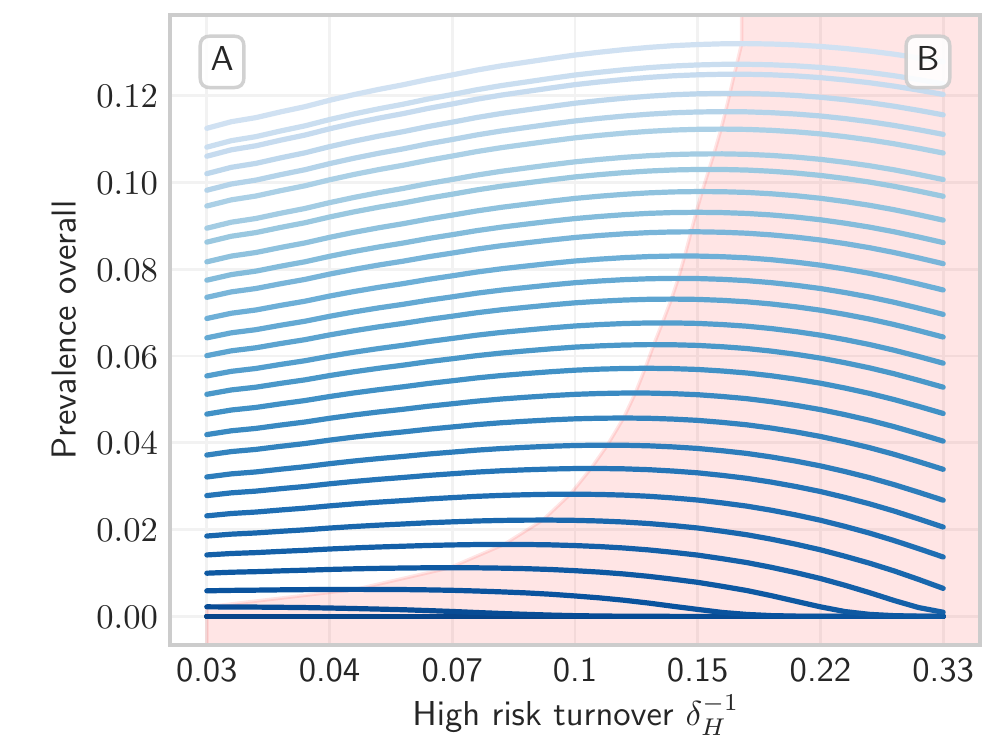}
      \caption{Prevalence overall}
      \label{fig:2d-prev-all}
    \end{subfigure}\\
    \begin{subfigure}[t]{0.8\linewidth}
      \includegraphics[width=\linewidth]{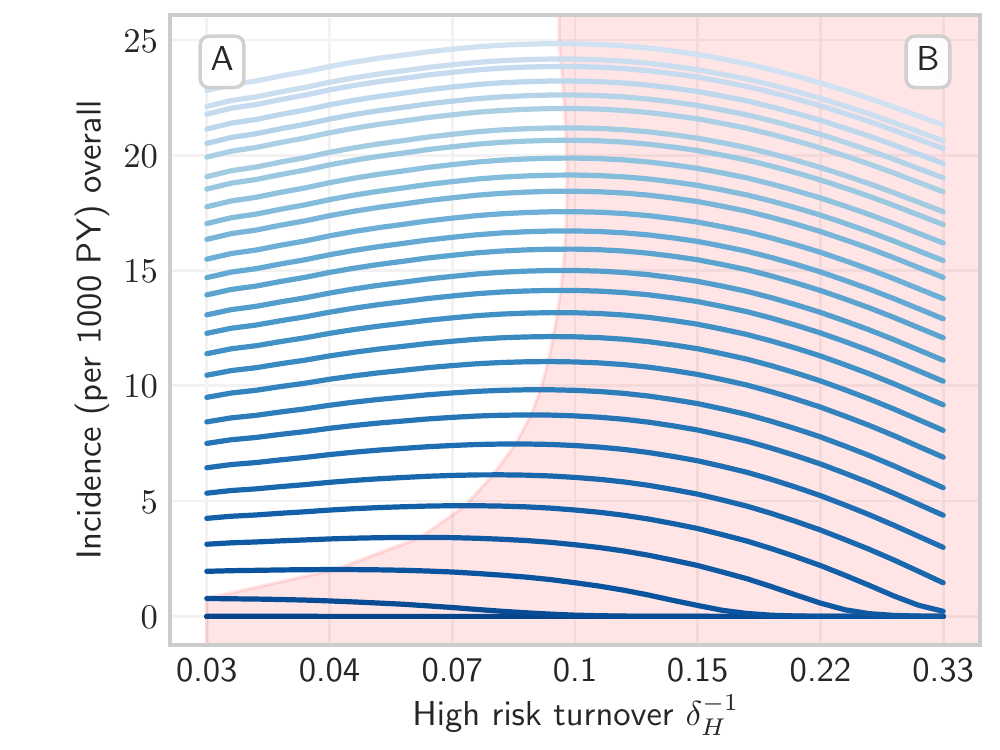}
      \caption{Incidence overall $\lambda$}
      \label{fig:2d-incidence-all}
    \end{subfigure}\\
    \endgroup
  \end{minipage}
  \caption{Relationship between turnover rate and
    equilibrium STI prevalence in high, medium, and low risk groups,
    as well as overall STI prevalence and incidence,
    and average $C$ among infectious individuals,
    for a range of treatment rates $\tau$.
    Darker blue indicates higher treatment rate.
    The threshold turnover rate separating regions~A~and~B
    decreases with treatment rate, meaning that
    increasing turnover becomes more likely to decrease equilibrium prevalence
    as treatment rate increases.}
  \label{fig:2d}
  \footnotesize\input{x-axis.tex}
\end{figure}

In order to examine the effect of treatment rate $\tau$ on
the results of Experiment~1
-- the influence of turnover on equilibrium prevalence --
we recreated Figures~\ref{fig:prevalence}~and~\ref{fig:incidence-factors}
for a range of treatment rates $\tau \in [0.05, 1.0]$.
The results are shown in Figure~\ref{fig:2d}.
\clearpage
\end{document}